\documentclass[11pt, a4paper]{article}
\pdfoutput=1 

\usepackage[height=8.85in,width=6.6in]{geometry}

\usepackage{graphicx,rotating}     
\usepackage[bookmarksopen,colorlinks=true,linkcolor=light_blue,
citecolor=light_pink,urlcolor=dark_red,linktoc=all]{hyperref}

\usepackage{amsmath}
\usepackage{mathtools}
\usepackage{amsfonts}
\usepackage{amssymb}
\usepackage{slashed}
\usepackage{braket}
\usepackage{bm}
\usepackage{cite}
\usepackage{comment}
\usepackage{xcolor}
\usepackage[utf8]{inputenc}
\usepackage[footnotesize]{caption}
\usepackage{bbold}

\usepackage{fourier}

\newcommand{\dd}{\mathrm{d}}

\newcommand{\abs}[1]{\left\vert {#1} \right\vert}

\makeatletter
\@addtoreset{equation}{section}

\makeatother

\usepackage{multicol}
\definecolor{dark_red}{rgb}{0.7, 0., 0.}
\definecolor{light_pink}{rgb}{1,0.4,0.4}
\definecolor{light_blue}{rgb}{0.284602,0.317763,0.963947}
\definecolor{cred}{RGB}{180,50,40} 
\definecolor{darkgreen}{RGB}{0, 100, 0}
\definecolor{desy_blue}{HTML}{009EE2}
\definecolor{desy_orange}{HTML}{FD8800}
\definecolor{forestgreen}{HTML}{228B22}
\definecolor{ochre}{HTML}{CCAA2B}

\begin{document}

\hypersetup{pageanchor=false}
\begin{titlepage}

\begin{center}

\hfill DESY 21-006\\
\hfill CERN-TH-2021-010 \\

\vskip 1.in

{\Huge \bfseries 
Axion assisted Schwinger effect} \\
\vskip .8in

{\Large Valerie Domcke$^{a,b}$, Yohei Ema$^c$, Kyohei Mukaida$^a$}

\vskip .3in
\begin{tabular}{ll}
$^a$& \!\!\!\!\!\emph{Theoretical Physics Department, CERN, 1211 Geneva 23, Switzerland}\\
$^b$& \!\!\!\!\!\emph{Laboratory for Particle Physics and Cosmology, Institute of Physics, }\\[-.3em]
& \!\!\!\!\!\emph{School of Basic Sciences, EPFL, 1015 Lausanne, Switzerland}\\ 
$^c$& \!\!\!\!\!\emph{DESY, Notkestra{\ss}e 85, D-22607 Hamburg, Germany}
\end{tabular}

\end{center}
\vskip .6in

\begin{abstract}
\noindent
We point out an enhancement of the pair production rate of charged fermions in a strong electric field in the presence of time dependent classical axion-like background field, which we call \textit{axion assisted Schwinger effect}. While the standard Schwinger production rate is proportional to  $\exp(- \pi (m^2 + p_T^2)/E)$, with $m$ and $p_T$ denoting the fermion mass and its momentum transverse to the electric field $E$, the axion assisted Schwinger effect  can be enhanced at large momenta to $\exp(- \pi m^2/E)$. The origin of this enhancement is a coupling between the fermion spin and its momentum, induced by the axion velocity. 
As a non-trivial validation of our result, we 
show its invariance under field redefinitions associated with a chiral rotation and
successfully reproduce the chiral anomaly equation in the presence of helical electric and magnetic fields. We comment on implications of this result for axion cosmology, focussing on axion inflation and axion dark matter detection. 
\end{abstract}

\end{titlepage}

\tableofcontents
\thispagestyle{empty}
\renewcommand{\thepage}{\arabic{page}}
\renewcommand{\thefootnote}{$\natural$\arabic{footnote}}
\setcounter{footnote}{0}
\newpage
\hypersetup{pageanchor=true}

\section{Introduction}
\label{sec:intro}

Schwinger production~\cite{Heisenberg:1935qt,Schwinger:1951nm} describes the quantum mechanical, non-perturbative production of pairs of particles and antiparticles in a strong electric field. The production rate is exponentially suppressed by the mass gap in the dispersion relation, 
\begin{align}
 \text{particle production} \sim \exp\left( - \frac{\pi \left(m^2 + p_T^2\right)}{g |Q| E}\right)\,,
\end{align}
for a particle of mass $m$ and charge $Q$, traveling with a momentum $p_T$ transverse to an electric field with magnitude $E$ and gauge coupling $g$. Not surprisingly, the spontaneous production of particles with large momenta $p_T$ is exponentially suppressed.

In this paper, we study the production of charged particles in a strong electric field in the presence of a time-dependent homogeneous background pseudoscalar field $\phi$, with $\nabla \phi = 0$ and $\dot \phi \neq 0$. In analogy to the QCD axion~\cite{Peccei:1977hh,Peccei:1977ur,Weinberg:1977ma,Wilczek:1977pj}, we will refer to $\phi$ as axion-like particle, or axion for short. 
In the presence of such an axion background field, a chiral current coupling (or equivalently a pseudoscalar coupling) induces an additional spin precession operator in the non-relativistic Hamiltonian  of a fermion $\psi$~\cite{Pospelov:2008jk}:\footnote{
For a discussion of the spatial components of this operator, referred to as `axion wind', see \textit{e.g.}, Refs.~\cite{Stadnik:2013raa,Graham:2013gfa}.
}
\begin{align}
	\frac{ \partial_\mu \phi }{f_a} \bar \psi \gamma^\mu \gamma_5 \psi
	\quad \text{or} \quad
	\frac{\phi}{f_a} m \bar \psi i \gamma_5 \psi
	\quad \longrightarrow \quad
	\frac{\dot \phi}{f_a} \frac{\bm{\Pi} \cdot \bm{\sigma}}{m}\,.
	\label{eq:operators}
\end{align}
where the gauge invariant momentum for a fermion is defined as $\bm{\Pi} = \bm{p} - g Q \bm{A}$ with $\bm{A}$ denoting the vector potential,
and $\bm\sigma$ denoting the Pauli matrices.
This spin precession leads to an additional contribution to the fermion energy, with a sign depending on the sign of the axion velocity. For one of the fermion modes, it reduces the mass gap in the dispersion relation, cancelling the contribution from the transverse momentum for suitable values of $\dot \phi$. Correspondingly, the particle production rate is exponentially enhanced for $p_T \neq 0$,\footnote{
The maximal particle production rate given in Eq.~\eqref{eq:maxPP} occurs for constructive interference of the two saddle points of the adiabatic approximation, the full expression for the rate oscillates as a function of $\dot \phi$ as discussed in the main text.
}
\begin{align}
 \text{maximal particle production} \sim  \exp\left( - \frac{\pi m^2}{g |Q| E}\right) \quad \quad \text{for} \quad  \dot \phi^2/f_a^2 \gtrsim \frac{\pi m^2 p_T^2}{g |Q| E}\,.
 \label{eq:maxPP}
\end{align}
Somewhat counter-intuitively, for sufficiently large axion velocities, this \textit{axion assisted Schwinger effect} is thus not exponentially suppressed at large momenta, implying a fundamentally different resulting fermion spectrum. 
This is the main result of the present paper.

To obtain this result, we solve the Dirac equation in the presence of an electric background field, including the most general shift-symmetric dimension five couplings of an axion-like field to a fermion and gauge field that do not  explicitly break $CP$. The axion velocity and the electric field imply a time-dependent background for the fermions, leading to the particle production described by time-dependent Bogoliubov coefficients. We verify the intuition behind key aspects of our computation using the non-relativistic effective field theory of fermions. Our results are intrinsically invariant under (chiral) fermion rotations, reflecting the basis invariance of all observables.

The non-vanishing axion velocity moreover leads to a magnetic field parallel to the electric field, \textit{i.e.}, a helical gauge field configuration, if the axion couples to the electromagnetic fields.
This is, in particular, relevant if we identify the axion $\phi$ as the inflaton particle driving cosmic inflation~\cite{Turner:1987bw,Garretson:1992vt,Anber:2006xt}.
Motivated by this, we also study the axion assisted Schwinger effect 
in a background of parallel electric and magnetic fields.
In this case, the fermion dispersion relation is quantized in terms of the Landau levels~\cite{Nikishov:1969tt,Bunkin:1969if}, each corresponding to a particular transverse momentum. 
We confirm that the enhancement mechanism applies also in this case. 
The helical gauge field background sources a chiral asymmetry as indicated by the chiral anomaly~\cite{Adler:1969gk,Bell:1969ts}.
As a non-trivial consistency check, we reproduce the chiral anomaly equation from our result.

Our analysis builds on earlier studies of particle production in (helical) electromagnetic fields, reproducing the chiral anomaly equation and the Schwinger production rate~\cite{Nielsen:1983rb,Warringa:2012bq,Domcke:2018eki,Copinger:2018ftr,Domcke:2019qmm}. This paper extends these analyses by including a dynamical axion field with general couplings to the fermions and gauge fields, including in particular both operators in Eq.~\eqref{eq:operators}. The corresponding results of Refs.~\cite{Nielsen:1983rb,Warringa:2012bq,Domcke:2018eki,Copinger:2018ftr,Domcke:2019qmm} are obtained as a particular limit of the more general results presented here. Moreover our analysis shares some similarities with the so-called `dynamically assisted Schwinger mechanism'~\cite{Schutzhold:2008pz,Dumlu:2010ua}, which describes an enhancement of the Schwinger pair production rate in electric fields with a non-trivial time dependence.

The remainder of this paper is organized as follows. In Sec.~\ref{sec:woB}, we compute the particle production due to the axion assisted Schwinger effect in the presence of an electric field. We generalize this analysis in Sec.~\ref{sec:wB} to include an (anti-)parallel magnetic field, in particular verifying the chiral anomaly equation and providing an estimate for the induced fermion current. We comment on implications for axion cosmology, in particular on axion inflation and on axion dark matter detection in Sec.~\ref{sec:applications} before concluding in Sec.~\ref{sec:conclusions}. We have relegated several technical but important details to six appendices.  App.~\ref{app:notation} fixes our notation and conventions. Apps.~\ref{app:particles} and \ref{app:particlesB} provide the details on deriving the equations of motion for the Bogoliubov coefficients used in Secs.~\ref{sec:woB} and \ref{sec:wB}, respectively. App.~\ref{app:bilinear} is dedicated to our analytical result for particle production due to the axion assisted Schwinger effect. 
Our results are interpreted in the language of non-relativistic effective field theory in App.~\ref{app:nreft}.
Finally, we briefly review the WKB estimate of Schwinger particle production in App.~\ref{app:WKB}.

\section{Axion assisted Schwinger effect in an electric field}
\label{sec:woB}

\subsection{Dirac equation with classical background fields}
\label{sec:dirac}

Let us consider an axion-like particle $\phi$, coupled through shift-symmetric dimension five operators to an Abelian gauge boson $A_\mu$ and a Dirac fermion $\psi$. Our goal is to study non-perturbative fermion production by explicitly solving the Dirac equation,
\begin{align}
	\left[i \slashed{D} - m \, e^{2i c_m  \gamma_5 \phi/f_a} + c_5 \frac{\partial_\mu \phi}{f_a}
	\gamma^\mu \gamma_5\right] \psi = 0 \,,
	\label{eq:dirac}
\end{align}
in a classical background of a homogeneous axion field (with $\dot \phi \neq 0$) in the presence of a strong electric field.
Here $\slashed{D} = (\partial_\mu + i g Q A_\mu) \gamma^\mu$ denotes the covariant derivative, $m$ the conformal fermion mass\footnote{In an FRW background with scale factor $a(t)$, the conformal mass $m$ is related to the physical mass $m_\text{ph}$ as $m = m_\text{ph} a$.}, $g$ the gauge coupling constant, $Q$ the fermion charge, $f_a$ the axion decay constant, and $c_m$,$c_5$ are dimensionless couplings constants.
We have further assumed a Friedmann-Lema\^itre-Robertson (FRW) background metric, and Eq.~\eqref{eq:dirac} is expressed in conformal coordinates, implying that all contractions of Lorentz indices are taken with respect to the flat metric $\eta_{\mu \nu}$. For later convenience we introduce
\begin{align}
 \theta_m = c_m \phi/f_a  \quad \text{and} \quad \theta_5 = c_5 \phi/f_a  \,,
 \label{eq:thetas}
\end{align}
see App.~\ref{app:notation} for more details on our notation and conventions.

Without loss of generality, we choose the direction of the electric field along the $z$-axis, \textit{i.e.}, we work in a classical background field configuration described by
\begin{align}
	A^{\mu} = \left(0, 0, 0, A_z(t)\right),
	\qquad
	\phi = \phi(t),
\end{align}
where the electric field is given by $E = -\dot{A}_z \geq 0$. 
Introducing the generalized momentum $\Pi_z = p_z - g Q A_z$,
we can write the equation of motion~\eqref{eq:dirac} in Fourier space as
\begin{align}
	0 = \begin{pmatrix}
	i \partial_0 + \Pi_z - \dot \theta_5  & p_x - ip_y & -m e^{2i\theta_{m}} & 0 \\
	p_x + ip_y & i \partial_0 - \Pi_z - \dot \theta_5 & 0 & -m e^{2i\theta_{m}} \\
	- m e^{-2i\theta_{m}} & 0 & i\partial_0 - \Pi_z + \dot \theta_5 & -(p_x - ip_y) \\
	0 & -m e^{-2i\theta_{m}} & -(p_x + ip_y) & i\partial_0 + \Pi_z + \dot \theta_5
	\end{pmatrix}
\psi \,.
	\label{eq:eom}
\end{align}
We now decompose the fermion modes as 
\begin{align}
 \psi  =  e^{i \gamma_5 \theta_5} \sum_{\lambda = 1, 2} 
	\left[
	\alpha_\lambda u_\lambda \exp\left(-i \Omega t\right)
	+ \beta_\lambda v_\lambda \exp\left(+i \Omega t\right)
	\right] \,, \label{eq:psi2}
\end{align}
with $u_\lambda$ ($v_\lambda$) denoting the (anti-)particle eigenfunctions of the fermionic part of the Hamiltonian for constant $A_z$ and $\phi$, respectively, see App.~\ref{app:particles} for details.
Inserting this into Eq.~\eqref{eq:eom} yields the dispersion relation
\begin{align}
 \Omega = \sqrt{\Pi_z^2 + m_T^2}\,,
 \label{eq:energy-levels}
\end{align}
with the effective mass $m_T = \sqrt{p_x^2 + p_y^2 + m^2}$ now including the transverse momentum.
Note that, although the dispersion relation~\eqref{eq:energy-levels} is independent of $\theta_5$ and $\theta_m$, the eigenvectors $u_\lambda$, $v_\lambda$ depend on $\theta_{5 + m} \equiv \theta_5 + \theta_m$. The appearance of only this linear combination can be traced back to the symmetry of our setup under chiral fermion rotations.
This in particular implies that in the limit $m \rightarrow 0$ any dependence on $\theta_{5}$ and $\theta_m$ must drop out, since it can be removed from Eq.~\eqref{eq:dirac} by a suitable fermion rotation.

\paragraph{Bogoliubov coefficients.} We now turn on the time-dependence of $A_z$ and ${\phi}$, thus promoting the  coefficients $\alpha_\lambda$ and $\beta_\lambda$ to time dependent Bogoliubov coefficients.
Then Eq.~\eqref{eq:eom} requires that
\begin{align}
	0 = \sum_{\lambda = 1,2} 
	\left[ 
	\left(\dot{\alpha}_\lambda u_\lambda + \alpha_\lambda \dot{u}_\lambda\right)
	\exp\left(-i \int \dd t \, \Omega\right)
	+ \left(\dot{\beta}_\lambda v_\lambda + \beta_\lambda \dot{v}_\lambda\right)
	\exp\left(i \int \dd t \, \Omega\right)
	\right]\,.
\end{align}
By exploiting the orthogonality among the eigenvectors, we obtain
\begin{align}
	\dot{\alpha}_\lambda &=
	-\sum_{\lambda' = 1,2}
	\left\{
	u^\dagger_\lambda \dot{u}_{\lambda'} \alpha_{\lambda'}
	+
	u^\dagger_\lambda \dot{v}_{\lambda'} e^{2i\int \dd t \, \Omega} \beta_{\lambda'}
	\right\}\,,
	\label{eq:eom_alpha} \\
	\dot{\beta}_\lambda &=
	-\sum_{\lambda' = 1,2}
	\left\{
	v^\dagger_\lambda \dot{u}_{\lambda'} e^{-2i\int \dd t \, \Omega} \alpha_{\lambda'}
	+
	v^\dagger_\lambda \dot{v}_{\lambda'} \beta_{\lambda'}
	\right\}\,.
	\label{eq:eom_beta}
\end{align}
Using the relations for the scalar products of the eigenvectors $u_\lambda$ and $v_\lambda$ and their time derivatives given in App.~\ref{app:particles}, we obtain the equations of motion for the  Bogoliubov coefficients,
\begin{align}
	\begin{pmatrix}
	\dot{\alpha}_1 \\
	\dot{\alpha}_2 \\
	\dot{\beta}_1 \\
	\dot{\beta}_2 
	\end{pmatrix}
	=&
	\left[
	i\dot{\theta}_{5+m}
	\begin{pmatrix}
	- \frac{m}{m_T}\frac{\Pi_z}{\Omega} & \frac{p_T}{m_T} & \frac{m}{\Omega}e^{2i\Theta} & 0 \\
	\frac{p_T}{m_T} & \frac{m}{m_T}\frac{\Pi_z}{\Omega} & 0 & -\frac{m}{\Omega}e^{2i\Theta} \\
	\frac{m}{\Omega}e^{-2i\Theta} & 0 & \frac{m}{m_T}\frac{\Pi_z}{\Omega} & \frac{p_T}{m_T} \\
	0 & -\frac{m}{\Omega}e^{-2i\Theta} & \frac{p_T}{m_T} & -\frac{m}{m_T}\frac{\Pi_z}{\Omega}
	\end{pmatrix}
	+ \frac{m_T \dot{\Pi}_z}{2\Omega^2}
	\begin{pmatrix}
	0 & 0 & -e^{2i\Theta} & 0 \\
	0 & 0 & 0 & -e^{2i\Theta} \\
	e^{-2i\Theta} & 0 & 0 & 0 \\
	0 & e^{-2i\Theta} & 0 & 0
	\end{pmatrix}
	\right]
	\begin{pmatrix}
	{\alpha}_1 \\
	{\alpha}_2 \\
	{\beta}_1 \\
	{\beta}_2 
	\end{pmatrix}\,,
	\label{eq:alphabeta}
\end{align}
where we denote $\Theta = \int \dd t \, \Omega$. Here and henceforth we ignore the time dependence of the scale factor, which is justified as long as particle production rate, driven by the time dependence of $A_z$ and $\phi$, is much larger than the Hubble expansion rate.

\paragraph{Quantization and initial conditions.} In terms of these Bogoliubov coefficients, the quantized fermionic mode function is given by
\begin{align}
	\psi = e^{i \gamma_5 \theta_5} 
	\sum_{\lambda, \lambda' = 1,2} \left[ u_\lambda e^{-i\Theta}
	\left(\alpha_\lambda^{(\lambda')} b_{\lambda'} - 
	\left(-1\right)^{\lambda+\lambda'}{\beta_{\lambda}^{(\lambda')}}^* d_{\lambda'}^\dagger \right)
	+
	v_\lambda e^{i\Theta}
	\left(\beta_\lambda^{(\lambda')} b_{\lambda'} 
	+ \left(-1\right)^{\lambda+\lambda'}{\alpha_{\lambda}^{(\lambda')}}^* d_{\lambda'}^\dagger \right)
	\right]\,,
	\label{eq:quantization_woB}
\end{align}
with $b_\lambda, b_\lambda^\dagger, d_\lambda$ and $d_\lambda^\dagger$ denoting the usual fermionic annihilation and creation operators.
Here the minus signs arise from the observation that $\pm \left(-\beta_1^*, \beta_2^*, \alpha_1^*, -\alpha_2^*\right)^\mathrm{T}$ satisfies the same equation of motion as $\left(\alpha_1, \alpha_2, \beta_1, \beta_2\right)^\mathrm{T}$, and the superscript indicates the initial conditions for $\alpha$ and $\beta$, that is,
\begin{align}
	\alpha_{1}^{(1)} &= 1\,,
	\quad
	\alpha_{2}^{(1)} = \beta_{1}^{(1)} = \beta_{2}^{(1)} = 0 \quad \text{and}
	\quad
	\alpha_{2}^{(2)} = 1\,,
	\quad
	\alpha_{1}^{(2)} = \beta_{1}^{(2)} = \beta_{2}^{(2)} = 0\,, 
	\label{eq:init_cond_woB}
\end{align}
at the initial time $t = -\infty$.
In other words, there are two types of particles that are the remnants of the two helicities,
and we take both particles absent for the initial state.
In our final results we need to sum over both particles since they both contribute to physical quantities.
Note that, due to the degeneracy in the eigenvalue $\Omega$, the labelling of the states $\lambda = \{1,2\}$ is arbitrary, and any two orthogonal linear combinations would lead to the same final results. In particular, (non-perturbative) particle production corresponds to a non-zero value of $\sum_{\lambda, \lambda'} |\beta_{\lambda}^{(\lambda')}|^2$ arising from the time-dependence in the background $A_z$ and $\phi$.

The quantized fermionic mode function~\eqref{eq:quantization_woB} can be equally expressed in terms of the fermionic annihilation and creation operators
$B_\lambda, B_\lambda^\dagger, D_\lambda$ and $D_\lambda^\dagger$ at any given time $t$, defined as
\begin{align}
	B_\lambda &= \sum_{\lambda' = 1, 2}\left(\alpha_\lambda^{(\lambda')} b_{\lambda'} - 
	\left(-1\right)^{\lambda+\lambda'}{\beta_{\lambda}^{(\lambda')}}^* d_{\lambda'}^\dagger \right)\,, \\
	D_\lambda^\dagger &= \sum_{\lambda' = 1, 2}\left(\beta_\lambda^{(\lambda')}b_{\lambda'} + 
	\left(-1\right)^{\lambda + \lambda'}{\alpha_{\lambda}^{(\lambda')}}^*d^\dagger_{\lambda'}\right)\,. 
\end{align}
They satisfy the standard anti-commutation relations,
\begin{align}
	\left\{B_{\lambda_1}(\vec p), B_{\lambda_2}^\dagger(\vec p')\right\} &= \left\{D_{\lambda_1}(\vec p), D_{\lambda_2}^\dagger(\vec p')\right\} = \left(2\pi\right)^3\delta_{\lambda_1 \lambda_2}\delta^{(3)}\left(\vec{p} - \vec{p}'\right)\,, \nonumber \\
	\left\{B_{\lambda_1}(\vec p), D_{\lambda_2}^\dagger(\vec p') \right\} &= \left\{B_{\lambda_1}(\vec p), D_{\lambda_2}(\vec p') \right\} = 0\,,
\end{align}
where here we have made the dependence on the momentum explicit.
These relations indicate that the Bogoliubov coefficients satisfy
\begin{align}
	\sum_\lambda 
	\left[
	\alpha_{\lambda_1}^{(\lambda)}{\alpha_{\lambda_2}^{(\lambda)}}^*
	+ \left(-1\right)^{\lambda_1+\lambda_2}\beta_{\lambda_2}^{(\lambda)}
	{\beta_{\lambda_1}^{(\lambda)}}^* 
	\right]
	&= \delta_{\lambda_1 \lambda_2}\,, 
	\label{eq:conserved_quantity1} \\
	\sum_\lambda 
	\left[
	\alpha_{\lambda_1}^{(\lambda)}{\beta_{\lambda_2}^{(\lambda)}}^*
	- \left(-1\right)^{\lambda_1+\lambda_2}\alpha_{\lambda_2}^{(\lambda)}
	{\beta_{\lambda_1}^{(\lambda)}}^* 
	\right]
	&= 0\,,
	\label{eq:conserved_quantity2}
\end{align}
independently of the time $t$. We show that these relations follow from Eqs.~\eqref{eq:alphabeta} 
and~\eqref{eq:init_cond_woB} in App.~\ref{app:bilinear}.

\subsection{Particle production}
\label{subsec:pp_woB}

\begin{figure}[t]
	\centering
 	\includegraphics[width=0.495\linewidth]{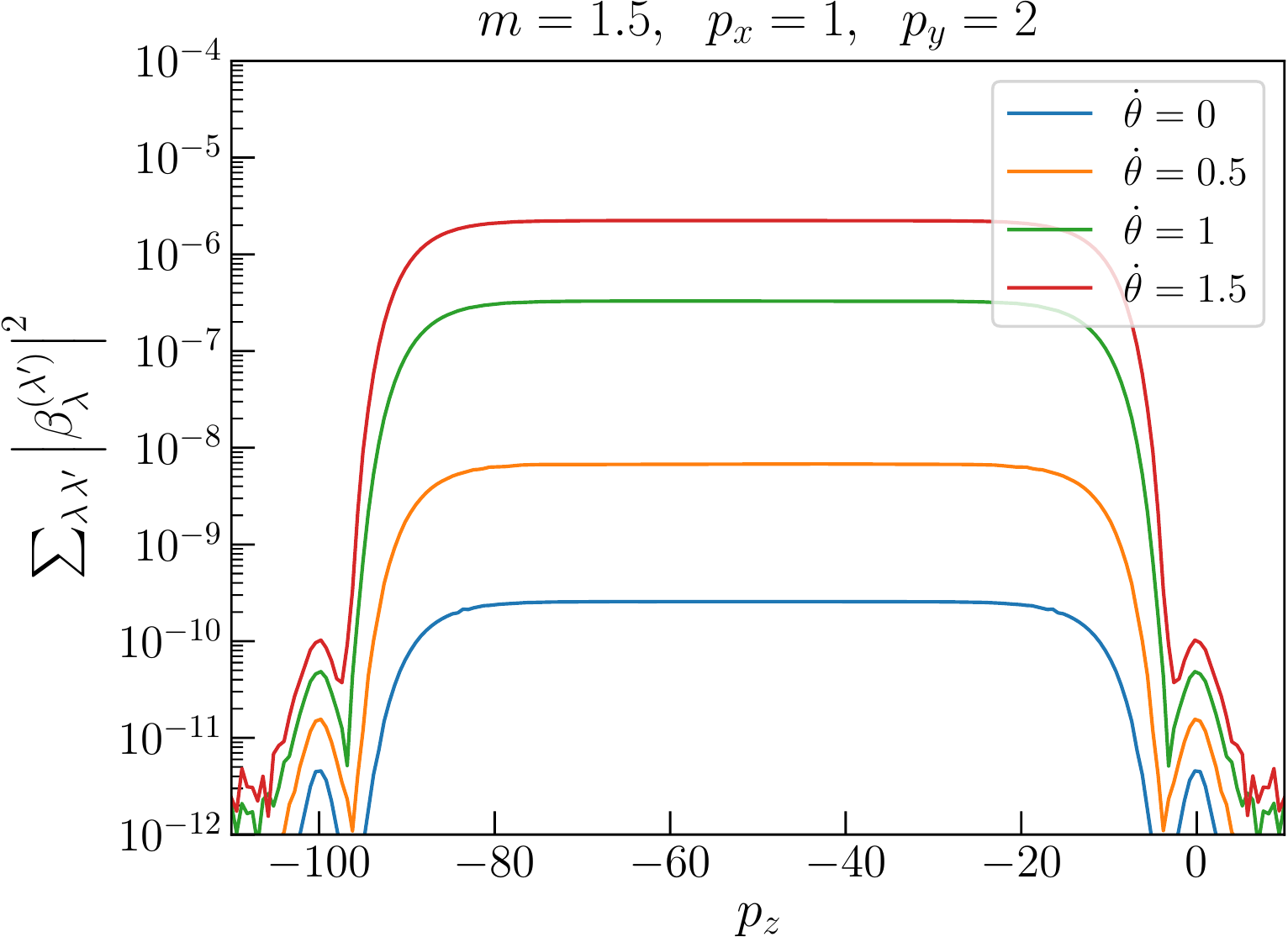}
 	\includegraphics[width=0.495\linewidth]{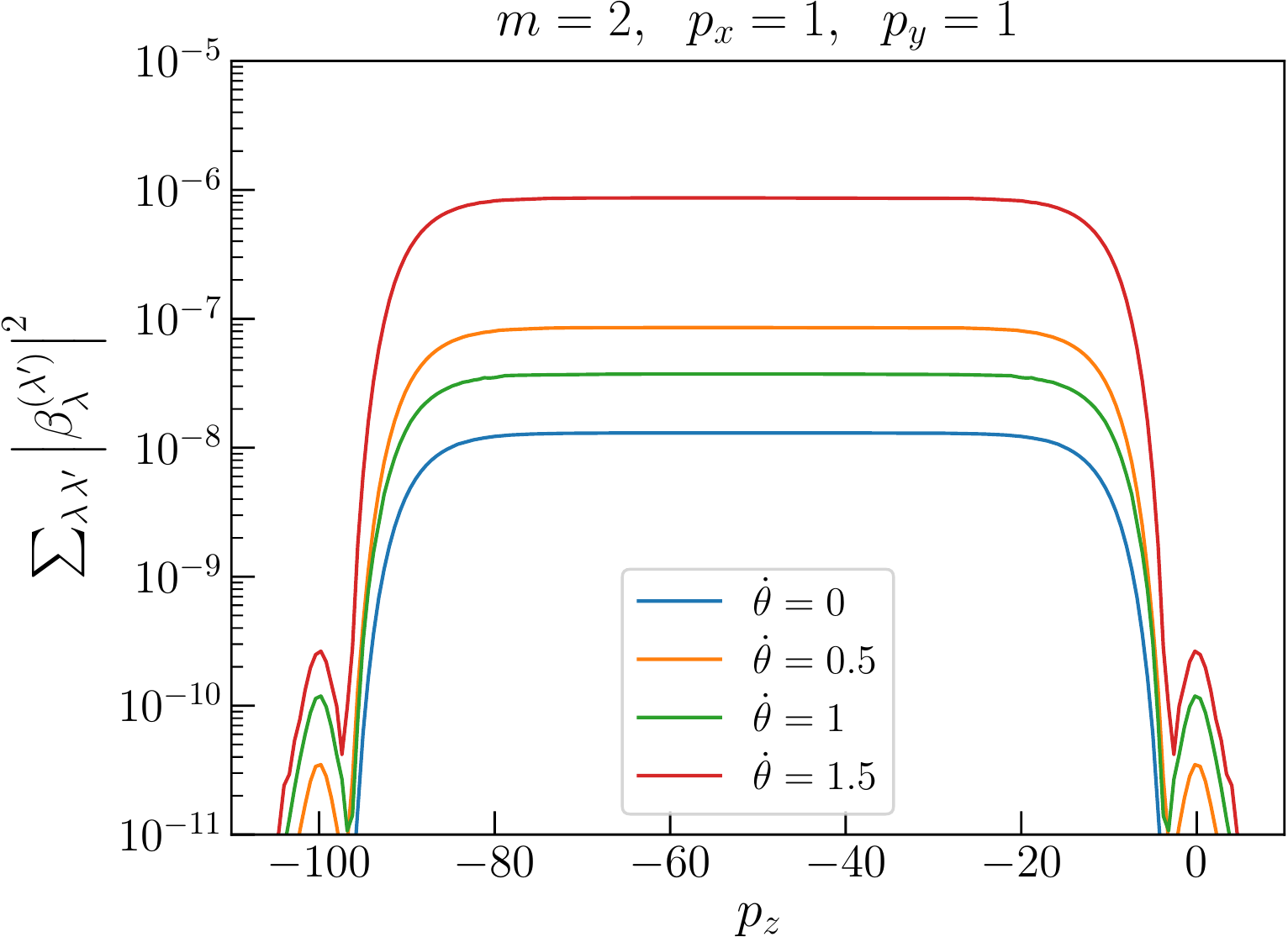}
 	\includegraphics[width=0.495\linewidth]{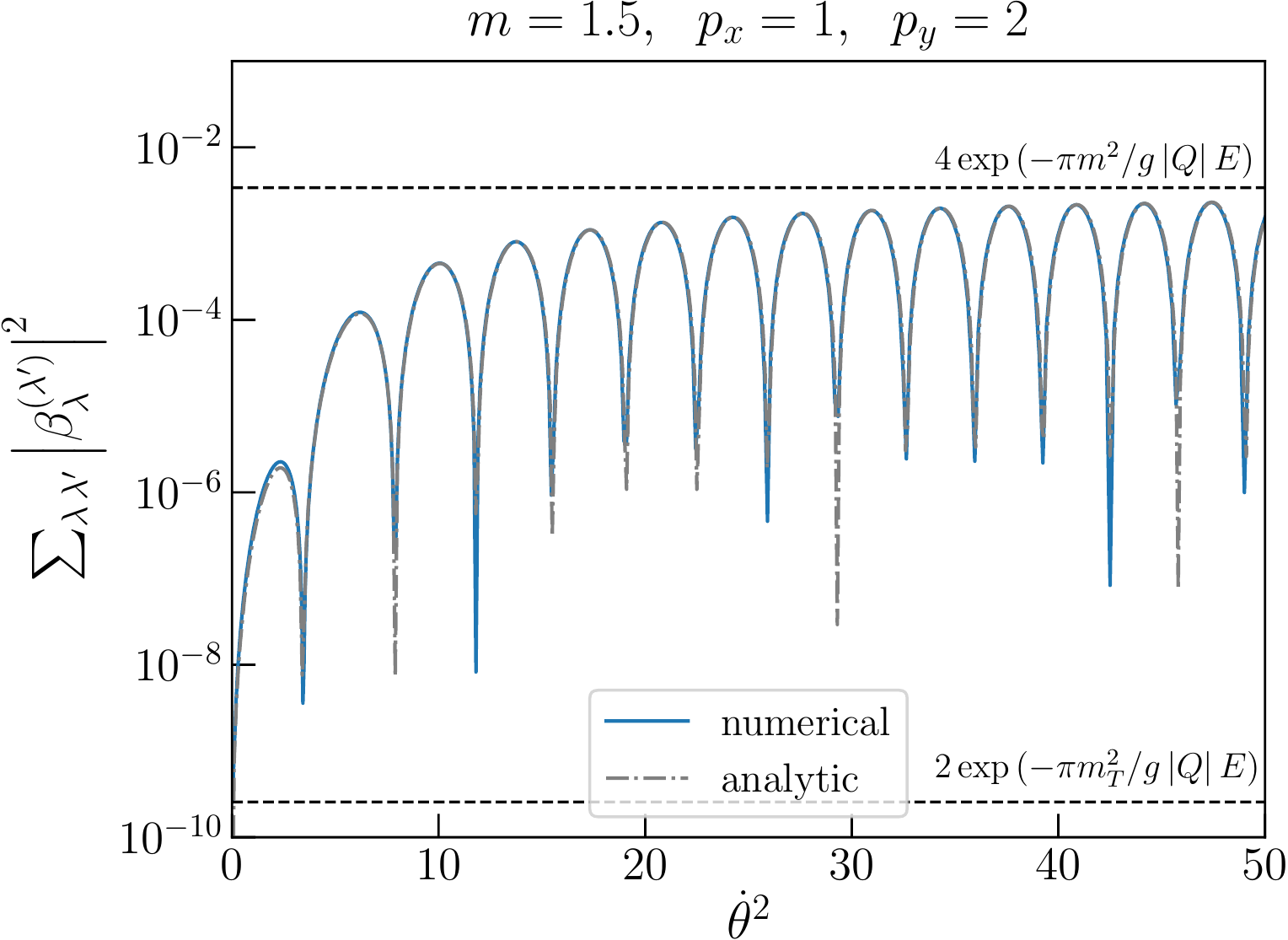}
 	\includegraphics[width=0.495\linewidth]{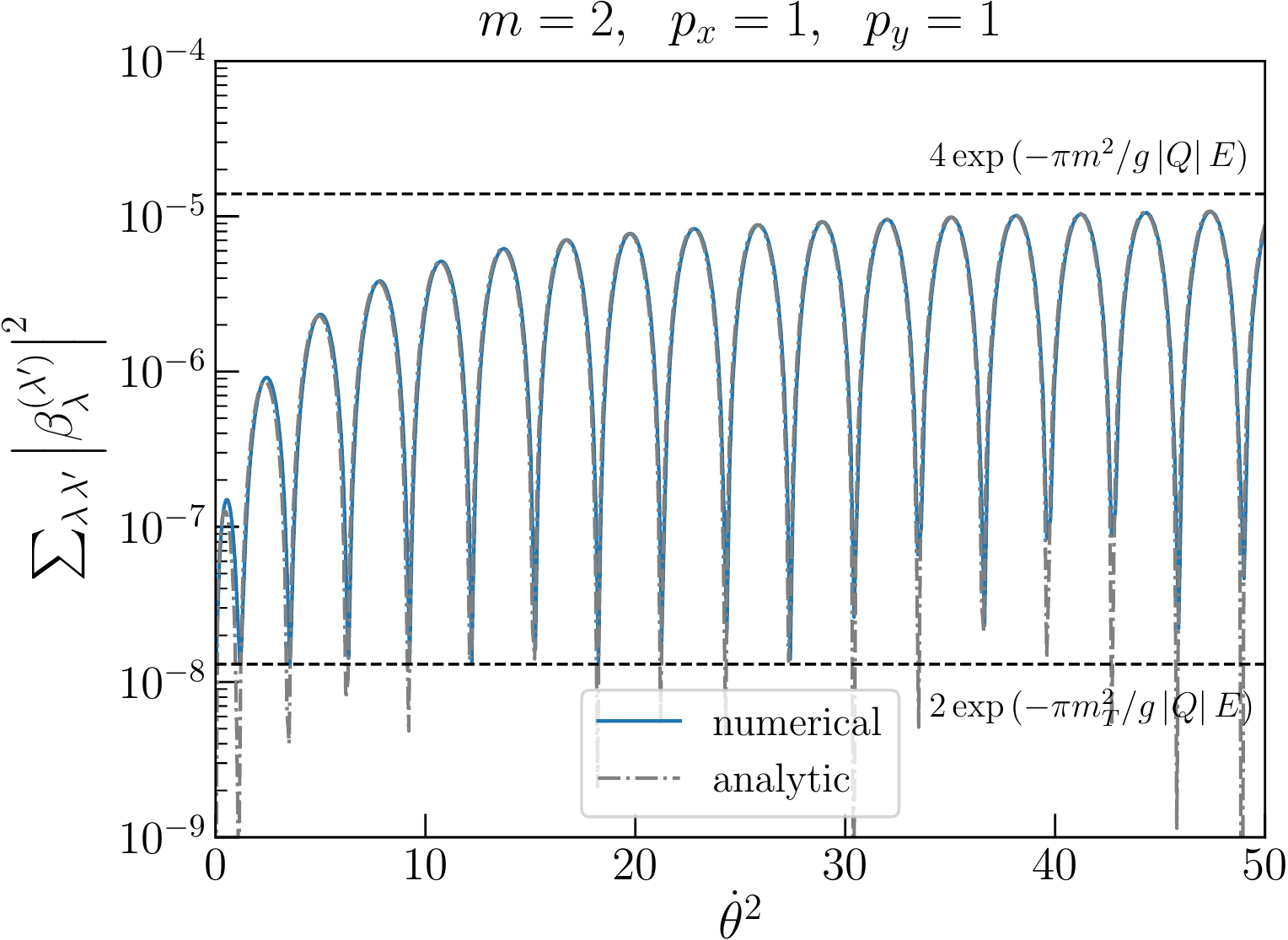}
	\caption{\small
	Upper panels: the spectrum $\sum_{\lambda \lambda'} \abs{\beta_\lambda^{(\lambda')}}^2$
	as a function of $p_z$ for several different values of $\dot{\theta}$.
	Lower panels: the height of the plateau of the spectrum evaluated at $p_z = -50$
	as a function of $\dot{\theta}$.
	The blue solid lines are the full numerical results,
	while the gray dashed lines are the analytical formula~\eqref{eq:saddle_pt_approx}.
	The parameters are shown in the unit $g\abs{Q}E = 1$ and for all figures in the paper we chose $Q > 0$.} 
	\label{fig:woB}
\end{figure}

For concreteness, let us consider that $\dot \phi$ and $\dot A_z$ are turned on and off adiabatically at $t_\text{min}$ and $t_\text{max}$ respectively, \textit{i.e.},
\begin{align}
	\dot{\theta}_{5+m} & = \frac{\dot{\theta}}{4}
	\left[1 + \tanh\left(\frac{t-t_\mathrm{min}}{T}\right)\right]\left[1-\tanh\left(\frac{t-t_\mathrm{max}}{T}\right)\right] \,, 
	\label{eq:axion_time_dep} \\
	\dot A_z & = - \frac{E}{4}
	\left[1 + \tanh\left(\frac{t-t_\mathrm{min}}{T}\right)\right]\left[1-\tanh\left(\frac{t-t_\mathrm{max}}{T}\right)\right] \,.
	\label{eq:Az_time_dep}
\end{align}
Here $\dot \theta$ and $E$ denote the constant amplitude of these functions for $t_\text{min} \ll t \ll t_\text{max}$ and $T$ denotes the characteristic time-scale for switching the $\dot \phi$ and $\dot A_z$ on and off.
We can now proceed to solve Eq.~\eqref{eq:alphabeta} numerically. The result, depicted in Fig.~\ref{fig:woB}, displays a remarkable property: The exponential suppression of the non-perturbative Schwinger particle production is not given by $\exp( - \pi m_T^2/(g |Q| E))$, as one would expect from the gap in the dispersion relation~\eqref{eq:energy-levels}, but instead the suppression is governed by the bare mass $m$, 
\begin{align}
  \text{particle production} \sim \sum_{\lambda, \lambda'} \abs{\beta_{\lambda}^{\left(\lambda'\right)}}^2 \propto \exp\left( -\frac{ \pi m^2}{g |Q| E}\right) \,.
\end{align}
This corresponds to an exponential enhancement of the Schwinger production rate for $p_{x,y} \neq 0$, possible only in the presence of a suitable $\dot \theta \neq 0$, \textit{i.e.}, in the presence of a moving axion background field.
We call this exponential enhancement of the particle production the \emph{axion assisted Schwinger effect}.
The remainder of this section is dedicated to explaining this result.

\paragraph{Numerical results.} We first explain our numerical results displayed in Fig.~\ref{fig:woB} in more detail.
In the upper panels, we plot the spectrum of the produced particles
for several different values of $\dot{\theta}$
with two sets of $m$, $p_x$ and $p_y$.
We take $t_\mathrm{min} = 0$, $t_\mathrm{max} = 100/\sqrt{g\abs{Q}E}$ and $T = \sqrt{50/(g\abs{Q}E)}$,
and evaluated the spectrum at $t = 150/\sqrt{g\abs{Q}E}$.
The resulting spectrum features an approximate plateau for $-\tau < p_z/gQE < 0$ with $\tau = t_\mathrm{max} - t_\mathrm{min}$.
The width of the plateau thus depends linearly on the duration of the non-zero electric field $\tau$ which implies that for large values of $\tau$, any transient effects at $t_\text{min, max}$ become irrelevant.
The height and the width of the plateau in the case $\dot{\theta} = 0$ can be intuitively understood as follows.
When the electric field is imposed, the particle is accelerated and $\Pi_z$ increases with time.
The gap between the particle and antiparticle energy levels is minimized when $\Pi_z = 0$,
and hence particle production is most efficient at this point.
Therefore, only the modes which cross $\Pi_z = 0$, \textit{i.e.} $-\tau < p_z/gQE < 0$,
are efficiently enhanced, corresponding to the plateau of the spectrum.
The gap at $\Pi_z = 0$ is given by $m_T = \sqrt{p_T^2 + m^2}$, and hence the height of the plateau is suppressed by $\exp(-\pi m_T^2/g\abs{Q}E)$.\footnote{
	One can make the qualitative argument here more rigorous
	with the help of the WKB analysis,
	see App.~\ref{app:WKB} for a brief review and references. 
}

As one can see in the upper panels of Fig.~\ref{fig:woB}, once we turn on $\dot{\theta}_{5+m}$, 
the height of the plateau depends strongly on this parameter.
This dependence is shown explicitly  in the respective lower panels, where
we plot the occupation number for $p_z/\sqrt{g\abs{Q}E} = -50$, 
corresponding to the center of the plateau, as a function of $\dot{\theta}$.
Two important features stand out.
First, the particle production is drastically enhanced as $\dot{\theta}$ increases,
and the {envelope of the} suppression factor asymptotically approaches to $\exp(-\pi m^2/g\abs{Q}E)$.
In other words, the part of the gap in the dispersion relation due to $p_T$ is overcome 
and the spectrum is correspondingly enhanced by $\exp(\pi p_T^2/g\abs{Q}E)$.
Second, on top of the exponential enhancement, the height of the plateau oscillates with $\dot{\theta}$.
As we explain below, we may interpret this oscillation as an interference effect of two saddle points.
We note in passing that we have checked that the particle production is independent of $\dot{\theta}_{5+m}$
if the fermion is massless. This should be the case since $\theta_{5+m}$ can be rotated away if the fermion is massless
and hence $\theta_{5+m}$ is physical only when $m \neq 0$.

\paragraph{Non-relativistic limit.} We now proceed to interpret our numerical results. 
For this purpose, we study the dispersion relation of the Bogoliubov coefficients.
As we saw in Sec.~\ref{sec:dirac}, the eigenvalues $\Omega$ of the Hamiltonian are independent of $\dot{\theta}_{5+m}$.
However, since the equation of motion of the Bogoliubov coefficients 
depends on $\dot{\theta}_{5+m}$, their time evolution is not simply governed by $\Omega$.
Indeed, by taking the non-relativistic limit,
we show in App.~\ref{app:nreft} that 
the non-zero axion velocity induces the following operators (in the particle sector):
\begin{align}
	\mathcal{L}_\eta
	= \eta^\dag i \partial_0 \eta
	+ \frac{1}{2m} \eta^\dag \left( \bm{\Pi}^2 - 2 \dot \theta_{5 + m} \bm{\Pi} \cdot \bm{\sigma} 
	+ \dot \theta_{5 + m}^2 \right) \eta 
	+ \mathcal{O}\left(\frac{1}{m^2}\right)\, ,
\end{align}
where $\eta$ is a two-component spinor corresponding to the positive frequency part,
$\bm{\sigma}$ is the Pauli matrix and $\bm{\Pi} = \left(p_x, p_y, \Pi_z\right)$.
Thus, the axion induces a coupling between the spin and momentum.
Accordingly we obtain for the time-dependent eigenfrequency,
\begin{align}
	\tilde \Omega^{\pm}_\mathrm{NR} = \frac{1}{2m}\left(\sqrt{\Pi_z^2 + p_T^2} \pm \dot{\theta}_{5+m}\right)^2,
	\label{eq:eigenvalue_NR}
\end{align}
by diagonalizing the equation of motion of $\eta$.\footnote{
{The eigenvalues of the fermion equation of motion depend on the choice of basis for $\psi$. For the basis choice in Eq.~\eqref{eq:uv}, which is the eigenbasis of the Hamiltonian~\eqref{eq:H_app},  $\tilde \Omega^{\pm}_\text{NR}$ coincides with Eq.~\eqref{eq:energy-levels}.}
}
These eigenvalues have an interesting property. For definiteness, let us take $\dot{\theta}_{5+m} > 0$
and consider the minimum value of $\tilde \Omega^-_\mathrm{NR}$ with respect to $\Pi_z$.
The minimum value is $\tilde \Omega^-_\mathrm{NR} = p_T^2/(2 m)$ at $\Pi_z = 0$ 
for $\dot{\theta}_{5+m} = 0$.
Once $\dot{\theta}_{5+m}$ is turned on, however, the minimum value of $\tilde \Omega^-_\mathrm{NR}$ is smaller than $p_T^2/(2 m)$,
and eventually becomes zero when $\dot{\theta}_{5+m} > p_T$.
In other words, the spin-momentum interaction induced by the axion velocity
compensates the gap from the transverse momentum.
We thus naturally expect that the exponential suppression from $p_T$
is compensated by the axion velocity, which is exactly what we find in Fig.~\ref{fig:woB}.

\paragraph{Semi-analytical results.} The above observation relies on the non-relativistic limit, but this limit is not essential.
Indeed, as we see in App.~\ref{app:bilinear}, 
the full relativistic equation of motion of the Bogoliubov coefficients has the following eigenvalues (among others),
\begin{align}
	\tilde \Omega^\pm(t) = \sqrt{\left(\sqrt{\Pi_z^2 + p_T^2}\pm\dot{\theta}_{5+m}\right)^2 + m^2}\,,
	\label{eq:smallest_eigenvalue}
\end{align}
which reduce to Eq.~\eqref{eq:eigenvalue_NR} in the non-relativistic limit.
As an empirical proof that these modes play the essential role, 
we find that the following integral approximates the numerical results well {for $m, p_T \gtrsim \sqrt{g\abs{Q}E}$}:\footnote{
	This formula works for a sizable value of $\dot{\theta}$.
	It does not reproduce the correct expression in the limit $\dot{\theta} \rightarrow 0$
	since the full result reduces to $2\exp\left(-\pi m_T^2/g\abs{Q}E\right)$,
	while this formula vanishes in this limit.
	It is still enough for our purpose since our main interest is the exponential enhancement
	due to a sizable $\dot{\theta}$.
}
\begin{align}
	\sum_{\lambda, \lambda'} \abs{\beta_{\lambda}^{\left(\lambda'\right)}}^2 &\simeq \abs{\exp\left[2i\int^{\Pi_{++}}_0\frac{\dd \Pi_z}{g\abs{Q}E}\,\tilde \Omega^-\right]}^2
	+ \abs{\exp\left[2i\int^{\Pi_{--}}_0\frac{\dd \Pi_z}{g\abs{Q}E}\,\tilde \Omega^-\right]}^2
	- 2\,\mathrm{Re}\left[\exp\left[2i\int^{\Pi_{++}}_{\Pi_{-+}} \frac{\dd \Pi_z}{g\abs{Q}E}\,\tilde \Omega^- \right]\right]\,,
	\label{eq:saddle_pt_approx}
\end{align}
where  
$\dot{\theta}_{5+m}$ is replaced by $\dot{\theta}$ in $\tilde{\Omega}^-$ in this expression, and
\begin{align}
	\Pi_z = \Pi_{\sigma \sigma'},
	\quad
	\Pi_{\sigma \sigma'} = \sigma \sqrt{\left(\dot{\theta} + \sigma' im\right)^2 - p_T^2},
	\quad
	\sigma, \sigma' = \pm,
\end{align}
correspond to the points $\tilde \Omega^- = 0$ in the complex $\Pi_z$ plane.
This is depicted as the gray dashed curves in the lower panels of Fig.~\ref{fig:woB},
and shows excellent agreement with the full numerical result displayed by the blue solid curves.
Although we could not derive this expression from first principles
and hence this is an empirical formula,
it is motivated by the saddle point approximation of an integral
controlled by $\tilde \Omega^-$ as we discuss in App.~\ref{app:bilinear}.
Note that $\tilde \Omega^-$ has two pairs of saddle points in the complex $\Pi_z$ plane.
The last term in Eq.~\eqref{eq:saddle_pt_approx} originates from an interference 
of these two saddle points, and this interference term induces the oscillatory behavior 
in the lower panels of Fig.~\ref{fig:woB} (see also App.~\ref{app:WKB}).
We further note that the formula~\eqref{eq:saddle_pt_approx} overestimates the spectrum for small values of $m$ and/or $p_T$.
In particular, the axion assisted Schwinger effect dies out as $m\rightarrow 0$,
which is not captured by Eq.~\eqref{eq:saddle_pt_approx} but is displayed clearly in our numerical results (see App.~\ref{app:bilinear}).
The case of small $p_T$ is not particularly relevant to our discussion, 
since the axion assisted Schwinger effect is most prominent when $p_T$ is large.
It would however be certainly interesting if one could derive an analytical formula that works 
for both small and large values of $m$ and $p_T$, which we leave as a future work.
See App.~\ref{app:bilinear} for more details on the range of validity and limitations of Eq.~\eqref{eq:saddle_pt_approx}.

In short, we interpret our numerical results as follows.
The axion motion induces the spin-momentum interaction.
It modifies the dispersion relation and in particular can compensate the energy gap from the transverse momentum $p_T$ (see Eq.~\eqref{eq:smallest_eigenvalue}).
The remaining minimal gap is given solely by the intrinsic mass $m$ in the large $\dot{\theta}_{5+m}$ limit,
and hence the particle production is suppressed only by $\exp(-\pi m^2/g\abs{Q}E)$. In other words,
the particle production is exponentially enhanced by $\exp(\pi p_T^2/g\abs{Q}E)$.
{The UV cut-off of this process is set by $\dot \theta_{5+m}$, as will be discussed in more detail in Sec.~\ref{subsec:pp_wB}.}

\section{Axion assisted Schwinger effect in an electric and magnetic field}
\label{sec:wB}

\subsection{Dirac equation with classical background fields}
\label{sec:dirac_wB}

This section generalizes the analyses performed in Sec.~\ref{sec:woB} by including a constant magnetic field, aligned (anti-)parallel to the electric field. Such a configuration is generated dynamically through a tachyonic instability in the axion gauge field coupling, leading to an exponential enhancement of one of the two gauge field helicities~\cite{Turner:1987bw,Garretson:1992vt,Anber:2006xt}. We will thus consider the background field configuration
\begin{align}
	A^{\mu} = \left(0, 0, B x, A_z(t)\right),
	\qquad
	\phi = \phi(t),
\end{align}
where the magnetic field $B$ is constant and as before the electric field is given by $-\dot{A}_z$.
Here, our starting point will be the most general action coupling axions, fermions and gauge bosons through dimension five operators, while preserving the shift-symmetry of the axion and $CP$-invariance,
\begin{align}
	S = \int \dd^4 x &\left[\frac{a^2}{2}\left(\partial \phi\right)^2
	- \frac{1}{4}F^{\mu\nu}F_{\mu\nu}
	+ \bar{\psi} \left(i \slashed{D} - ma e^{2i c_m \phi/f_a \gamma_5} \right)\psi 
	+ c_A \frac{\alpha}{4\pi f_a} \phi F_{\mu\nu} \tilde{F}^{\mu\nu}
	+ c_5 \frac{\partial_\mu \phi}{f_a} \bar{\psi} \gamma^\mu \gamma_5 \psi
	\right],
	\label{eq:action_FRW}
\end{align}
This in particular extends the discussion of Ref.~\cite{Domcke:2019qmm}, which assumed for simplicity $\theta_5 + \theta_m = 0$. 

The introduction of the magnetic field induces a spiralling motion of the charged fermions.
This spiralling motion is quantum-mechanically quantized and the dispersion relations are consequently described by discrete Landau levels (see Ref.~\cite{Domcke:2019qmm} for an explicit construction). The equation of motion for the lowest Landau level reads
\begin{align}
	 \quad 0 = \left[i \mathbb{I}_2 \partial_0 + \left(s\Pi_z - \dot \theta_5 \right)
	\begin{pmatrix} 1 & 0 \\ 0 & -1 \end{pmatrix} - m \begin{pmatrix} 0 & e^{2i\theta_m} \\ e^{-2i\theta_m} & 0 \end{pmatrix} 
	\right]
	\begin{pmatrix} \psi_{0}^{\left({L}\right)} \\ \psi_{0}^{\left({R}\right)} \end{pmatrix}\,,
	\label{eq:eom_lll}
\end{align}
where $s = \text{sign}(QB)$.
We note that in this lowest Landau level, the effect of $\theta_5$ is degenerate with $\Pi_z$. In particular, if $\dot{\phi}$ is constant, we can absorb it by shifting the momentum $p_z$. Hence, since the computation of particle production involves an integral over $p_z$, we can anticipate that the particle production in the lowest Landau level will not be affected by the axion motion.

Similarly, introducing the magnetic mass $m_B^2 = 2n g |QB|$, the equation of motion for the higher Landau levels reads
\begin{align}
	 \quad 0 = \begin{pmatrix}
	i \partial_0 + s\Pi_z - \dot \theta_5 & i m_B & -m e^{2i\theta_m} & 0 \\
	-i m_B & i \partial_0 - s\Pi_z - \dot \theta_5 & 0 & -m e^{2i\theta_m} \\
	- m e^{-2i\theta_m} & 0 & i\partial_0 - s\Pi_z + \dot \theta_5& -i m_B \\
	0 & -m e^{-2i\theta_m} & i m_B & i\partial_0 + s\Pi_z +\dot \theta_5
	\end{pmatrix} \psi_n(t)\,,
	\label{eq:eom_hll}
\end{align}
where $n = 1,2,..$ labels the Landau levels and $\psi_n$ is a vector containing the four fermionic modes, \textit{i.e.}, (anti-)particles of both helicities, which mix for a given Landau level, see Ref.~\cite{Domcke:2019qmm} for details. 

As in Sec.~\ref{sec:dirac} we may now expand the fermionic mode functions as
\begin{align}
 \psi_0 =  e^{- i \sigma_3 \theta_5} \left[\alpha_0 u_0 \exp\left(-i \Omega_0 t\right)
	+ \beta_0 v_0 \exp\left(+i \Omega_0 t\right)\right]\,,
	\label{eq:lll_prefactor}
\end{align}
for the lowest Landau level, and
\begin{align}
 \psi_n = e^{i \gamma_5 \theta_5} \sum_{\lambda = 1, 2} 
	\left[
	\alpha_{n,\lambda} u_{n,\lambda} \exp\left(-i \Omega_n t\right)
	+ \beta_{n,\lambda} v_{n,\lambda} \exp\left(+i \Omega_n t\right)
	\right]\,,
\end{align}
for the higher Landau levels, with $u_0$, $v_0$ and $u_{n,\lambda}$, $v_{n,\lambda}$ denoting the eigenfunctions of the fermionic part of the Hamiltonian, see App.~\ref{app:particlesB} for details.
Inserting this into Eqs.~\eqref{eq:eom_lll} and \eqref{eq:eom_hll} yields the dispersion relations
\begin{align}
 \Omega_0 = \sqrt{\Pi_z^2 + m^2} \quad \text{and} \quad \Omega_n = \sqrt{\Pi_z^2 + m_T^2}\,,
 \label{eq:energy-levels0}
\end{align}
with the effective mass labelling the Landau level, $m_T = \sqrt{m_B^2 + m^2}$.

\paragraph{Bogoliubov coefficients.} We again turn on the time-dependence of $A_z$ and ${\phi}$, thus promoting the  coefficients $\alpha_\lambda$ and $\beta_\lambda$ to time dependent Bogoliubov coefficients.
Proceeding as in Sec.~\ref{sec:dirac}, the equations of motion are given by
\begin{align}
	\dot{\alpha}_0 &= i\dot{\theta}_{5+m}  \frac{s {\Pi}_z}{\Omega_0} \alpha_0 
	- \left(s\frac{m \dot{{\Pi}}_z}{2\Omega_0^2} + i \dot{\theta}_{5+m}  \frac{m}{\Omega_0}\right) e^{2i\Theta_0} \beta_0, 
	\label{eq:eom_lll_alpha} \\
	\dot{\beta}_0 &= -i\dot{\theta}_{5+m}  \frac{s {\Pi}_z}{\Omega_0} \beta_0 
	+ \left(s\frac{m \dot{{\Pi}}_z}{2\Omega_0^2} - i \dot{\theta}_{5+m}  \frac{m}{\Omega_0}\right) e^{-2i\Theta_0} \alpha_0,
	\label{eq:eom_lll_beta}
\end{align}
for the lowest Landau level, and
{\small
\begin{align}
	\begin{pmatrix}
	\dot{\alpha}_{n, 1} \\
	\dot{\alpha}_{n, 2} \\
	\dot{\beta}_{n, 1} \\
	\dot{\beta}_{n, 2} 
	\end{pmatrix}
	=&
	\left[
	i\dot{\theta}_{5+m}
	\begin{pmatrix}
	- \frac{m}{m_T}\frac{s\Pi_z}{\Omega_n} & \frac{m_B}{m_T} & \frac{m e^{2i\Theta_n}}{\Omega_n} & 0 \\
	\frac{m_B}{m_T} & \frac{m}{m_T}\frac{s\Pi_z}{\Omega_n} & 0 & -\frac{me^{2i\Theta_n} }{\Omega_n}\\
	\frac{m e^{-2i\Theta_n}}{\Omega_n} & 0 & \frac{m}{m_T}\frac{s\Pi_z}{\Omega_n} & \frac{m_B}{m_T} \\
	0 & -\frac{me^{-2i\Theta_n}}{\Omega_n} & \frac{m_B}{m_T} & -\frac{m}{m_T}\frac{s\Pi_z}{\Omega_n}
	\end{pmatrix}
	+ \frac{ s m_T \dot{\Pi}_z}{2\Omega_n^2}
	\begin{pmatrix}
	0 & 0 & -e^{2i\Theta_n} & 0 \\
	0 & 0 & 0 & -e^{2i\Theta_n} \\
	e^{-2i\Theta_n} & 0 & 0 & 0 \\
	0 & e^{-2i\Theta_n} & 0 & 0
	\end{pmatrix}
	\right]
	\begin{pmatrix}
	{\alpha}_{n, 1} \\
	{\alpha}_{n, 2} \\
	{\beta}_{n, 1} \\
	{\beta}_{n, 2} 
	\end{pmatrix},
	\label{eq:alphabeta_hll}
\end{align}
}for the higher Landau levels,
where $\Theta_0 = \int \dd t \, \Omega_0$ and $\Theta_n = \int \dd t \, \Omega_n$.
In particular, we see that Eq.~\eqref{eq:alphabeta_hll} 
is equivalent to its counterpart without the magnetic field~\eqref{eq:alphabeta}
after replacing $p_T$ with  $m_B$.
The only difference is that the transverse momentum is quantized due to the magnetic field 
in Eq.~\eqref{eq:alphabeta_hll}, while it is continuous in Eq.~\eqref{eq:alphabeta}.

\paragraph{Quantization and initial conditions.} In terms of these Bogoliubov coefficients, 
the quantized mode function is given by
\begin{align}
	\psi_0 = e^{- i \sigma_3 \theta_5} \left[
	u_0 e^{-i\Theta_0}\left(\alpha_0 b_0 - \beta_0^* d_0^\dagger\right)
	+ v_0 e^{i\Theta_0}\left(\beta_0 b_0 + \alpha_0^* d_0^\dagger\right)\right]\,,
\end{align}
for the lowest Landau level, and
\begin{align}
	\psi_n = e^{i \gamma_5 \theta_5}\sum_{\lambda, \lambda' = 1,2} \left[ u_{n, \lambda} e^{-i\Theta_n}
	\left(\alpha_{n, \lambda}^{(\lambda')} b_{n, \lambda'} - 
	\left(-1\right)^{\lambda+\lambda'}{\beta_{n, \lambda}^{(\lambda')}}^* d_{n, \lambda'}^\dagger \right)
	+
	v_\lambda e^{i\Theta_n}
	\left(\beta_{n, \lambda}^{(\lambda')} b_{n, \lambda'} 
	+ \left(-1\right)^{\lambda+\lambda'}{\alpha_{n, \lambda}^{(\lambda')}}^* d_{n, \lambda'}^\dagger \right)
	\right]\,,
\end{align}
for the higher Landau levels,
where $b_0, d_0, b_{n,\lambda}, d_{n,\lambda}$ are the fermion creation and annihilation operators.
The initial conditions are given by
\begin{align}
	\alpha_0 = 1\,,
	\quad
	\beta_0 = 0\,,
\end{align}
for the lowest Landau level and
\begin{align}
	\alpha_{n, 1}^{(1)} &= 1\,,
	\quad
	\alpha_{n, 2}^{(1)} = \beta_{n, 1}^{(1)} = \beta_{n, 2}^{(1)} = 0\,, \\
	\alpha_{n, 2}^{(2)} &= 1\,,
	\quad
	\alpha_{n, 1}^{(2)} = \beta_{n, 1}^{(2)} = \beta_{n, 2}^{(2)} = 0\,, 
\end{align}
for the higher Landau level, respectively, at the initial time $t = -\infty$.

As in the case without a {magnetic} field, one can define the fermionic annihilation and creation operators
at any given time $t$ as
\begin{align}
	B_0 &= \alpha_0 b_0 - \beta_0^* d_0^\dagger\,,
	\quad
	D_0^\dagger = \beta_0 b_0 + \alpha_0^* d_0^\dagger\,, 
\end{align}
for the lowest Landau level, and
\begin{align}
	B_{n, \lambda} &= \sum_{\lambda' = 1, 2}\left(\alpha_{n, \lambda}^{(\lambda')} b_{n, \lambda'} - 
	\left(-1\right)^{\lambda+\lambda'}{\beta_{n, \lambda}^{(\lambda')}}^* d_{n, \lambda'}^\dagger\right)\,, \\
	D_{n, \lambda}^\dagger &= \sum_{\lambda' = 1, 2}\left(\beta_{n, \lambda}^{(\lambda')} b_{n, \lambda'} 
	+ \left(-1\right)^{\lambda+\lambda'}{\alpha_{n, \lambda}^{(\lambda')}}^* d_{n, \lambda'}^\dagger \right)\,,
\end{align}
for the higher Landau levels.
They satisfy the standard anti-commutation relations, indicating that
\begin{align}
	\abs{\alpha_0}^2 + \abs{\beta_0}^2 &= 1\,,
\end{align}
for the lowest Landau level, and
\begin{align}
	\sum_\lambda 
	\left[
	\alpha_{n, \lambda_1}^{(\lambda)}{\alpha_{n, \lambda_2}^{(\lambda)}}^*
	+ \left(-1\right)^{\lambda_1+\lambda_2}\beta_{n, \lambda_2}^{(\lambda)}
	{\beta_{n, \lambda_1}^{(\lambda)}}^* 
	\right]
	&= \delta_{\lambda_1 \lambda_2}\,, \\
	\sum_\lambda 
	\left[
	\alpha_{n, \lambda_1}^{(\lambda)}{\beta_{n, \lambda_2}^{(\lambda)}}^*
	- \left(-1\right)^{\lambda_1+\lambda_2}\alpha_{n, \lambda_2}^{(\lambda)}
	{\beta_{n, \lambda_1}^{(\lambda)}}^* 
	\right]
	&= 0\,,
\end{align}
for the highest Landau level,
for all times $t$.
The former trivially follows from Eqs.~\eqref{eq:eom_lll_alpha} and~\eqref{eq:eom_lll_beta},
while the latter is shown to be satisfied in App.~\ref{app:bilinear}.

\subsection{Anomaly equation}

The anomaly equation relates the chiral fermion current with the Chern-Simons term as~\cite{Adler:1969gk,Bell:1969ts}
\begin{align}
	\partial_\mu J^\mu_5 = - \frac{g^2Q^2}{8\pi^2}\epsilon^{\mu\nu\rho\sigma}F_{\mu\nu}F_{\rho\sigma}
	+ 2im \bar{\psi} e^{2i\theta_m \gamma_5} \gamma_5 \psi,
	\quad
	J^\mu_5 = \bar{\psi} \gamma^\mu \gamma_5 \psi.
\end{align}
An important property of the system in this section is that the Chern-Simons density is non-vanishing,
$F\tilde{F} = - 4 EB \neq 0$.
Therefore  a non-trivial consistency check of our computation 
is to correctly reproduce the  anomaly equation.
Ref.~\cite{Domcke:2019qmm} shows that the anomaly equation holds in the case $\theta_{5} + \theta_{m} = 0$,
and we now generalize this result including a non-vanishing $\theta_{5} + \theta_{m}$.

In the following we focus on the spatially averaged version of the anomaly equation, which reads
\begin{align}
	\dot{q}_5 = \frac{g^2 Q^2 E B}{2\pi^2} 
	+ 2 i m \left\langle  \bar{\psi} e^{2i\theta_m \gamma_5} \gamma_5 \psi \right\rangle.
\end{align}
where
\begin{align}
	q_5 &\equiv \frac{1}{2\mathrm{Vol}\left(\mathbb{R}^3\right)}\int \dd^3x
	\left\langle 
	\left[ {\psi}^\dagger, \gamma_5 \psi\right]
	\right\rangle,
	\quad
	\left\langle  \bar{\psi} e^{2i\theta_m \gamma_5} \gamma_5 \psi \right\rangle
	\equiv \frac{1}{2\mathrm{Vol}\left(\mathbb{R}^3\right)}\int \dd^3x
	\left\langle 
	\left[ {\psi}^\dagger, \gamma^0 \gamma_5 e^{2i\theta_m \gamma_5} \psi\right]
	\right\rangle,
\end{align}
with the expectation value $\langle \cdots \rangle$ taken with respect to the initial vacuum state.
We consider contributions from the lowest and higher Landau levels separately.
In particular, we will see that only the lowest Landau level contributes to the Chern-Simons term.
See App.~\ref{app:particlesB} for some useful relations that are used in the computation below.

\paragraph{Lowest Landau level.}
The chiral charge from the lowest Landau level is given by\footnote{
	We dropped a regulator in this expression.
	One can show, as in Ref.~\cite{Domcke:2019qmm},  that the results do not depend on the choice of a regulator function
	as long as it depends only on $\Omega_0$.
}
\begin{align}
	q_{5,0} = \frac{g\abs{QB}}{4\pi^2}\int \dd p_z 
	\left[
	\frac{2s{\Pi}_z}{\Omega_0} \left\lvert \beta_0 \right\rvert^2
	+ \frac{m}{\Omega_0}\left(\alpha_0 \beta_0^*e^{-2i\Theta_0} + \alpha^*_0 \beta_0 e^{2i\Theta_0} \right)
	\right],
\end{align}
where the subscript ``$0$" indicates the contribution of the lowest Landau level.
With Eqs.~\eqref{eq:eom_lll_alpha} and~\eqref{eq:eom_lll_beta},
its time derivative reads
\begin{align}
	\dot{q}_{5,0} = \frac{g\abs{QB}}{4\pi^2} \int \dd p_z
	\left[
	s\frac{m^2 \dot{{\Pi}}_z}{\Omega_0^3}
	- 2im \left(\alpha_0 \beta_0^* e^{-2i\Theta_0} - \alpha_0^* \beta_0 e^{2i\Theta_0}\right)
	\right].
\end{align}
The first term is easily integrated,
and the second term corresponds to the mass term,
\begin{align}
	\left.\left\langle \bar{\psi} e^{2i\theta_m \gamma_5} \gamma_5 \psi \right\rangle\right\rvert_{\mathrm{LLL}}
	&\equiv \frac{1}{2\mathrm{Vol}\left(\mathbb{R}_3\right)}\int \dd^3 x \left\langle 
	\left.\left[\bar{\psi}, e^{2i\theta_m \gamma_5} \gamma_5 \psi \right] \right\rangle \right\rvert_{\mathrm{LLL}}
	= - \frac{g\abs{QB}}{4\pi^2} \int \dd p_z \left[ \alpha_0 \beta_0^* e^{-2i\Theta_0} - \alpha_0^* \beta_0 e^{2i\Theta_0} \right],
\end{align}
where ``LLL" stands for the lowest Landau level.
As a result, we obtain
\begin{align}
	\dot{q}_{5,0} = \frac{g^2 Q^2EB}{2\pi^2} 
	+ 2i m \left.\left\langle \bar{\psi} e^{2i\theta_m \gamma_5} \gamma_5 \psi \right\rangle\right\rvert_{\mathrm{LLL}}.
	\label{eq:anomaly_lll}
\end{align}
Thus, the Chern-Simons term is supplied by the lowest Landau level.
Below we confirm that the higher Landau levels
do not induce additional contributions to the Chern-Simons term.

\paragraph{Higher Landau level.}
The chiral charge from the higher Landau levels is given by
\begin{align}
	q_{5, n} = \frac{g\abs{QB}}{4\pi^2}\int \dd p_z \sum_{\lambda} 
	&\left\{
	\frac{m}{m_T}\frac{s\Pi_z}{\Omega_n}
	\left[
	\abs{\alpha_{n, 1}^{\left(\lambda\right)}}^2
	- \abs{\beta_{n, 1}^{\left(\lambda\right)}}^2
	- \abs{\alpha_{n, 2}^{\left(\lambda\right)}}^2
	+ \abs{\beta_{n, 2}^{\left(\lambda\right)}}^2
	\right]
	\right. \nonumber \\ &\left.
	-\frac{m_B}{m_T}
	\left[
	{\alpha_{n, 1}^{\left(\lambda\right)}}^* \alpha_{n, 2}^{\left(\lambda\right)} 
	+ {\beta_{n, 1}^{\left(\lambda\right)}}^* \beta_{n, 2}^{\left(\lambda\right)} 
	+ \left(\mathrm{c.c.}\right)
	\right]
	- \frac{m}{\Omega_n}
	\left[
	\left({\alpha_{n, 1}^{\left(\lambda\right)}}^* \beta_{n, 1}^{\left(\lambda\right)} 
	- {\alpha_{n, 2}^{\left(\lambda\right)}}^* \beta_{n, 2}^{\left(\lambda\right)} \right) e^{2i\Theta_n}
	+ \left(\mathrm{c.c.}\right)
	\right]
	\right\},
	\label{eq:axial_current_hll}
\end{align}
where the subscript ``$n$" indicates the contribution of the $n$th Landau level for $n \geq 1$.
By taking the time derivative and using Eq.~\eqref{eq:alphabeta_hll}, we obtain
\begin{align}
	\dot{q}_{5,n} = -2im \frac{g\abs{QB}}{4\pi^2}
	\int \dd p_z \sum_{\lambda}
	\left[
	\left({\alpha_{n, 1}^{\left(\lambda\right)}}^* \beta_{n, 1}^{\left(\lambda\right)} 
	- {\alpha_{n, 2}^{\left(\lambda\right)}}^* \beta_{n, 2}^{\left(\lambda\right)} \right) e^{2i\Theta_n}
	- \left({\alpha_{n, 1}^{\left(\lambda\right)}} {\beta_{n, 1}^{\left(\lambda\right)}}^*
	- {\alpha_{n, 2}^{\left(\lambda\right)}} {\beta_{n, 2}^{\left(\lambda\right)}}^* \right) e^{-2i\Theta_n}
	\right].
\end{align}
It is straightforward to show that the right-hand-side corresponds to the mass term,
\begin{align}
	\dot{q}_{5,n} = 2im 
	\left. \left\langle \bar{\psi} e^{2i\gamma_5 \theta_m}\gamma_5 \psi \right\rangle \right\rvert_{\mathrm{HLL},n},
\end{align}
where ``HLL" stands for the higher Landau levels.
Therefore the higher Landau levels do not contribute to the Chern-Simons term.
This completes the proof of the anomaly equation.

\subsection{Particle production}
\label{subsec:pp_wB}
We now study the particle production for the lowest and higher Landau levels separately,
and estimate the induced current.
We impose the electric field and the axion dynamics as in Eqs.~\eqref{eq:axion_time_dep} and~\eqref{eq:Az_time_dep}.

\begin{figure}[t]
	\centering
 	\includegraphics[width=0.495\linewidth]{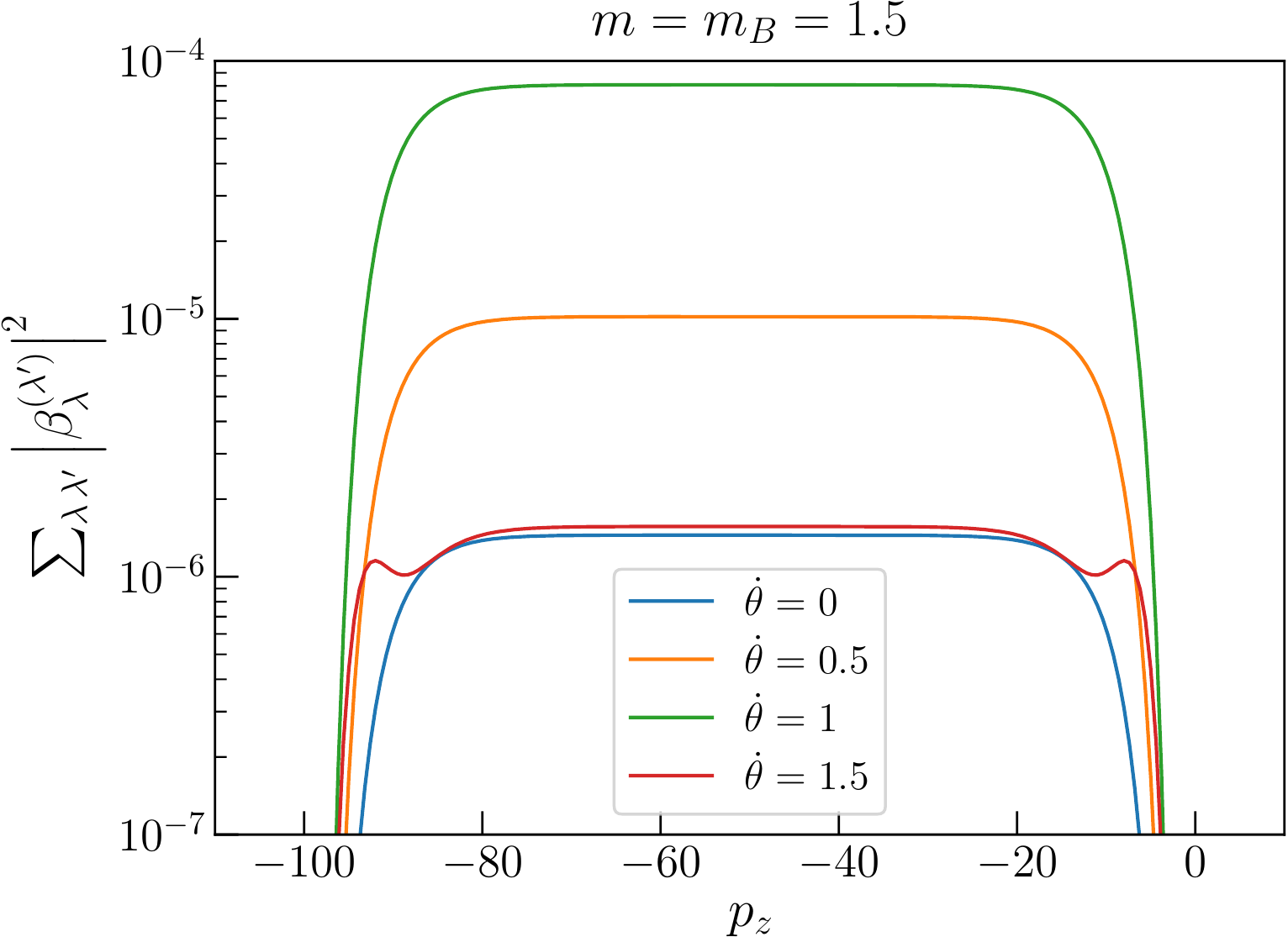}
 	\includegraphics[width=0.495\linewidth]{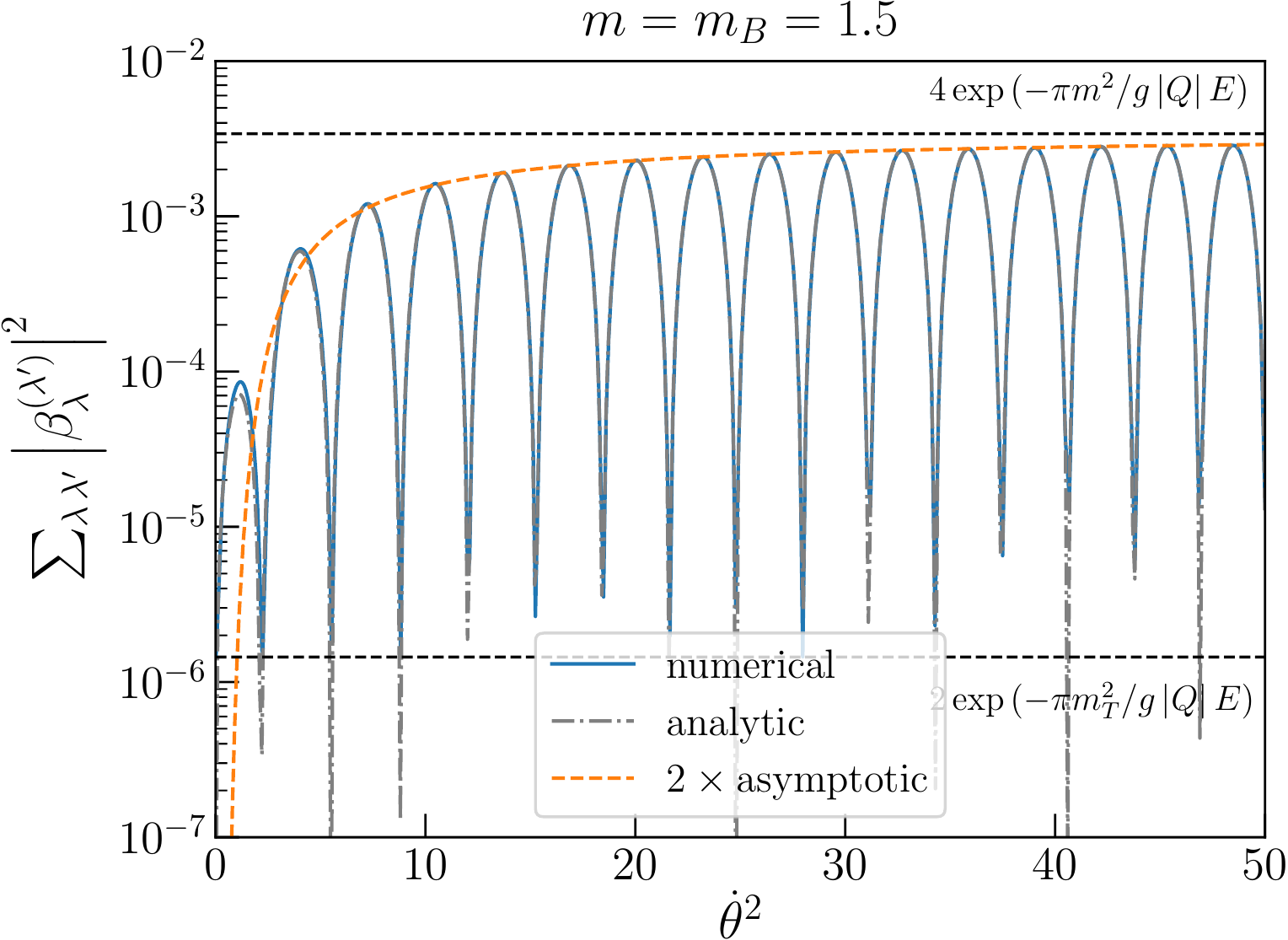}
	\caption{\small Left panel: the spectrum $\sum_{\lambda \lambda'} \abs{\beta_\lambda^{(\lambda')}}^2$
	as a function of $p_z$ for several different values of $\dot{\theta}$.
	Right panel: the height of the plateau of the spectrum evaluated at $p_z = -50$
	as a function of $\dot{\theta}$.
	The blue solid line is the full numerical results,
	the gray dashed line is the analytical formula~\eqref{eq:saddle_pt_approx}
	after replacing $p_T$ by $m_B$,
	and the orange dashed line {shows our analytical result for the envelope, given by} twice the asymptotic formula~\eqref{eq:asymptotic}.
	The parameters are shown in the unit $g\abs{Q}E = 1$.}
	\label{fig:HLL}
\end{figure}

\paragraph{Lowest Landau level.}

We have checked numerically that the spectrum of the lowest Landau level does not depend on $\dot{\theta}$.
This result is easily understood based on our discussion in Sec.~\ref{subsec:pp_woB} as follows.
The lowest Landau level corresponds to the mode that moves parallel to the magnetic field,
or equivalently has a vanishing transverse momentum.
In this case, the minimum of $\Omega_-$ is not affected by the presence of $\dot{\theta}_{5+m}$,
and hence no exponential enhancement of the particle production is expected to occur.
The non-vanishing axion velocity simply leads to the replacement
$p_z$ by $p_z - \dot{\theta}_{5+m}$
in the eigenvalue,
and this is absorbed by a constant shift of $p_z$ for a constant $\dot{\theta}$.
As a result there is no effect on the particle production of the lowest Landau level.

\paragraph{Higher Landau levels.}
In Fig.~\ref{fig:HLL}, we plot our numerical results for the particle production for the higher Landau levels.
It shows that the spectrum is again exponentially enhanced when $\dot{\theta}$ is large.
As we noted above, the equation of motion of the Bogoliubov coefficients for the higher Landau levels
are equivalent to Eq.~\eqref{eq:alphabeta} after replacing $p_T$ with $m_B$.
Therefore, we can interpret this result in exactly the same way as we did in Sec.~\ref{subsec:pp_woB}.
In particular, the enhancement of the particle production is well approximated by Eq.~\eqref{eq:saddle_pt_approx}
after replacing $p_T$ with $m_B$ as we show in the right panel.
The axion-induced spin-momentum interaction does not care 
whether the transverse momentum is continuous or discretized,
and the axion assisted Schwinger effect is at work for both cases.

\paragraph{Induced current.}
We finally estimate the current of the produced fermions. 
As discussed in Sec.~\ref{sec:applications}, this is a key quantity to determine the backreaction of the fermion production on the background gauge fields in a dynamical system.
In our setup, only the $z$-component of the induced current is non-vanishing,
and is given as
\begin{align}
	\langle J_{z} \rangle \equiv \frac{1}{2\mathrm{Vol}\left(\mathbb{R}^3\right)}\int \dd^3 x 
	\left\langle \left[\bar{\psi}, \gamma^3 \psi\right] \right\rangle  
	 = \frac{gQB}{4\pi^2} &
	\int \dd p_z \left\{  \left[2\frac{s\Pi_z}{\Omega_0}\abs{\beta_0}^2
	+ \frac{m_T}{\Omega_n}
	\left({\alpha_0}^*\beta_0
	+ \mathrm{h.c.}\right) \right] \right. \nonumber \\ 
	& \left. + 
	\sum_{n, \lambda, \lambda'}
	\left[
	2\frac{s\Pi_z}{\Omega_n}\abs{\beta_{n, \lambda}^{\left(\lambda'\right)}}^2
	+ \frac{m_T}{\Omega_n}
	\left({\alpha_{n, \lambda}^{\left(\lambda'\right)}}^*\beta_{n, \lambda}^{\left(\lambda'\right)}
	+ \mathrm{h.c.}\right)
	\right]\right\}\,,
\end{align}
We focus on the first term since it is proportional to the duration $\tau$ of the non-zero electric field, and hence dominates for large $\tau$.
As we saw before, the spectrum develops a plateau approximated as
\begin{align}
	\sum_{\lambda, \lambda'}\abs{\beta_{n, \lambda}^{\left(\lambda'\right)}}^2
	\simeq 
	\abs{\bar{\beta}_n}^2\Theta\left(\mathrm{sgn}(Q) p_z + g\abs{Q}E \tau\right) \Theta\left(-\mathrm{sgn}(Q)p_z\right)\,,
\end{align}
where we denote the height of the plateau as $\abs{\bar{\beta}_n}^2$, $\Theta$ is the Heaviside theta function
and $\mathrm{sgn}(Q) = Q/\abs{Q}$.
Therefore the induced current is estimated as
\begin{align}
	gQ \langle J_{z} \rangle \simeq 
	\tau \times \frac{ \left(g \abs{Q}\right)^3}{2\pi^2}E \abs{B} \sum_{n}\abs{\bar{\beta}_n}^2,
\end{align}
where we assumed $\tau \gg m_T/(g\abs{Q}E)$.

If there is no coupling to the axion, $\dot{\theta}_{5+m} = 0$, the height is given by
\begin{align}
	\abs{\bar{\beta}_0}^2 \simeq \exp\left[-\frac{\pi m^2}{g\abs{Q}E}\right],
	\quad
	\abs{\bar{\beta}_n}^2 \simeq 2\exp\left[-\frac{\pi \left(m^2 + 2 n g\abs{QB}\right)}{g\abs{Q}E}\right],
\end{align}
and hence the induced current is estimated as
\begin{align}
	gQ \langle J_{z} \rangle \simeq 
	\tau \times \frac{ \left(g \abs{Q}\right)^3}{2\pi^2}E \abs{B} \coth\left(\frac{\pi \abs{B}}{E}\right)
	e^{-\frac{\pi m^2}{g\abs{Q}E}}\,,
	\quad
	\mathrm{for}
	\quad
	\dot{\theta}_{5+m} = 0\,,
\end{align}
reproducing our previous result~\cite{Domcke:2019qmm}.

Once we turn on the axion coupling, the spectrum for the higher Landau level is exponentially enhanced 
and $\abs{\bar{\beta}_n}^2$ is estimated by Eq.~\eqref{eq:saddle_pt_approx} after replacing $p_T$ with $m_B$.
Unfortunately, Eq.~\eqref{eq:saddle_pt_approx} is still complicated enough so that
we could not obtain the induced current analytically.
Therefore, we just make a crude estimation of the induced current by further simplifying Eq.~\eqref{eq:saddle_pt_approx}.
First, we approximate the oscillatory behavior of $\abs{\bar{\beta}_n}^2$ with respect to $\dot{\theta}$ by simply inserting
 half the envelope of the oscillation.
We also focus on the modes that satisfy $m_B \gtrsim g\abs{Q}E$ 
since otherwise the axion assisted Schwinger effect is not so drastic {(see App.~\ref{app:bilinear} for a discussion of the $m \ll g|Q|E$ limit)}.
With these simplifications, we find
\begin{align}
	\abs{\bar{\beta}_n}^2 
	\sim 
	\displaystyle 2\exp\left[-\frac{\pi m^2}{g\abs{Q}E}
	\left(1 + \frac{m_B^2}{2\dot{\theta}^2}\right)\right]
	\quad \mathrm{for} \quad
	\dot{\theta}^2 \gg m^2, m_B^2\,.
	\label{eq:asymptotic}
\end{align}
{As shown in Fig.~\ref{fig:HLL}, this formula describes the envelope well in the asymptotic regime.}
We see that the  suppression factor approaches $\exp(-\pi m^2/(g\abs{Q}E))$, 
{\textit{i.e.}, $\exp( - \pi m^2 m_B^2/(2 g |Q| E \dot \theta^2) \simeq 1$,} when 
\begin{align}
	\frac{\pi m^2 m_B^2}{g\abs{Q}E} \lesssim \dot{\theta}^2.
	\label{eq:approx_saturation}
\end{align}
Thus, we may estimate the induced current as
\begin{align}
	gQ \langle J_{z} \rangle \sim 
	\tau \times \frac{ \left(g \abs{Q}\right)^3}{2\pi^2}E^2
	e^{-\frac{\pi m^2}{g\abs{Q}E}}\times
	\mathrm{max}\left[\frac{\abs{B}}{E} \coth\left(\frac{\pi \abs{B}}{E}\right),
	\frac{\dot{\theta}^2}{\pi m^2}\right]\,,
	\label{eq:induced_current}
\end{align}
where we simply count the number of the modes that satisfy Eq.~\eqref{eq:approx_saturation}
and introduce the ``max" function so that it reduces to the previous result when $\dot{\theta}_{5+m}$ is small.
One can see  that the induced current is enhanced as $\dot{\theta}_{5+m}$ increases.
In particular, for $E \sim |B|$, the axion assisted Schwinger effect leads to an enhancement of the induced currently by a factor of roughly $\dot \theta^2/m^2$ compared the standard result in the absence of the axion field.
Here we emphasize again that our estimation above is quite rough,
and we leave a more precise estimation to future work.

\section{Implications for axion cosmology}
\label{sec:applications}

\paragraph{Axion inflation.} Identifying the inflaton, \textit{i.e.}, the particle responsible for driving cosmic inflation, with an axion-like particle with shift-symmetric dimension five couplings to gauge fields and fermions, naturally ensures a sufficiently flat scalar potential as required for slow-roll inflation~\cite{Freese:1990rb,Nomura:2017ehb}. The coupling to gauge fields induces a tachyonic instability in one of the helicities of the vector potential, leading to the production of a strong, large-scale helical gauge field configuration during inflation, driven by the kinetic energy of the axion field~\cite{Turner:1987bw,Garretson:1992vt,Anber:2006xt}.\footnote{
As in the rest of this paper, we focus on Abelian gauge fields here. Non-Abelian gauge field configurations can also be sourced by a non-vanishing axion velocity~\cite{Adshead:2012kp,Maleknejad:2012fw,Domcke:2018rvv,Domcke:2019lxq}, however due to the self-interactions of the non-Abelian gauge fields, the fermion backreaction originating from the induced fermion current is less relevant in this case~\cite{Domcke:2018gfr}. Moreover, for a discussion of the gravitational production of neutral fermions in axion inflation, see Ref.~\cite{Adshead:2018oaa}. 
} The production of fermions in this helical gauge field background is well described by the analysis of Sec.~\ref{sec:wB} as long as the axion velocity varies only slowly, implying approximately constant physical electric and magnetic field strengths.\footnote{
For very strong gauge field backgrounds with correspondingly strong backreaction effects on the axion equation of motion, this approximation becomes invalid due to resonance effects in the coupled axion gauge field system~\cite{Cheng:2015oqa,Notari:2016npn,DallAgata:2019yrr,Domcke:2020zez}. However, in the presence of light fermions, the gauge field production is inhibited, and hence the resonance effects are expected to be less relevant.
}

In the absence of charged fermions, the exponential gauge field production in axion inflation leads to striking signatures~\cite{Barnaby:2011qe}, including the generation of gravitational waves in the range of ground-and space-based interferometers~\cite{Cook:2011hg,Anber:2012du} and of primordial black holes~\cite{Linde:2012bt,Garcia-Bellido:2016dkw,Domcke:2017fix}. The dual production of helical gauge fields and charged fermions was first studied in Ref.~\cite{Domcke:2018eki} for massless fermions and extended in Ref.~\cite{Domcke:2019qmm} to massive fermions, for the particular parameter choice of $c_m + c_5 = 0$, corresponding to the absence of the last term in Eq.~\eqref{eq:action_FRW}. The fermion production and the resulting induced current lead to the formation of electric and magnetic fields anti-aligned to the background fields, and thus to a reduction of the net gauge field background generated in axion inflation by several orders of magnitude. This dramatically changes the predictions of axion inflation.

Our new estimate for the induced current, Eq.~\eqref{eq:induced_current}, indicates that for suitable values of $\dot \theta$, the induced current is enhanced by roughly
\begin{align}
 \frac{\dot \theta^2}{m^2} \sim \epsilon \left(\frac{H}{m}\right)^2 \left(\frac{M_P}{f}\right)^2\,,
\end{align}
with $\epsilon = \dot \phi^2/(M_P^2 H^2)/2 < 1$ denoting the first slow-roll parameter. The expression~\eqref{eq:induced_current} is valid for $m^2 \geq g |Q| E$ (with an additional suppression arising for smaller masses , see App.~\ref{app:bilinear}), implying $(H/m)^2 \lesssim H^2/E$. In axion inflation, we typically expect $H^2/E \sim 10^{-4..-2}$~\cite{Domcke:2019qmm} and $(M_P/f)^2 \lesssim 10^3$~\cite{Barnaby:2011qe}, indicating a potentially sizable enhancement of the induced current.
This is in particular true towards the end of inflation, when $\epsilon \sim 1$. 
Consequently, we expect a further reduction of the gauge fields production compared to the analysis of Ref.~\cite{Domcke:2019qmm}.

Moreover, due to the slow variation of $\dot \phi$ over the course of slow-roll inflation, we expect to scan the oscillatory features of Fig.~\ref{fig:HLL}. For moderate values of $\dot \theta$ when only a small number of Landau levels contribute significantly to the induced current, this may lead to an oscillation in the fermion backreaction which could induce oscillations in gauge field background, and thus, through the gauge friction effect, in the inflaton velocity. Since the approximation of an adiabatically varying inflaton velocity and gauge field strength may fail in this case, a detailed analysis of this system requires non-linear methods such as lattice simulations. Qualitatively, resonance effects similar to Refs.~\cite{Cheng:2015oqa,Notari:2016npn,DallAgata:2019yrr,Domcke:2020zez} may occur, leading to characteristic `spikes' in the spectra of scalar and tensor perturbations. We leave a thorough investigation  to future work.

\paragraph{Axion dark matter.} Axion-like particles are intriguing dark matter candidates. Here we are particularly interested in ultralight (sub-eV) axions, which are stable on cosmological time scales and can be described by a coherently oscillating axion background field.\footnote{
After gravitational collapse and structure formation, this remains true on time-scales less than the coherence time, $\tau_\text{coh} \sim 10^6 \, (2 \pi/m_\phi)$, which accounts for the loss of perfect monochromaticity due to virialisation, see e.g.~\cite{Stadnik:2015upa}. 
}
In the rest-frame of an earth-based laboratory, this leads to an `axion wind', with characteristic frequencies associated with the motion with respect to DM halo of our galaxy as well as with the axion mass, $m_\phi$. For an overview of axion searches exploiting this effect, see Ref.~\cite{Stadnik:2015upa}. 

In principle, earth-based experiments aimed at detecting Schwinger production in strong electric fields may thus be sensitive to the axion-electron coupling through a distortion of the high-momentum tail of the electron spectrum, resulting from the enhanced electron production for $m_T > m_e$ depicted in Fig.~\ref{fig:woB}, with $m_e$ denoting the electron mass. However, since $\dot \theta < m_\phi \ll m_e \sim g E$, 
the enhancement is only mild (see Fig.~\ref{fig:woB}). Since moreover
Schwinger production has not yet been successfully observed in the laboratory due to the experimental challenges involved~\cite{Ringwald:2001cp}, this conceptually interesting observation seems of little practical use in the immediate future.

\section{Discussion and conclusions}
\label{sec:conclusions}

In this paper, we study Schwinger production of charged fermions in the background of a homogeneous axion field with a non-vanishing velocity.
By numerically solving the Dirac equation in background gauge and axion fields, we obtain time-dependent Bogoliubov coefficients describing the non-perturbative particle production.
We find that the Schwinger production rate is exponentially enhanced when the axion velocity is sufficiently large~\eqref{eq:maxPP} {and the transverse momentum of the produced particle is non-zero}, which we dub \textit{axion assisted Schwinger effect}.
We also provide a semi-analytic expression for the number densities of produced particles, which well explains our numerical results [see Eq.~\eqref{eq:saddle_pt_approx} and below].

Throughout this paper, we allow for general dimension-five couplings which preserve the axion shift symmetry.
By means of a chiral rotation, some couplings can be expressed by the others, implying that physical quantities should solely depend on their specific linear combinations invariant under it.
We have explicitly demonstrated that our result is invariant under this field redefinition.
In addition, in the presence of magnetic fields parallel to electric fields, a chiral asymmetry is sourced according to the chiral anomaly equation.
We confirm that our result reproduces the chiral anomaly.

The enhancement happens only if a specific combination of parameters $\dot \theta_{5+m}$, invariant under chiral rotations, is non-vanishing [see Eq.~\eqref{eq:thetas} and below Eq.~\eqref{eq:energy-levels}], 
implying that the enhancement dies out for massless fermions.
This is because one may perform a chiral rotation so that all the axion couplings are expressed as the axion Chern-Simons coupling $\phi F \tilde F$ in the case of massless fermions, rendering $\theta_{5+m}$  unphysical.
For this reason, the axion assisted Schwinger effect is more pronounced for a larger mass, while we should bear in mind that the overall production rate gets more suppressed as Eq.~\eqref{eq:maxPP}.
As a consistency check, we also confirm that our result reproduces the earlier studies~\cite{Nielsen:1983rb,Warringa:2012bq,Domcke:2018eki,Copinger:2018ftr,Domcke:2019qmm} in the limit of a vanishing $\theta_{5 + m}$ or massless fermions (see also Figs.~\ref{fig:woB}, \ref{fig:HLL}, and \ref{fig:numerics_mpT_small}).

Based on these results, we discuss phenomenological implications of the axion assisted Schwinger effect.
The enhanced production rate results in an enhancement of the induced current, which backreacts on the gauge field equation of motion.
This is, in particular, relevant for axion inflation which predicts the production of helical gauge fields.
Since the backreaction is enhanced by the axion assisted Schwinger effect, we expect a reduced helical gauge field production in parameter regimes where the axion assisted Schwinger production is relevant. 
Moreover, the enhanced production rate predicts a distortion of the high-momentum tail of the electron spectrum in laboratory experiments aiming at measuring the Schwinger mechanism, such as the X-ray laser XFEL and the extreme-light infrastructure ELI~\cite{Ringwald:2001cp,Dunne:2008kc,Tajima}.
We briefly discuss its implication in the case of axion dark matter, but the enhancement is small for the parameters of our interest.

\paragraph{Acknowledgements}
It is a pleasure to thank Ben Mares for  helpful discussions related to this project.
This work was partly funded by the Deutsche Forschungsgemeinschaft under Germany's Excellence Strategy - EXC 2121 ``Quantum Universe'' - 390833306.

\appendix

\section{Notation and conventions}
\label{app:notation}

In Eq.~\eqref{eq:action_FRW} we have introduced the comoving quantities $\psi, A_\mu$ and $g^{\mu \nu}$, related to the corresponding physical quantities (indicated by a hat) as
\begin{align}
 \psi & = a^{3/2} \hat \psi \,, \qquad A_\mu =  (A_0, - \bm A) = \hat A_\mu  \,, \quad A^\mu = a^2 (A_0, \bm A)  = a^2 \hat A^\mu \,. 
\end{align}
and correspondingly, with $\bm E = - \partial_0 \bm A$, $\bm B = \nabla \times \bm A$ in  temporal gauge $A_0 = 0$,
\begin{align}
  \hat{\bm E} = \bm E/a^2 \,, \qquad \hat{\bm B} = \bm B/a^2 \,.
\end{align}
The indices of the physical quantities are raised/lowered by the FRW metric $g^{\mu\nu}$ whereas the indices of the comoving quantities are raised/lowered by the flat metric $\eta_{\mu \nu} = \text{diag}(+,-,-,-) = g_{\mu \nu}/a^2$. {Note that for brevity, in Eq.~\eqref{eq:Az_time_dep} and thereafter, we denote the constant amplitude of the electric field imposed for some time $\tau$ with $E$.}

Using the chiral representation of the $\gamma$ matrices, $(\gamma^\mu) = (\gamma_0, \bm{\gamma})$ with
\begin{align}
 \{ \gamma^\mu ,\gamma^\nu \} = 2 \eta^{\mu \nu} \,, \quad  \gamma^0 = \begin{pmatrix}
             0 & 1 \\ 1 &  0 
            \end{pmatrix} \,, \quad
 \bm{\gamma} = \begin{pmatrix}
                0 & \bm{\sigma} \\
                - \bm{\sigma} & 0 
               \end{pmatrix} \,, \quad
 \gamma_5 = \begin{pmatrix}
             -1 & 0 \\
             0 & 1
            \end{pmatrix}
\end{align}
the left(right-)handed component of the four-spinor $\psi = (\psi_L, \psi_R)$ is projected out by the projection operator $P_{L/R} = (1 \mp \gamma_5)/2$. The (dual) field strength tensor of the gauge field is given by
\begin{align}
 F_{\mu \nu} = \partial_\mu A_\nu - \partial_\nu A_\mu \,, \quad \widetilde F^{\mu \nu} = \frac{1}{2}\epsilon^{\mu \nu \rho \sigma} F_{\rho \sigma} \,,
\end{align}
with $\epsilon^{0123} = +1$.

\section{Particles and antiparticles}
\label{app:particles}

\paragraph{Hamiltonian.} To introduce the notion of particles and anti-particles, we derive the Hamiltonian density of our system. The conjugate momentum are given from Eq.~\eqref{eq:action_FRW} with $a = 1$ as
\begin{align}
	\pi_\phi &= \dot{\phi} + \frac{c_5}{f_a}\bar{\psi}\gamma^0 \gamma_5 \psi, \\
	\pi_\psi &= i\bar{\psi}\gamma^0, \\
	\pi_{A_i} &= -F^{0i} + c_A\frac{\alpha}{\pi f_a}\phi\,\epsilon^{0ijk}F_{jk}.
\end{align}
The Hamiltonian density is then given by
\begin{align}
	\mathcal{H} = \frac{1}{2}\dot{\phi}^2 + V(\phi) 
	- \bar{\psi}\left(i \gamma^i D_i - m e^{2i\theta_m \gamma_5}\right)\psi
	+ \frac{1}{2}\left(\vec{E}^2 + \vec{B}^2\right).
\end{align}
For our purpose the fermion part is important.
Imposing the equation of motion for $\psi$,
it is given by
\begin{align}
	\mathcal{H}_\psi = \psi^\dagger \left(i \partial_0 + \dot \theta_5 \gamma_5\right) \psi.
	\label{eq:H_app}
\end{align}

\paragraph{Wave function.} In the absence of any magnetic field, we decompose the fermion mode function in Fourier space as
\begin{align}
 \psi = \sum_{s,\sigma} \psi_s^{\sigma} \chi_s^{(\sigma)} \,,
 \label{eq:fermion_decomposition_E}
\end{align}
with
\begin{align}
	\chi_+^{\left({L}\right)} = \begin{pmatrix} 1 \\ 0 \\ 0 \\ 0 \end{pmatrix},~~
	\chi_-^{\left({L}\right)} = \begin{pmatrix} 0 \\ 1 \\ 0 \\ 0 \end{pmatrix},~~
	\chi_+^{\left({R}\right)} = \begin{pmatrix} 0 \\ 0 \\ 1 \\ 0 \end{pmatrix},~~
	\chi_-^{\left({R}\right)} = \begin{pmatrix} 0 \\ 0 \\ 0 \\ 1 \end{pmatrix}.
\end{align}
Here the superscript $(\sigma)$ labels left- and righthanded particles in the $m \rightarrow 0$ limit. The meaning of the subscript $s$ will be more transparent once we include magnetic fields, for the moment this just labels to two linearly independent modes associated with each value of $\lambda$. 

\paragraph{Positive and negative frequency modes.} Inserting Eq.~\eqref{eq:fermion_decomposition_E} into Eq.~\eqref{eq:H_app} gives
\begin{align}
	H_\psi = \int \dd^3 x \mathcal{H}_\psi
	= 
	\sum_{s} \int \frac{\dd^3 p}{\left(2\pi\right)^3}
	\left[\psi_{s}^\dagger 
	\left(i\partial_0 +\dot \theta_5 \gamma_5\right) \psi_{s}\right],
\end{align}
where we have used the notation $\psi_{s} \equiv \sum_{\sigma} \psi_{s}^{(\sigma)}\chi_{s}^{(\sigma)}$. We note that the time derivative is shifted by $i \dot \theta_5 \gamma_5$, and consequently the eigenstates $\tilde \psi$ of the Hamiltonian are given by extracting this phase factor, $\tilde \psi = \exp(- i \gamma_5 \theta_5) \psi$. In this basis, the equation of motion~\eqref{eq:eom} simplifies to 
\begin{align}
	0 = \begin{pmatrix}
	i \partial_0 + \Pi_z   & p_x - ip_y & -m e^{2i\theta_{5+m}} & 0 \\
	p_x + ip_y & i \partial_0 - \Pi_z  & 0 & -m e^{2i\theta_{5+m}} \\
	- m e^{-2i\theta_{5+m}} & 0 & i\partial_0 - \Pi_z  & -(p_x - ip_y) \\
	0 & -m e^{-2i\theta_{5+m}} & -(p_x + ip_y) & i\partial_0 + \Pi_z
	\end{pmatrix}
\tilde \psi \,.
	\label{eq:eom-s}
\end{align}
Inserting the ansatz $\tilde \psi \propto \exp(- i \Omega t)$, we find
\begin{align}
  \Omega = \sqrt{\Pi_z^2 + m_T^2} \,,
\end{align}
which in particular does not depend on $\dot \theta_5$ or $\dot \theta_m$. 

We now proceed to express our wave function as a decomposition of positive and negative frequency states, 
\begin{align}
 \psi =  e^{i \gamma_5 \theta_5} \sum_{\lambda = 1, 2} 
	\left[
	\alpha_\lambda u_\lambda \exp\left(-i \Omega t\right)
	+ \beta_\lambda v_\lambda \exp\left(+i \Omega t\right)
	\right] \,,
\end{align}
where $u_\lambda$, $v_\lambda$ are eigenvectors of the Hamiltonian~\eqref{eq:H_app} for constant $A_z$:
\begin{align}
	u_1 &= \frac{e^{-i\gamma_5 \theta_{5+m}}}{N}
	\begin{pmatrix}
	-(p_x - ip_y) m_T \\ \left(\Omega + \Pi_z\right)\left(m + m_T\right) \\
	(p_x - ip_y) \left(\Omega + \Pi_z\right) \\ m_T\left(m + m_T\right)
	\end{pmatrix},
	\quad
	u_2 = \frac{e^{-i\gamma_5 \theta_{5+m}}e^{-i\varphi_p}}{N}
	\begin{pmatrix}
	m_T\left(m + m_T\right) \\  -(p_x + ip_y)\left(\Omega + \Pi_z\right) \\
	 \left(\Omega + \Pi_z\right)\left(m+m_T\right) \\ (p_x + ip_y) m_T
	\end{pmatrix}, \nonumber \\
	v_1 &= \frac{e^{-i\gamma_5 \theta_{5+m}}}{N}
	\begin{pmatrix}
	-(p_x - ip_y) \left(\Omega + \Pi_z\right) \\  -m_T\left(m + m_T\right) \\
	-(p_x - ip_y)m_T \\ \left(\Omega + \Pi_z\right)\left(m + m_T\right)
	\end{pmatrix},
	\quad
	v_2 = \frac{e^{-i\gamma_5 \theta_{5+m}}e^{-i\varphi_p}}{N}
	\begin{pmatrix}
	 \left(\Omega + \Pi_z\right)\left(m + m_T\right)  \\  (p_x + ip_y) m_T \\
	 -m_T\left(m + m_T\right) \\ (p_x + ip_y)\left(\Omega + \Pi_z\right)
	\end{pmatrix},
	\label{eq:uv}
\end{align}
with the normalization factor
\begin{align}
	N = 2\sqrt{\Omega\left(\Omega + \Pi_z\right)\left(m + m_T\right)m_T}.
\end{align}
Here we define the transverse momentum
\begin{align}
	p_x + i p_y = p_T e^{i\varphi_p},
	\quad
	p_T = \sqrt{p_x^2 + p_y^2},
\end{align}
and choose the phases of $u_2$ and $v_2$ such that
the analogy to the case with the magnetic field becomes transparent.
Note that the eigenvalues $\Omega$ associated with $u_1$ and $u_2$ (and correspondingly $v_1$ and $v_2$) are degenerate. Here we have chosen $u_{n,2}$ and $v_{n,2}$ so as to match $u_0$, $v_0$ for $p_T \rightarrow 0$.

\paragraph{Some useful relations.} Computing the equations of motion for the coefficients $\alpha$ and $\beta$ requires the evaluation of inner products among the vectors  $u_\lambda$, $v_\lambda$ and their time derivatives in the presence of time-dependent $A_z$ and $\theta_{5+m}$. We give some useful relations in the following:
\begin{align}
	u_1^\dagger \dot{u}_1 &= 
	-u_2^\dagger \dot{u}_2 = -v_1^\dagger \dot{v}_1 = v_2^\dagger \dot{v}_2
	= i\dot{\theta}_{5+m} \frac{m}{m_T}\frac{\Pi_z}{\Omega}, \\
	u_1^\dagger \dot{u}_2 &= 
	v_1^\dagger \dot{v}_2 = 
	u_2^\dagger\dot{u}_1  = v_2^\dagger \dot{v}_1
	= -i\dot{\theta}_{5+m} \frac{p_T}{m_T}, \\
	u_1^\dagger \dot{v}_1 &= \frac{m_T}{2\Omega^2}\dot{\Pi}_z - i\dot{\theta}_{5+m}\frac{m}{\Omega},
	\quad
	v_1^\dagger \dot{u}_1 = -\frac{m_T}{2\Omega^2}\dot{\Pi}_z - i\dot{\theta}_{5+m}\frac{m}{\Omega}, \\
	u_2^\dagger \dot{v}_2 &= \frac{m_T}{2\Omega^2}\dot{\Pi}_z + i\dot{\theta}_{5+m}\frac{m}{\Omega},
	\quad
	v_2^\dagger \dot{u}_2 = -\frac{m_T}{2\Omega^2}\dot{\Pi}_z + i\dot{\theta}_{5+m}\frac{m}{\Omega}, \\
	u_1^\dagger \dot{v}_2 &= v_2^\dagger \dot{u}_1 = u_2^\dagger \dot{v}_1 = v_1^\dagger \dot{u}_2 = 0.
\end{align}

\section{Particles and antiparticles - with magnetic field}
\label{app:particlesB}

\paragraph{Wave function.} In the presence of (anti-)parallel electric and magnetic fields pointing in the $z$-direction, we can expand the fermion wave function as 
\begin{align}
	\psi(t,x;p_y,p_z) = \sum_{n, s', \sigma} \psi_{n, s'}^{\left(\sigma\right)}(t) h_{n}(\bar{x}_{s})\chi_{s'}^{\left(\sigma\right)},
	\label{eq:fermion_decomposition}
\end{align}
where we have performed a Fourier-transform with respect to  the $y$ and $z$ direction, $\bar{x}_s = \sqrt{g\abs{QB}} \left(x - s\frac{p_y}{g\abs{QB}}\right)$ and  $h_n$ is related to the Hermite function that satisfies
\begin{align}
	\hat{a} h_n = \sqrt{n}\, h_{n-1},
	\quad
	\hat{a}^\dagger h_n = \sqrt{n+1}\,h_{n+1},
\end{align}
with the ladder operators 
\begin{align}
	\hat{a} = \frac{1}{\sqrt{2}}\left(\partial_{\bar{x}_s} + \bar{x}_s\right),~~
	\hat{a}^\dagger = \frac{1}{\sqrt{2}}\left(-\partial_{\bar{x}_s} + \bar{x}_s\right)\,,
\end{align}
characterizing the Landau levels $n$ with $\partial_{\bar{x}_s} = \frac{1}{\sqrt{g\abs{QB}}} \partial_x$.
Recall that $s = \text{sign}(QB)$.
The mode with $2n + 1 - s s' = 0$, \textit{i.e.}, $n = 0, s' = s$ corresponds to the lowest Landau level,
while the others describe the higher Landau levels.

\paragraph{Positive and negative frequency modes.} As before, we obtain the eigenstates  $\tilde \psi$ of the Hamiltonian by extracting the $\theta_5$ phase factor, $\tilde \psi = \exp(- i \gamma_5 \theta_5) \psi$. In this basis, the equations of motion for the lowest and higher Landau levels, Eqs.~\eqref{eq:eom_lll} and \eqref{eq:eom_hll}, simplify to 
\begin{align}
	0 = \left[i \mathbb{I}_2 \partial_0 + s\Pi_z
	\begin{pmatrix} 1 & 0 \\ 0 & -1 \end{pmatrix} 
	- m \begin{pmatrix} 0 & e^{2i\theta_{5+m}} \\ 
	e^{-2i\theta_{5+m}} & 0 \end{pmatrix} 
	\right]
	\tilde{\psi}_0.
\end{align}
and
\begin{align}
	0 = \begin{pmatrix}
	i \partial_0 + s\Pi_z & i m_B & -m e^{2i\theta_{5+m}} & 0 \\
	-i m_B & i \partial_0 - s\Pi_z & 0 & -m e^{2i\theta_{5+m}} \\
	- m e^{-2i\theta_{5+m}} & 0 & i\partial_0 - s\Pi_z & -i m_B \\
	0 & -m e^{-2i\theta_{5+m}} & i m_B & i\partial_0 + s\Pi_z
	\end{pmatrix}
	\tilde{\psi}_n,
	\label{eq:eom_hll2}
\end{align}
{where we used the short-hand notation
\begin{align}
	\tilde \psi_0 &= \begin{pmatrix} \tilde{\psi}_{0,s}^{\left({L}\right)} \\ \tilde{\psi}_{0,s}^{\left({R}\right)} \end{pmatrix},
	\quad
	\tilde \psi_n = 
	\begin{pmatrix}
	\tilde{\psi}_{n+1, s}^{\left({L}\right)} \\
	\tilde{\psi}_{n, -s}^{\left({L}\right)} \\
	\tilde{\psi}_{n+1, s}^{\left({R}\right)} \\
	\tilde{\psi}_{n, -s}^{\left({R}\right)}
	\end{pmatrix}.
\end{align}
}
Inserting the ansatz $\tilde \psi \propto \exp(- i \Omega t)$, we find
\begin{align}
  \Omega_0 = \sqrt{\Pi_z^2 + m^2} \quad \text{and} \quad \Omega_n = \sqrt{\Pi_z^2 + m_T^2} \,,
\end{align}
which is in particular does not depend on $\dot \theta_5$ or $\dot \theta_m$.

We now proceed to express our wave function as a decomposition of positive and negative frequency states,
\begin{align}\tilde \psi_0 &=  
	\alpha_0 u_0 \exp\left(-i \Omega_0 t\right)
	+ \beta_0 v_0 \exp\left(+i \Omega_0 t\right),\\
 \tilde \psi_n &= \sum_{\lambda = 1, 2} 
	\left[
	\alpha_{n,\lambda} u_{n,\lambda} \exp\left(-i \Omega_n t\right)
	+ \beta_{n,\lambda} v_{n,\lambda} \exp\left(+i \Omega_n t\right)
	\right]\,, \quad n \geq 1\,,
\end{align}
where $u_0$, $v_0$ and $u_{n,\lambda}$, $v_{n,\lambda}$ are eigenvectors of the Hamiltonian~\eqref{eq:H_app} for constant $A_z$. 
For the lowest Landau level, this yields
\begin{align}
	u_0 = \frac{e^{-i\gamma_5 \left(\theta_m + \theta_5\right)}}
	{\sqrt{2\Omega_0 \left(\Omega_0 + s {\Pi}_z\right)}}
	\begin{pmatrix}
	m \\
	\Omega_0 + s {\Pi}_z
	\end{pmatrix},
	\quad
	v_0 = \frac{e^{-i\gamma_5 \left(\theta_m + \theta_5\right)}}
	{\sqrt{2\Omega_0 \left(\Omega_0 + s {\Pi}_z\right)}}
	\begin{pmatrix}
	\Omega_0 + s {\Pi}_z \\
	-m
	\end{pmatrix},
\end{align}
whereas for the higher Landau levels we find
\begin{align}
	u_{n,1} &= \frac{e^{-i\gamma_5 \theta_{5+m}}}{N}
	\begin{pmatrix}
	-i m_B m_T \\ \left(\Omega_n + s \Pi_z\right)\left(m + m_T\right) \\
	i \left(\Omega_n + s \Pi_z\right) m_B \\ m_T\left(m + m_T\right)
	\end{pmatrix},
	\quad
	u_{n,2} = \frac{i e^{-i\gamma_5 \theta_{5+m}}}{N}
	\begin{pmatrix}
	m_T\left(m + m_T\right) \\  i \left(\Omega_n + s \Pi_z\right) m_B \\
	 \left(\Omega_n + s \Pi_z\right)\left(m+m_T\right) \\ -i m_B m_T
	\end{pmatrix}, \\
	v_{n,1} &= \frac{e^{-i\gamma_5 \theta_{5+m}}}{N}
	\begin{pmatrix}
	-i \left(\Omega_n + s \Pi_z\right) m_B \\  -m_T\left(m + m_T\right) \\
	-i m_Bm_T \\ \left(\Omega_n + s \Pi_z\right)\left(m + m_T\right)
	\end{pmatrix},
	\quad
	v_{n,2} = \frac{i e^{-i\gamma_5 \theta_{5+m}}}{N}
	\begin{pmatrix}
	 \left(\Omega_n + s \Pi_z\right)\left(m + m_T\right)  \\  -i m_Bm_T \\
	 -m_T\left(m + m_T\right) \\ -i \left(\Omega_n + s \Pi_z\right) m_B
	\end{pmatrix},
\end{align}
where the normalization factor is
\begin{align}
	N = 2\sqrt{\Omega_n \left(\Omega_n + s \Pi_z\right) m_T\left(m + m_T\right)}.
\end{align}
Note that the eigenvalues $\Omega_n$ associated with $u_{n,1}$ and $u_{n,2}$ (and correspondingly $v_{n,1}$ and $v_{n,2}$) are degenerate. Here we have chosen $u_{n,2}$ and $v_{n,2}$ so as to match $u_0$, $v_0$ for $m_B \rightarrow 0$ for $s>0$. 

\paragraph{Some useful relations.} Computing the equations of motion for the Bogoliubov coefficients requires the evaluation of inner products among the vectors $u_0$, $v_0$ and $u_{n,\lambda}$, $v_{n,\lambda}$. We give some useful relations in the following (dropping for notational simplicity the index $n$ for the higher Landau levels):
\begin{align}
	u_0^\dagger \dot{u}_0 &= -i \dot{\theta}_{5+m} \frac{s {\Pi}_z}{\Omega_0},
	\quad
	v_0^\dagger \dot{v}_0 = i \dot{\theta}_{5+m} \frac{s {\Pi}_z}{\Omega_0}, \\
	u_0^\dagger \dot{v}_0 &= s \frac{m \dot{{\Pi}}_z}{2\Omega_0^2} + i\dot{\theta}_{5+m} \frac{m}{\Omega_0},
	\quad
	v_0^\dagger \dot{u}_0 = -s\frac{m \dot{{\Pi}}_z}{2\Omega_0^2} + i\dot{\theta}_{5+m} \frac{m}{\Omega_0},
\end{align}
for the lowest Landau level, and correspondingly
\begin{align}
	u_{1}^\dagger \dot{u}_{1} &= 
	-u_2^\dagger \dot{u}_2 = -v_1^\dagger \dot{v}_1 = v_2^\dagger \dot{v}_2
	= i\dot{\theta}_{5+m} \frac{m}{m_T}\frac{s\Pi_z}{\Omega_n}, \\
	u_1^\dagger \dot{u}_2 &= u_2^\dagger\dot{u}_1 
	= v_1^\dagger \dot{v}_2 = v_2^\dagger \dot{v}_1
	= -i\dot{\theta}_{5+m} \frac{m_B}{m_T}, \\
	u_1^\dagger \dot{v}_1 &= \frac{m_T}{2\Omega_n^2}s\dot{\Pi}_z - i\dot{\theta}_{5+m}\frac{m}{\Omega_n},
	\quad
	v_1^\dagger \dot{u}_1 = -\frac{m_T}{2\Omega_n^2}s\dot{\Pi}_z - i\dot{\theta}_{5+m}\frac{m}{\Omega_n}, \\
	u_2^\dagger \dot{v}_2 &= \frac{m_T}{2\Omega_n^2}s\dot{\Pi}_z + i\dot{\theta}_{5+m}\frac{m}{\Omega_n},
	\quad
	v_2^\dagger \dot{u}_2 = -\frac{m_T}{2\Omega_n^2}s\dot{\Pi}_z + i\dot{\theta}_{5+m}\frac{m}{\Omega_n}, \\
	u_1^\dagger \dot{v}_2 &= v_2^\dagger \dot{u}_1 = u_2^\dagger \dot{v}_1 = v_1^\dagger \dot{u}_2 = 0,
\end{align}
for the higher Landau levels.

In addition, the following relations are useful to show the anomaly equation.
For the lowest Landau level,
\begin{align}
	u_0^\dagger \gamma_5 u_0 &= - v_0^\dagger \gamma_5 v_0 = s\frac{{\Pi}_z}{\Omega_0},
	\quad
	v_0^\dagger \gamma_5 u_0 =  u_0^\dagger \gamma_5 v_0 = -\frac{m}{\Omega_0}, 
\end{align}
and for the higher Landau levels,
\begin{align}
	u_1^\dagger \gamma_5 u_1 
	&= - u_2^\dagger \gamma_5 u_2 
	= - v_1^\dagger \gamma_5 v_1
	= v_2^\dagger \gamma_5 v_2
	= -\frac{m}{m_T}\frac{s\Pi_z}{\Omega_n}, \\
	u_1^\dagger \gamma_5 u_2
	&= u_2^\dagger \gamma_5 u_1
	= v_1^\dagger \gamma_5 v_2
	= v_2^\dagger \gamma_5 v_1
	= \frac{m_B}{m_T}, \\
	u_1^\dagger \gamma_5 v_1 
	&= v_1^\dagger \gamma_5 u_1
	= -u_2^\dagger \gamma_5 v_2
	= -v_2^\dagger \gamma_5 u_2
	= \frac{m}{\Omega_n}, \\
	u_1^\dagger \gamma_5 v_2 
	&= v_2^\dagger \gamma_5 u_1
	= u_2^\dagger \gamma_5 v_1
	= v_1^\dagger \gamma_5 u_2
	= 0,
\end{align}
are useful relations for the evaluation of the chiral charge.  Moreover, for the lowest Landau level,
\begin{align}	
	u_0^\dagger \gamma^0 e^{2i\theta_m \gamma_5} \gamma_5 u_0 
	&= v_0^\dagger \gamma^0 e^{2i\theta_m \gamma_5} \gamma_5 v_0 = 0,
	\quad
	v_0^\dagger \gamma^0 e^{2i\theta_m \gamma_5} \gamma_5 u_0 = 
	-u_0^\dagger \gamma^0 e^{2i\theta_m \gamma_5} \gamma_5 v_0 = 1, 
\end{align}
and for the higher Landau levels,
\begin{align}
	u_i^\dagger \gamma^0 e^{2i\gamma_5 \theta_{5+m}} \gamma_5 u_j 
	&= v_i^\dagger \gamma^0 e^{2i\gamma_5 \theta_{5+m}} \gamma_5 v_j 
	=
	u_1^\dagger \gamma^0 e^{2i\gamma_5 \theta_{5+m}} \gamma_5 v_2
	= u_2^\dagger \gamma^0 e^{2i\gamma_5 \theta_{5+m}} \gamma_5 v_1 
	= 0, \\
	u_1^\dagger \gamma^0 e^{2i\gamma_5 \theta_{5+m}} \gamma_5 v_1 
	&= -v_1^\dagger \gamma^0 e^{2i\gamma_5 \theta_{5+m}} \gamma_5 u_1
	= -u_2^\dagger \gamma^0 e^{2i\gamma_5 \theta_{5+m}} \gamma_5 v_2 
	= v_2^\dagger \gamma^0 e^{2i\gamma_5 \theta_{5+m}} \gamma_5 u_2 
	= 1,
\end{align}
with $i,j \in \{1,2\}$,  are useful for evaluating the mass term.

\section{Bilinear forms of Bogoliubov coefficients and particle production}
\label{app:bilinear}

In this appendix, we consider products of the Bogoliubov coefficients.
The purpose of this appendix is two-fold.
First, we show the existence of the conserved quantities 
(Eqs.~\eqref{eq:conserved_quantity1} and~\eqref{eq:conserved_quantity2} in the case with only an electric field).
Second, we motivate the analytical formula that we use to estimate the spectrum~\eqref{eq:saddle_pt_approx}.
Although we could not derive this formula rigorously, we outline our computation that leads us to this formula,
with the hope that one may find the argument there useful to derive a more complete analytical formula in the future.
We consider the case with only an electric field in the following, 
but the results equally apply to the case with both an electric and magnetic field after replacing $p_T$ by $m_B$.

For our purpose, it is convenient to treat the Bogoliubov coefficients associated with the two sets of initial conditions (see Eq.~\eqref{eq:init_cond_woB}) in a unified way.
We thus define matrices as
\begin{align}
	\left(\alpha \right)_{\lambda \lambda'} &\equiv e^{-i\Theta}\alpha_{\lambda}^{(\lambda')}, \\
	\left(\beta \right)_{\lambda \lambda'} &\equiv e^{i\Theta}\beta_{\lambda}^{(\lambda')},
\end{align}
where the quantities in the left-hand-side are now understood as $2\times 2$ matrices.
These matrices satisfy
\begin{align}
	\dot{\alpha} &=
	\left(-i\Omega -i\dot{\theta}_{5+m}\frac{m}{m_T}\frac{\Pi_z}{\Omega}\sigma_3 
	+ i\dot{\theta}_{5+m}\frac{p_T}{m_T}\sigma_1\right) \alpha
	+
	\left(-\frac{m_T\dot{\Pi}_z}{2\Omega^2}
	+ i\dot{\theta}_{5+m} \frac{m}{\Omega} \sigma_3\right) \beta, 
	\label{eq:eom_alpha_matrix} \\
	\dot{\beta} &=
	\left(\frac{m_T\dot{\Pi}_z}{2\Omega^2}
	+ i\dot{\theta}_{5+m}\frac{m}{\Omega}\sigma_3 \right) \alpha
	+ \left( i\Omega
	+ i\dot{\theta}_{5+m} \frac{m}{m_T}\frac{\Pi_z}{\Omega}\sigma_3
	+ i\dot{\theta}_{5+m} \frac{p_T}{m_T}\sigma_1
	\right)\beta,
	\label{eq:eom_beta_matrix}
\end{align}
where $\sigma_i$ is the standard Pauli matrix,
 and their initial conditions are given by
\begin{align}
	\alpha = \mathbb{I}_{2}\,, 
	\quad
	\beta = 0\,.
\end{align}
We consider products of these matrices.
They are again $2\times 2$ matrices and hence can be expanded in terms of the Pauli matrix as
\begin{align}
	\alpha \alpha^\dagger &= R^0_\alpha \mathbb{I}_{2} + R^i_\alpha \sigma_i,
	\quad
	\beta \beta^\dagger = R^0_\beta\mathbb{I}_{2} + R^i_\beta \sigma_i,
	\quad
	\alpha \beta^\dagger = C^0 \mathbb{I}_{2} + C^i \sigma_i.
\end{align}
Note that $R^\mu_\alpha$ and $R^\mu_\beta$ are real while $C^\mu$ are complex.
The initial condition now reads $R_\alpha^0 = 1$ and $R_\alpha^i = R_\beta^\mu = C^\mu = 0$ at the initial time.

\paragraph{Conserved quantities.}
The conserved quantities~\eqref{eq:conserved_quantity1} and~\eqref{eq:conserved_quantity2}
are expressed as
\begin{align}
	R^0_\alpha + R^0_\beta = 1,
	\quad
	R^1_\alpha - R^1_\beta = R^2_\alpha + R^2_\beta = R^3_\alpha + R^3_\beta = C^1 = 0.
\end{align}
These relations can be shown as follows.
The equations of motion of $R_\alpha^\mu$, $R_\beta^\mu$ and $C^\mu$
are derived from Eqs.~\eqref{eq:eom_alpha_matrix} and~\eqref{eq:eom_beta_matrix} as
\begin{align}
	\dot{R}_\alpha^0 
	&= -\frac{m_T\dot{\Pi}_z}{2\Omega^2}\left(C^0 + {C^0}^*\right)
	- i\dot{\theta}_{5+m}\frac{m}{\Omega}\left(C^3 - {C^3}^*\right), \\
	\dot{R}_\alpha^1 
	&= -2\dot{\theta}_{5+m}\frac{m}{m_T}\frac{\Pi_z}{\Omega}R^2_\alpha
	- \frac{m_T\dot{\Pi}_z}{2\Omega^2}\left(C^1 + {C^1}^*\right)
	+ \dot{\theta}_{5+m}\frac{m}{\Omega}\left(C^2 + {C^2}^*\right), \\
	\dot{R}_\alpha^2 
	&= 2\dot{\theta}_{5+m}\left(\frac{m}{m_T}\frac{\Pi_z}{\Omega}R^1_\alpha
	+ \frac{p_T}{m_T}R^3_\alpha\right)
	- \frac{m_T\dot{\Pi}_z}{2\Omega^2}\left(C^2 + {C^2}^*\right)
	- \dot{\theta}_{5+m}\frac{m}{\Omega}\left(C^1 + {C^1}^*\right), \\
	\dot{R}_\alpha^3 
	&= -2\dot{\theta}_{5+m}\frac{p_T}{m_T}R^2_\alpha
	- \frac{m_T\dot{\Pi}_z}{2\Omega^2}\left(C^3 + {C^3}^*\right)
	- i\dot{\theta}_{5+m}\frac{m}{\Omega}\left(C^0 - {C^0}^*\right), \\
	\dot{R}_\beta^0 
	&= \frac{m_T\dot{\Pi_z}}{2\Omega^2}\left(C^0 + {C^0}^*\right)
	+ i\dot{\theta}_{5+m}\frac{m}{\Omega}\left(C^3 - {C^3}^*\right), \\
	\dot{R}_\beta^1 
	&= 2\dot{\theta}_{5+m}\frac{m}{m_T}\frac{\Pi_z}{\Omega}R^2_\beta
	+ \frac{m_T\dot{\Pi}_z}{2\Omega^2}\left(C^1 + {C^1}^*\right)
	+ \dot{\theta}_{5+m}\frac{m}{\Omega}\left(C^2 + {C^2}^*\right), \\
	\dot{R}_\beta^2 
	&= 2\dot{\theta}_{5+m}\left(-\frac{m}{m_T}\frac{\Pi_z}{\Omega}R^1_\beta
	+ \frac{p_T}{m_T}R^3_\beta\right)
	+ \frac{m_T\dot{\Pi}_z}{2\Omega^2}\left(C^2 + {C^2}^*\right)
	- \dot{\theta}_{5+m}\frac{m}{\Omega}\left(C^1 + {C^1}^*\right), \\
	\dot{R}_\beta^3 
	&= -2\dot{\theta}_{5+m}\frac{p_T}{m_T}R^2_\beta
	+ \frac{m_T\dot{\Pi}_z}{2\Omega^2}\left(C^3 + {C^3}^*\right)
	+ i\dot{\theta}_{5+m}\frac{m}{\Omega}\left(C^0 - {C^0}^*\right), \\
	\dot{C}^0
	&= - 2i\Omega C^0 - 2i\dot{\theta}_{5+m}\frac{m}{m_T}\frac{\Pi_z}{\Omega}C^3
	+ \frac{m_T \dot{\Pi}_z}{2\Omega^2}\left(R_\alpha^0 - R_\beta^0\right)
	- i\dot{\theta}_{5+m}\frac{m}{\Omega}\left(R_\alpha^3 - R_\beta^3\right), \\
	\dot{C}^1
	&= -2i\Omega C^1
	+ \frac{m_T \dot{\Pi}_z}{2\Omega^2}\left(R_\alpha^1 - R_\beta^1\right)
	+ \dot{\theta}_{5+m}\frac{m}{\Omega}\left(R_\alpha^2 + R_\beta^2\right), \\
	\dot{C}^2
	&= -2i\Omega C^2
	+ 2\dot{\theta}_{5+m}\frac{p_T}{m_T}C^3
	+ \frac{m_T\dot{\Pi}_z}{2\Omega^2}\left(R_\alpha^2 - R_\beta^2\right)
	- \dot{\theta}_{5+m}\frac{m}{\Omega}\left(R_\alpha^1 + R_\beta^1\right), \\
	\dot{C}^3
	&= -2i\Omega C^3
	-2i\dot{\theta}_{5+m}\frac{m}{m_T}\frac{\Pi_z}{\Omega}C^0
	- 2\dot{\theta}_{5+m}\frac{p_T}{m_T}C^2
	+ \frac{m_T \dot{\Pi}_z}{2\Omega^2}\left(R^3_\alpha - R^3_\beta\right)
	- i\dot{\theta}_{5+m}\frac{m}{\Omega}\left(R_\alpha^0 - R_\beta^0\right).
\end{align}
One can see that
\begin{align}
	\dot{R}_\alpha^0 + \dot{R}_\beta^0 = 0,
\end{align}
and it follows from the initial condition that $R_\alpha^0 + R_\beta^0 = 1$.
One can also see that the equations of motion of $R^1_\alpha - R^1_\beta$,  
$R^2_\alpha + R^2_\beta$,  $R^3_\alpha + R^3_\beta$ and $C^1$ form a closed sub-system,
indicating that they all vanish as they vanish at the initial time. 
Note that the equations are linear in $R_\alpha^\mu$, $R_\beta^\mu$ and $C^\mu$.

\paragraph{Particle production.}
The occupation number of produced particles is given by
\begin{align}
	\sum_{\lambda, \lambda'} \abs{\beta_{\lambda}^{\lambda'}}^2
	= \mathrm{Tr}\left[\beta \beta^\dagger\right] = 2 R_\beta^0.
\end{align}
After exploiting the conserved quantities, we can reduce the equations of motion as
\begin{align}
	\dot{v} &= M v + c, \label{eq:v-eom} \\
	v &= \begin{pmatrix}
	R_\beta^0 & R_\beta^1 & R_\beta^2 & R_\beta^3 &
	\mathrm{Re}C^0 & \mathrm{Im}C^0 & \mathrm{Re}C^2 & \mathrm{Im}C^2 &
	\mathrm{Re}C^3 & \mathrm{Im}C^3
	\end{pmatrix}^T, \\
	c &= \begin{pmatrix}
	0 & 0 & 0 & 0 & \frac{m_T \dot{\Pi}_z}{2\Omega^2} &
	0 & 0 & 0 & 0 & -\frac{\dot{\theta} m}{\Omega}
	\end{pmatrix}^T, \\
	\hspace{-1cm}M &= \begin{pmatrix}
	0 & 0 & 0 & 0 & \frac{m_T \dot{\Pi}_z}{\Omega^2} & 0 & 0 & 0 & 0 & -\frac{2\dot{\theta} m}{\Omega} \\ 
	0 & 0 & \frac{2 \dot{\theta} m}{m_T}\frac{\Pi_z }{\Omega} & 0 & 0 & 0 & \frac{2\dot{\theta} m}{\Omega} & 0 & 0 & 0 \\
	0 & -\frac{2 \dot{\theta} m}{m_T}\frac{\Pi_z }{\Omega} & 0 & \frac{2\dot{\theta} p_T}{m_T} & 
	0 & 0 & \frac{m_T\dot{\Pi}_z}{\Omega^2} & 0 & 0 & 0 \\
	0 & 0 & -\frac{2\dot{\theta} p_T}{m_T} & 0 & 0 & -\frac{2\dot{\theta} m}{\Omega} & 0 & 0 & \frac{m_T\dot{\Pi}_z}{\Omega^2} & 0 \\
	-\frac{m_T\dot{\Pi}_z}{\Omega^2} & 0 & 0 & 0 & 0 & 2\Omega & 0 & 0 & 0 & \frac{2 \dot{\theta} m}{m_T}\frac{\Pi_z }{\Omega} \\
	0 & 0 & 0 & \frac{2\dot{\theta} m}{\Omega} & -2\Omega & 0 & 0 & 0 & -\frac{2 \dot{\theta} m}{m_T}\frac{\Pi_z }{\Omega} & 0 \\
	0 & -\frac{2\dot{\theta} m}{\Omega} & -\frac{m_T\dot{\Pi}_z}{\Omega^2} & 0 & 0 & 0 & 0 & 2\Omega & \frac{2\dot{\theta} p_T}{m_T} & 0 \\
	0 & 0 & 0 & 0 & 0 & 0 & -2\Omega & 0 & 0 & \frac{2\dot{\theta} p_T}{m_T} \\
	0 & 0 & 0 & -\frac{m_T\dot{\Pi}_z}{\Omega^2} & 0 & \frac{2 \dot{\theta} m}{m_T}\frac{\Pi_z }{\Omega} & -\frac{2\dot{\theta} p_T}{m_T} &
	0 & 0 & 2\Omega \\
	\frac{2\dot{\theta} m}{\Omega} & 0 & 0 & 0 & -\frac{2 \dot{\theta} m}{m_T}\frac{\Pi_z }{\Omega} & 0 & 0 & 
	-\frac{2\dot{\theta} p_T}{m_T} & -2\Omega & 0
	\end{pmatrix},
\end{align}
where we omit the subscript $5+m$ here and here only for notational simplicity.
We can formally solve this equation as follows.
We define the transfer function as
\begin{align}
	U(t, t_0) \equiv \mathcal{T}\exp\left[\int_{t_0}^t \dd t'\,M(t')\right],
\end{align}
where $\mathcal{T}$ indicates the time-ordered product.
This transfer function satisfies
\begin{align}
	\frac{\dd}{\dd t}U(t, t_0) = M(t) U(t, t_0),
\end{align}
and is unitary,
\begin{align}
	U^\dagger(t, t_0) = U^{-1}(t, t_0),
\end{align}
since $M$ is real and anti-symmetric.
With the help of this function, we can formally express the solution of the equation of motion~\eqref{eq:v-eom} as
\begin{align}
	v(t = \infty) = \int_{-\infty}^{\infty} \dd t\, U(\infty, t) c(t),
\end{align}
where we used $v(-\infty) = 0$.

\paragraph{Adiabatic approximation.}
We may evaluate this expression in the adiabatic approximation.
For this purpose, 
we denote the eigensystem of $M$ at a given time $t$ as
\begin{align}
	M(t) \lvert n; t\rangle = i \lambda_n(t) \lvert n; t\rangle,
\end{align}
and assume that $\lvert n;t\rangle$ is an orthonormal basis.
Note that $\lambda_n$ is real since $M$ is anti-hermitian.
In order to perform the adiabatic approximation, we may expand $U$ as
\begin{align}
	U(t, t_0) = \sum_{n, m} d_{nm}(t, t_0) e^{i \int^t_{t_0} \dd t'\, \lambda_n (t')}
	\lvert n; t\rangle \langle m; t_0\rvert,
\end{align}
and also $v$ and $c$ as
\begin{align}
	v(t) = \sum_n v_n(t) \lvert n; t\rangle,
	\quad
	c(t) = \sum_n c_n(t) \lvert n; t\rangle.
\end{align}
Note that $d_{nm}$ denotes the transition from an $m$-state to an $n$-state,
and hence the off-diagonal components are expected to be exponentially suppressed
in the adiabatic limit. The equation of motion of $d_{nm}$ reads
\begin{align}
	\dot{d}_{nm}(t, t_0) = -\sum_l e^{i \int^t_{t_0}\dd t'\left(\lambda_l - \lambda_m\right)}
	\left(\langle n; t\vert \frac{\dd}{\dd t}\vert l; t\rangle \right)
	d_{lm}(t, t_0),
	\label{eq:eom_dnm}
\end{align}
where the time derivative acts on $t$, not on $t_0$.

If there is no level crossing\footnote{
	In the case of our interest, there are actually level crossings as one increases $\dot{\theta}_{5+m}$.
	We ignore this subtlety here, justified a posteriori by a numerical verification of our result.
} and $m$ is large compared to the electric field, 
the adiabatic limit would be a good approximation.
To the leading order in the adiabatic limit, no transition among different eigenstates occurs. 
This means in that 
\begin{align}
	d_{nm}^{(0)} \propto \delta_{nm},
\end{align}
where the superscript indicates that this is the leading order in the adiabatic approximation.
By substituting it to Eq.~\eqref{eq:eom_dnm},
we obtain
\begin{align}
	d_{nm}^{(0)}(t, t_0) = e^{-\gamma_n(t, t_0)}\delta_{nm},
	\quad
	\gamma_n(t, t_0) = \int^t_{t_0} \dd t' \langle n; t'\vert \frac{\dd}{\dd t'}\vert n; t'\rangle.
\end{align}
This phase factor $\gamma_n(t, t_0)$ is called the geometrical phase, or the Berry phase~\cite{Berry:1984jv}.
In the adiabatic limit, one must pay an exponential suppression for each transition,
so at the next-to-leading order we can approximate the right hand side of Eq.~\eqref{eq:eom_dnm}
by $d_{nm}^{(0)}$.
Thus, we obtain
\begin{align}
	d_{nm}^{(1)}(t, t_0) = - \int_{t_0}^{t}\dd t' e^{i \int^{t'}_{t_0} \dd t'' \left(\lambda_m - \lambda_n\right) - \gamma_m(t', t_0)}
	\langle n; t'\vert \frac{\dd}{\dd t'}\vert m; t'\rangle 
	\quad
	\mathrm{for}~~m\neq n.
\end{align}
Thus, we may obtain $v_n(\infty)$ up to the next-to-leading order as
\begin{align}
	v_n(\infty) &\simeq \int_{-\infty}^{\infty}\dd t e^{i \int_t^{\infty}\dd t' \lambda_n(t') - \gamma_n(\infty, t)} c_n(t) \nonumber \\
	&+ \sum_{m\neq n} \int_{-\infty}^{\infty}e^{i \int_t^{\infty}\dd t' \lambda_n(t')}
	\left[
	\int_t^{\infty} \dd t' e^{i \int_t^{t'}\dd t''\left(\lambda_m - \lambda_n\right) - \gamma_m(t',t)}
	\langle n;t'\vert \frac{\dd}{\dd t'}\vert m; t'\rangle
	\right] c_m(t).
	\label{eq:adiabatic_approx}
\end{align}
We can repeat this computation and derive the higher order expressions.

\begin{figure}[t]
	\centering
 	\includegraphics[width=0.495\linewidth]{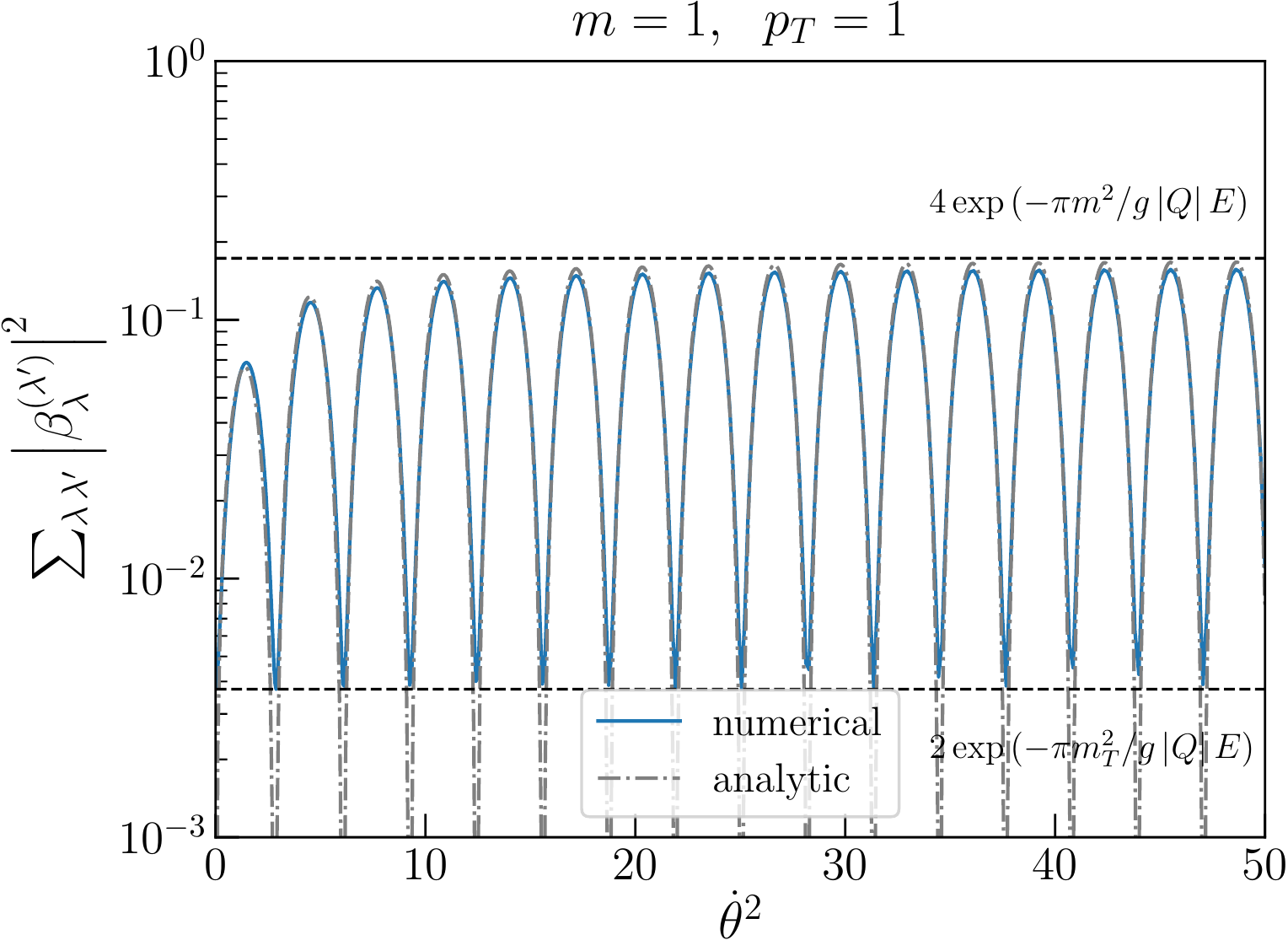}
 	\includegraphics[width=0.495\linewidth]{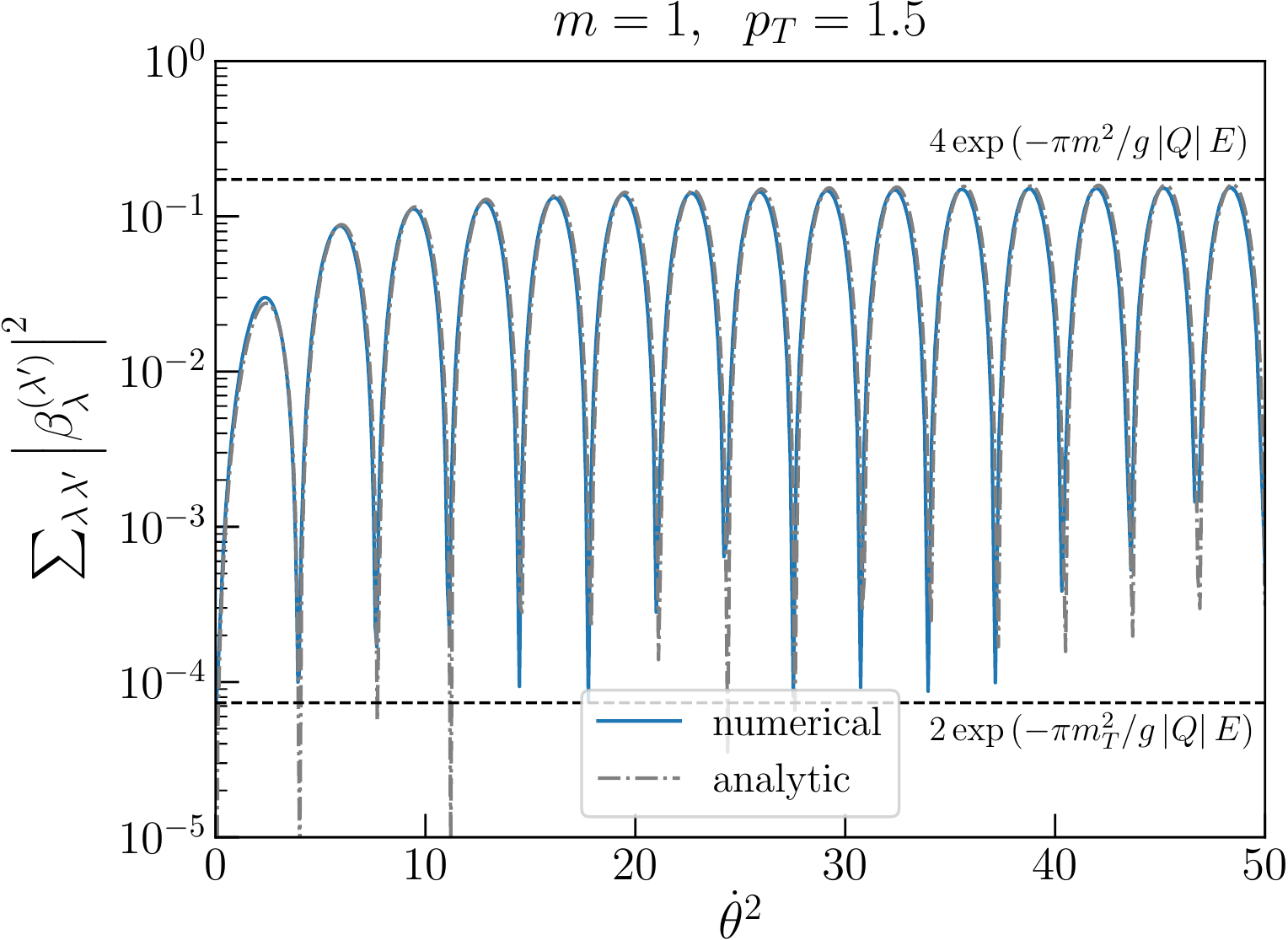}
 	\includegraphics[width=0.495\linewidth]{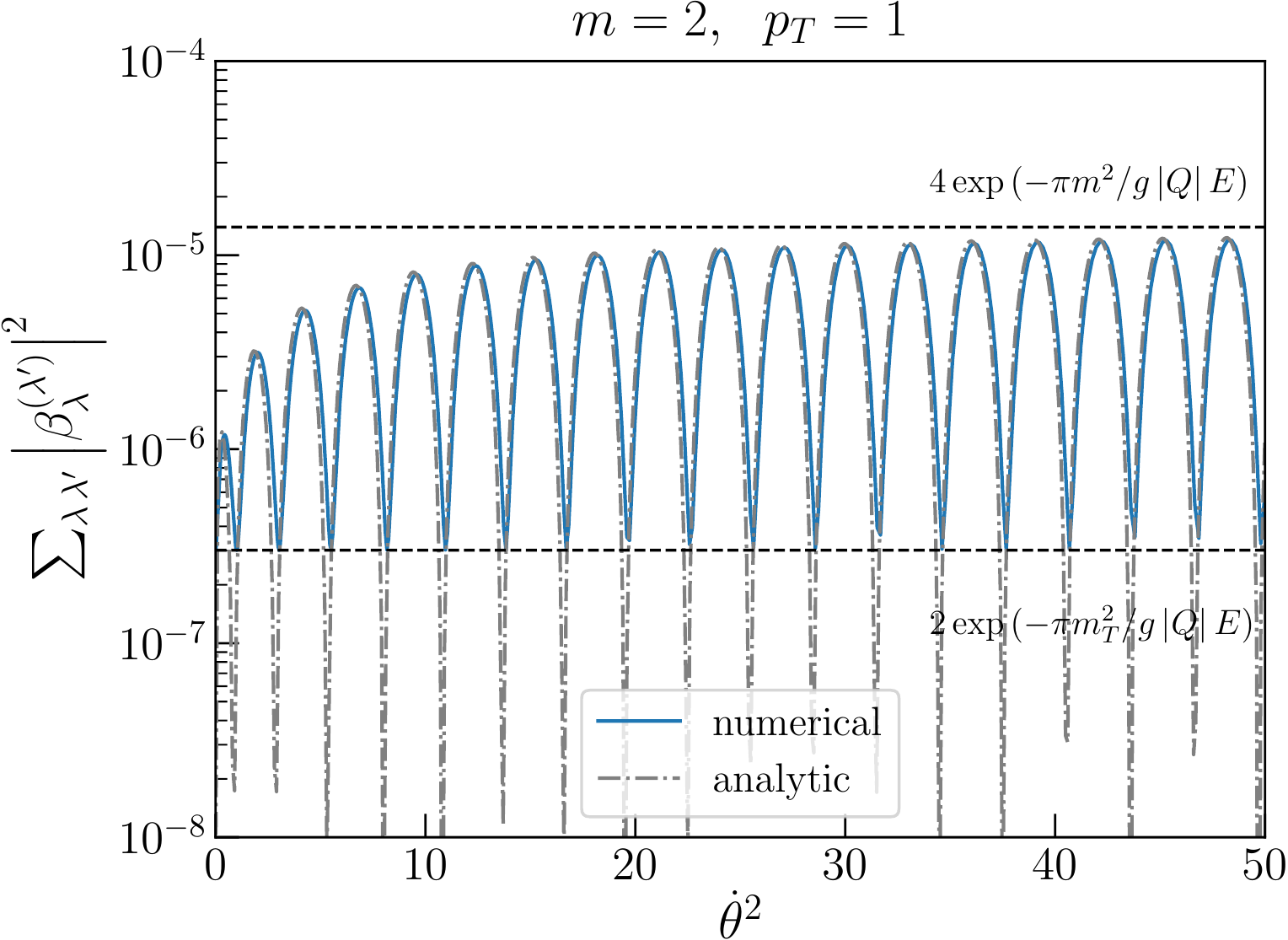}
 	\includegraphics[width=0.495\linewidth]{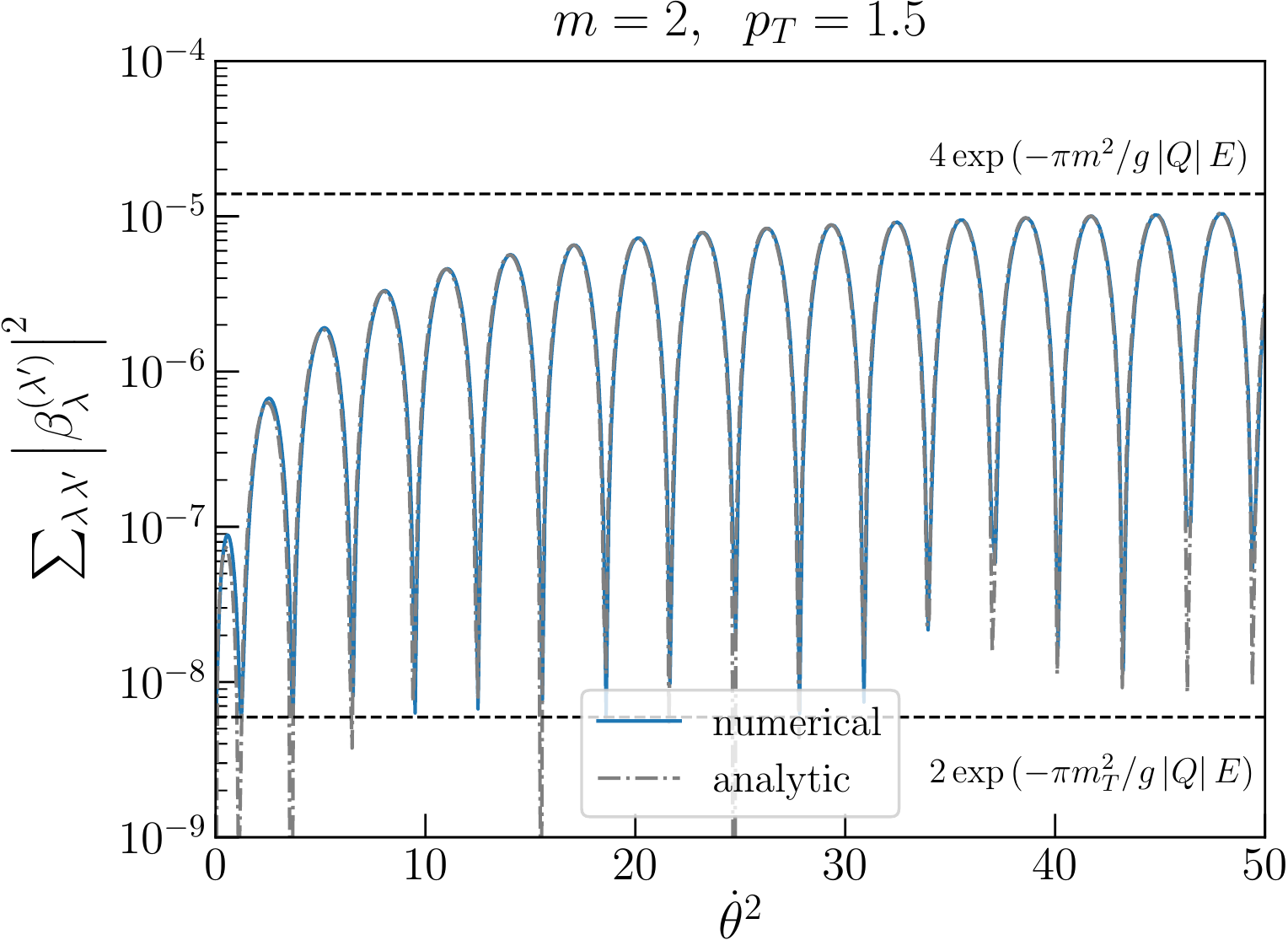}
	\caption{\small The height of the plateau of the spectrum  of $\sum_{\lambda \lambda'} |\beta_\lambda^{(\lambda')}|^2$ evaluated at $p_z = -50$
	as a function of $\dot{\theta}$ for a variety of model parameters.
	The blue solid lines are the full numerical results,
	while the gray dashed lines are Eq.~\eqref{eq:approx_full_system_app}.
	The parameters are shown in the unit $g\abs{Q}E = 1$.}
	\label{fig:saddle_pt_large}
\end{figure}
\begin{figure}[t]
	\centering
 	\includegraphics[width=0.495\linewidth]{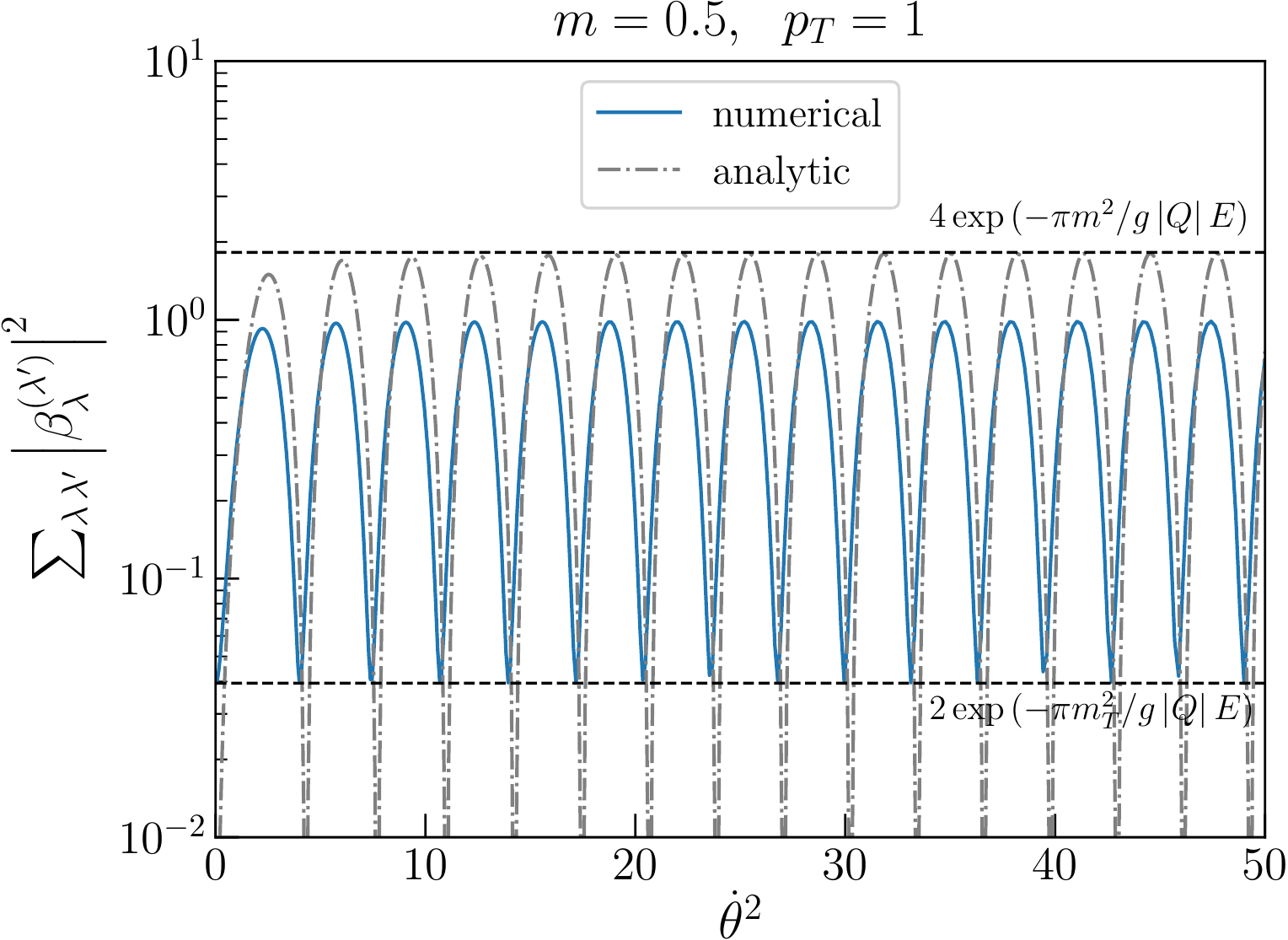}
 	\includegraphics[width=0.495\linewidth]{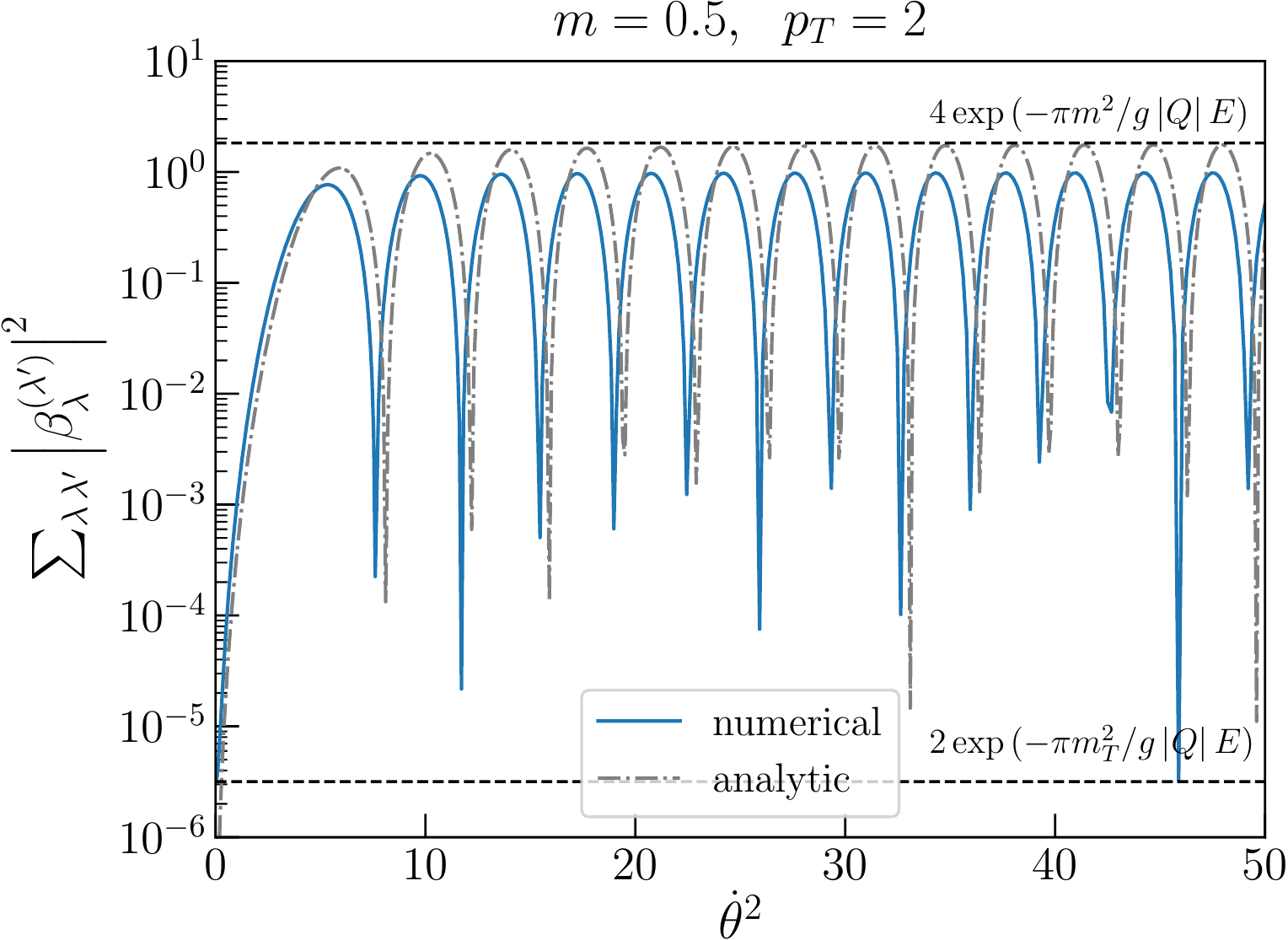}
 	\includegraphics[width=0.495\linewidth]{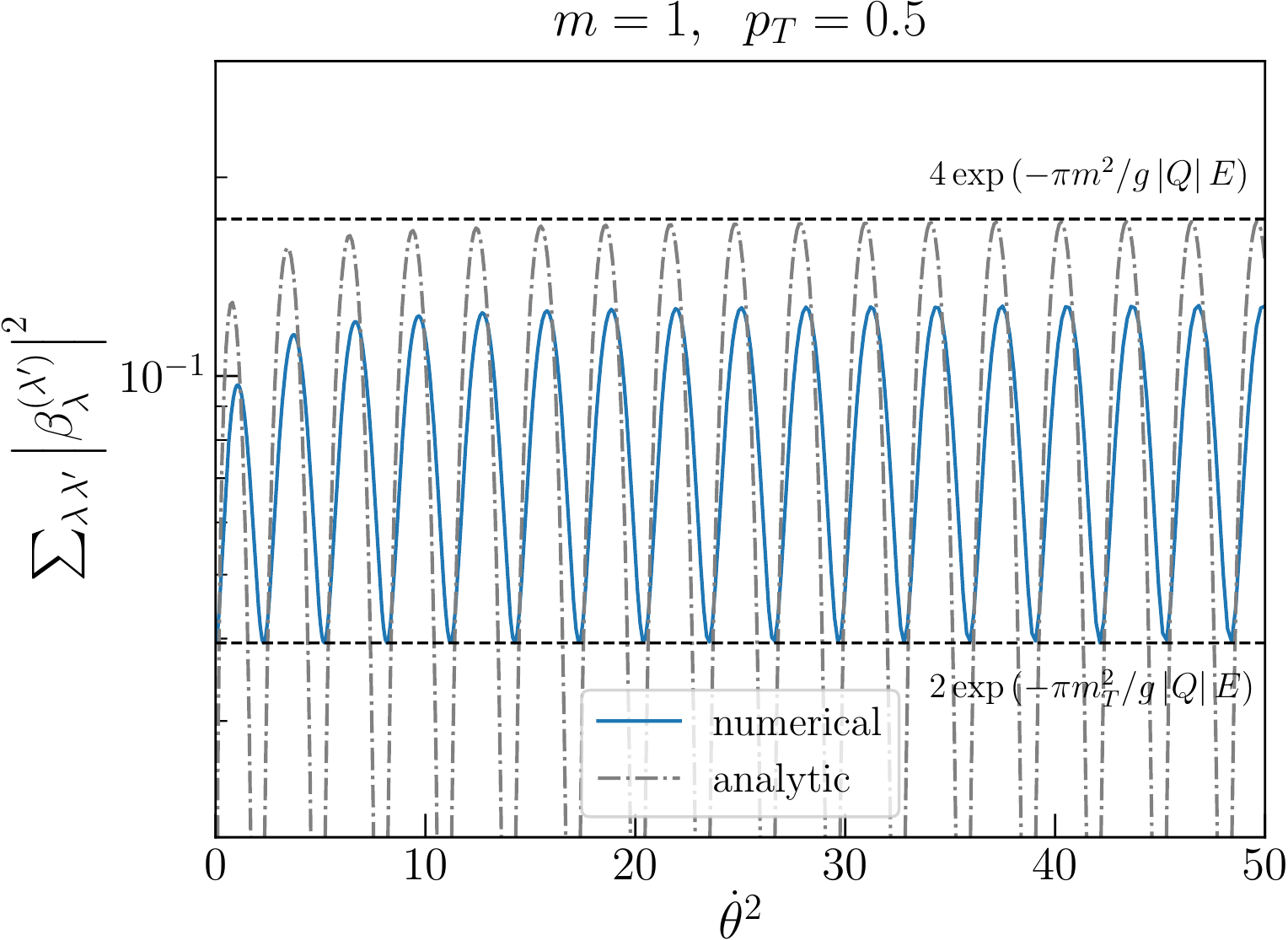}
 	\includegraphics[width=0.495\linewidth]{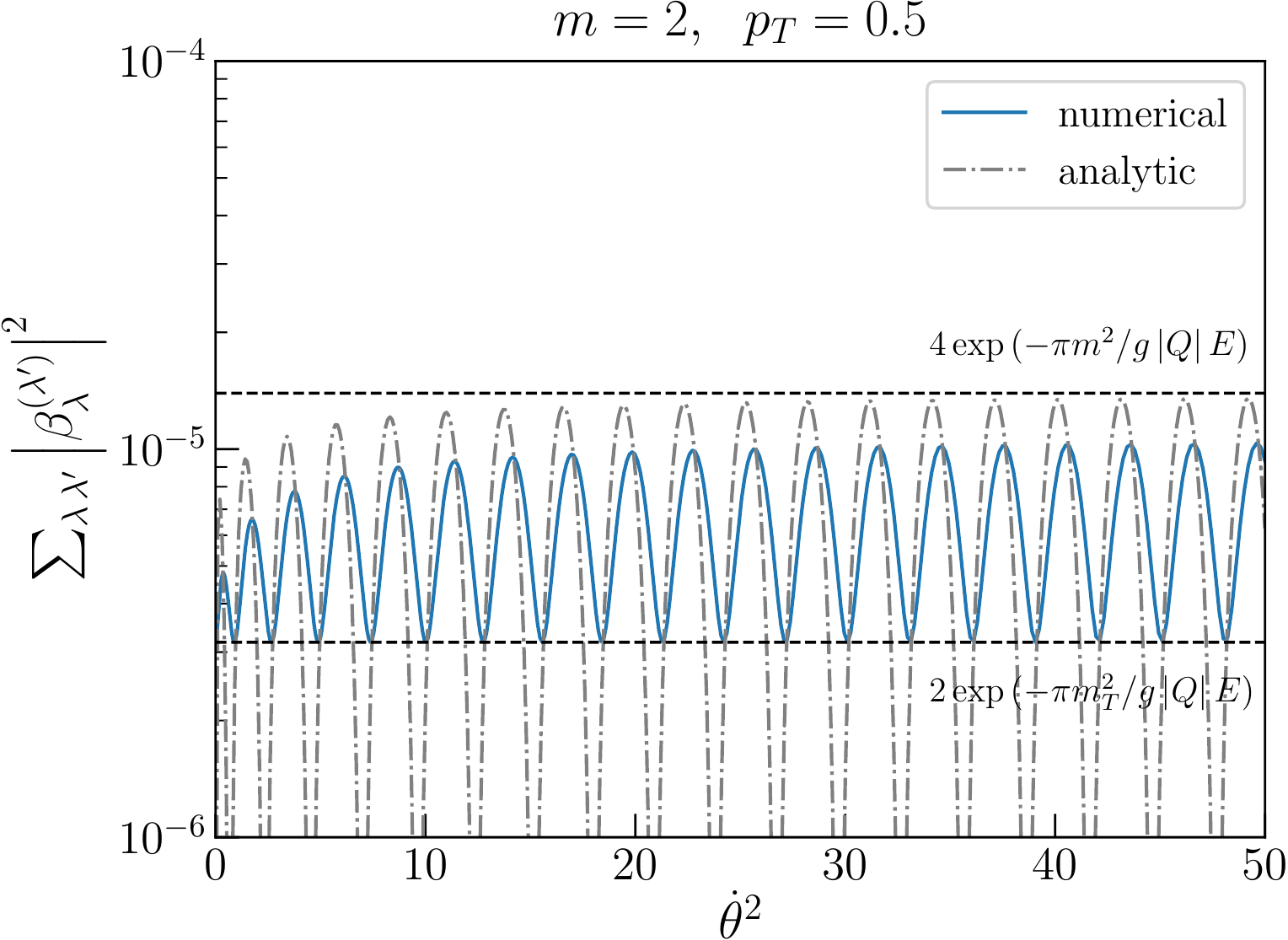}
	\caption{\small As Fig.~\ref{fig:saddle_pt_large}
	but with different model parameters, {showing the limitations of the analytical expression~\eqref{eq:approx_full_system_app} for $m^2, p_T^2 < g |Q| E$.}}
	\label{fig:saddle_pt_small}
\end{figure}

\paragraph{Empirical analytical formula.}
Unfortunately, we could not evaluate the expression~\eqref{eq:adiabatic_approx} analytically in the case of our interest.
Nevertheless we can learn properties of our system from this expression.
In particular, we see that $R_\beta^0$ is obtained after 
convoluting the source term with the phase factor $e^{2i\int \dd t \lambda_n(t)}$.
If $\abs{\lambda_n}$ is large, the integral is suppressed since the integrand oscillates rapidly.
Thus, we expect that the smallest eigenvalue is the most important (see also App.~\ref{app:WKB}).
One can see that the smallest eigenvalue for a sizable $\dot{\theta}_{5+m}$ is given by
\begin{align}
	\Omega^\pm(t) = \sqrt{\left(\sqrt{\Pi_z^2 + p_T^2}\pm \dot{\theta}_{5+m}\right)^2 + m^2},
\end{align}
if we ignore the terms proportional to $\dot{\Pi}_z$ in $M$,
where we take $\Omega^-$ for $\dot{\theta}_{5+m} > 0$ and $\Omega^+$ for $\dot{\theta}_{5+m} < 0$.
If this mode is the most important, we may expect that $R_\beta^0$ is related to the following integral:
\begin{align}
	I = \int_{-\infty}^{\infty} \dd t\,f(t) \exp\left[2i\int \dd t' \Omega^-(t')\right],
\end{align}
with some function $f(t)$, where we assume $\dot{\theta}_{5+m} > 0$ to be specific. 
This integral can be evaluated by the saddle point approximation.
The saddle points of the phase factor are given by
\begin{align}
	\Pi_z = \Pi_{\sigma \sigma'},
	\quad
	\Pi_{\sigma \sigma'} = \sigma \sqrt{\left(\dot{\theta}_{5+m} + \sigma' im\right)^2 - p_T^2},
\end{align}
where $\sigma, \sigma' = \pm$.
We take the saddle points in the upper half plane, and then the integral is given by
\begin{align}
	I &\sim \exp\left[2i\int^{\Pi_{++}}\frac{\dd \Pi_z}{g\abs{Q}E}\,\Omega^-\right] 
	+ \exp\left[2i\int^{\Pi_{--}}\frac{\dd\Pi_z}{g\abs{Q}E}\,\Omega^-\right],
\end{align}
where we changed the integral variable from $t$ to $\Pi_z$, 
and we ignored all the prefactors.
Since $R^0_\beta$ is positive definite, we may expect that
it is given by the integral squared as
\begin{align}
	2R^0_\beta &\simeq \abs{\exp\left[2i\int^{\Pi_{++}}_0\frac{\dd\Pi_z}{g\abs{Q}E}\,\Omega^-\right]}^2
	+ \abs{\exp\left[2i\int^{\Pi_{--}}_0\frac{\dd\Pi_z}{g\abs{Q}E}\,\Omega^-\right]}^2
	- 2\,\mathrm{Re}\left[\exp\left[2i\int^{\Pi_{++}}_{\Pi_{-+}} \frac{\dd\Pi_z}{g\abs{Q}E}\,\Omega^- \right]\right],
	\label{eq:approx_full_system_app}
\end{align}
where we chose the minus sign in the interference term
simply because it describes the numerical results well.\footnote{
See also Ref.~\cite{Dumlu:2010ua}, which traces the minus sign in the interference term of back to the fermionic nature of the produced particles in the context of the dynamically assisted Schwinger mechanism.
}
We show in Figs.~\ref{fig:woB} and~\ref{fig:HLL} that this formula
reproduces the full numerical results extraordinary well for several model parameters.
Since our setup only contains fairly few parameters, we can empirically gain confidence in the expression~\eqref{eq:approx_full_system_app} by systematically varying all these parameters and comparing with the full numerical result. 
For large $m$ and $p_T$, we find excellent agreement for the height of the plateau of the spectrum, despite the rather heuristic approach, as depicted in Fig.~\ref{fig:saddle_pt_large} for a variety of model parameters.
Eq.~\eqref{eq:approx_full_system_app} deviates from the full results when $m$ and/or $p_T$ is small
as shown in Fig.~\ref{fig:saddle_pt_small}.
We note that the case of small $p_T$ is however not particularly relevant for our discussion since the axion assisted Schwinger effect is most prominent when $p_T$ is large.
It would be certainly interesting if one could derive an analytical formula that works 
for both small and large values of $m$ and $p_T$,
which we leave as a future work.

\paragraph{Small mass/transverse momentum limit.}
\begin{figure}[t]
	\centering
 	\includegraphics[width=0.495\linewidth]{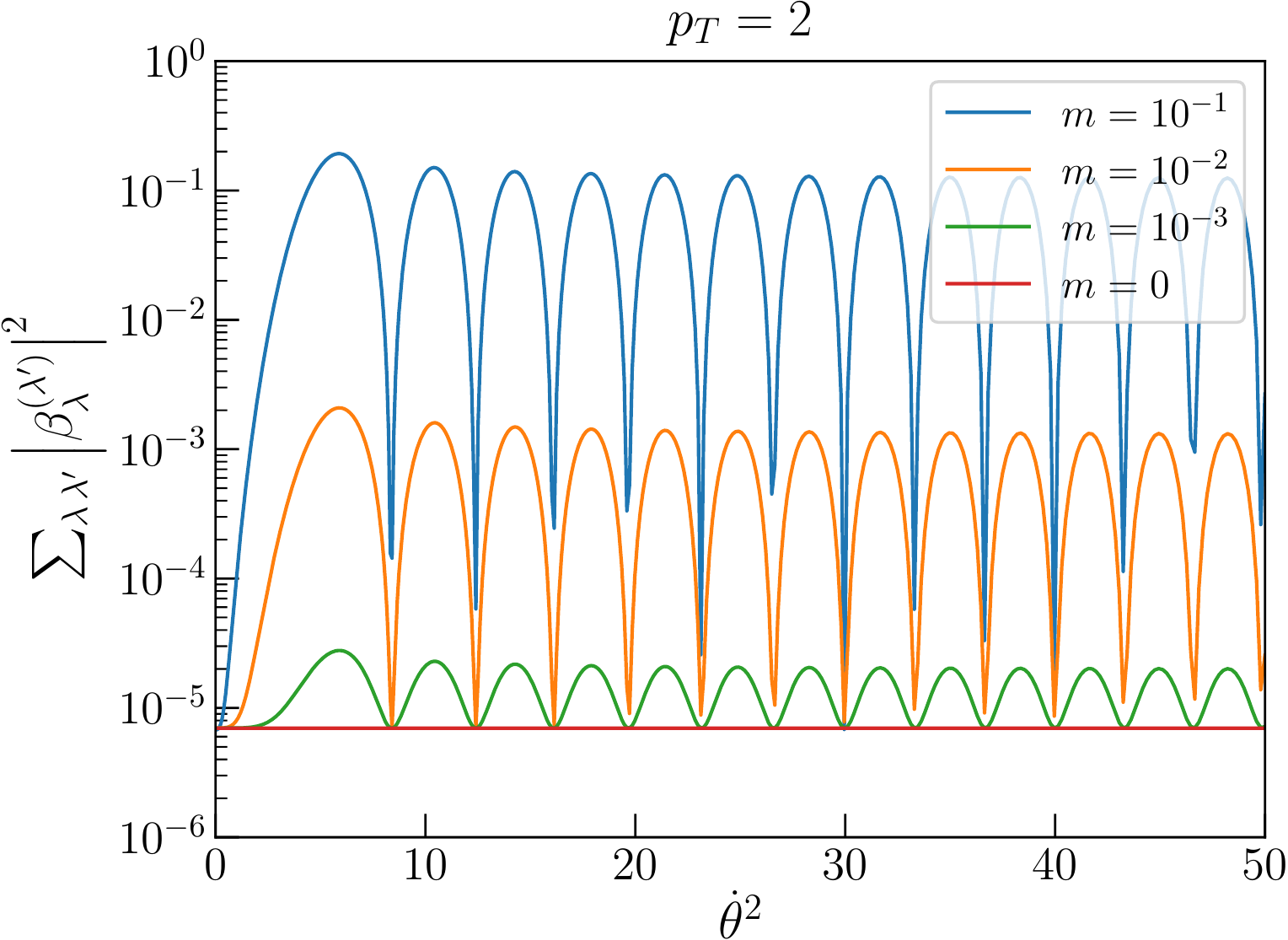}
 	\includegraphics[width=0.495\linewidth]{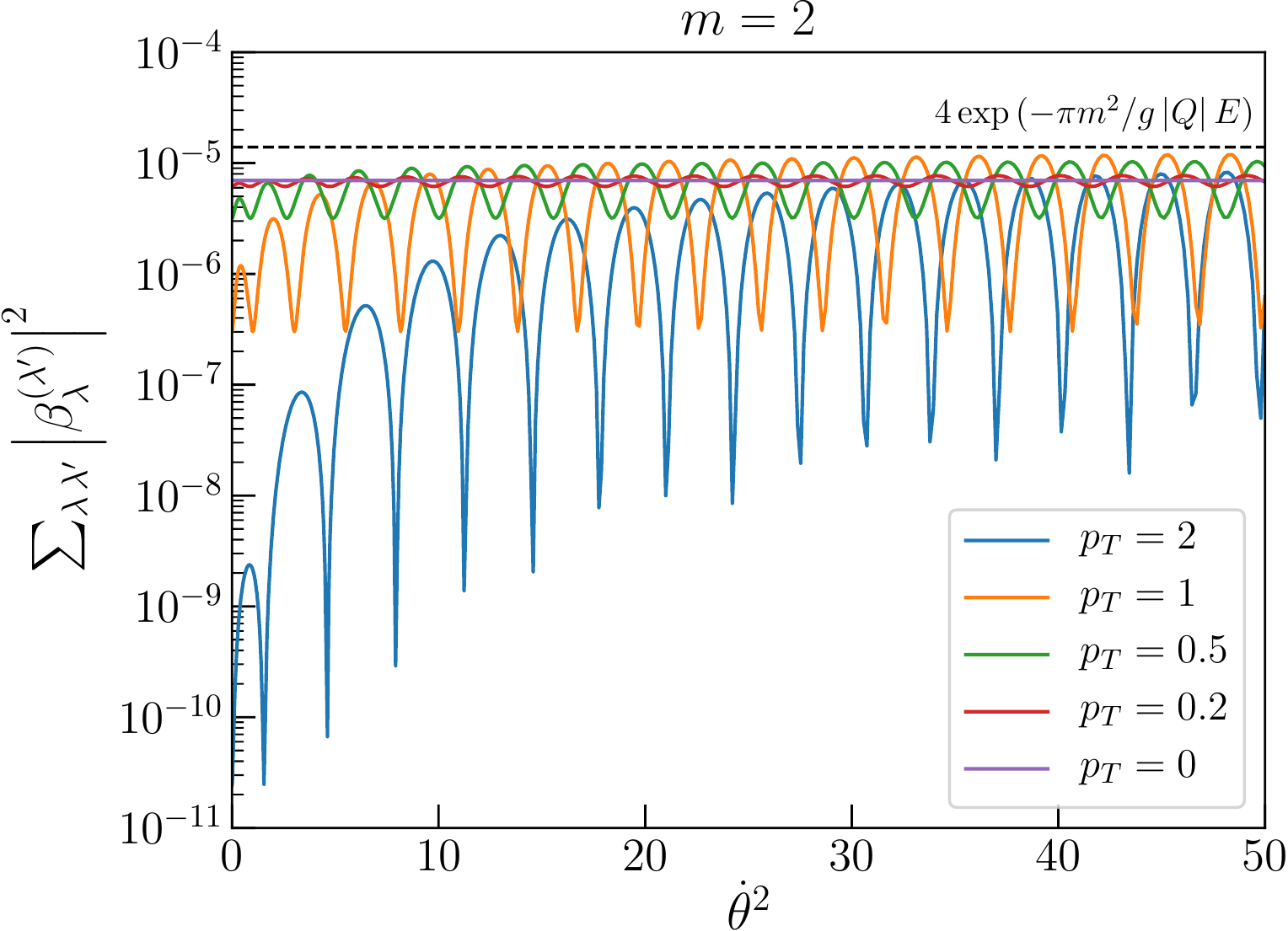}
	\caption{\small The height of the plateau of the spectrum 
	of $\sum_{\lambda \lambda'} |\beta_\lambda^{(\lambda')}|^2$ evaluated at $p_z = -50$
	as a function of $\dot{\theta}$ for small values of $m$ (left) and $p_T$ (right).
	The parameters are shown in the unit $g\abs{Q}E = 1$.}
	\label{fig:numerics_mpT_small}
\end{figure}
We finally comment on the limit $m\rightarrow 0$.
Our numerical result shows that the enhancement becomes less significant
and eventually dies out as $m$ gets smaller, as shown in the left panel of Fig.~\ref{fig:numerics_mpT_small}.
{In particular, we observe a scaling well approximated as  $\sum_{\lambda \lambda'} |\beta_\lambda^{(\lambda')}|^2 \propto m^2/(g |Q| E)$ for $m^2 \ll g |Q| E$.}
Since $\theta_{5+m}$ is unphysical when $m = 0$, this result is consistent with a continuous $m \rightarrow$ limit.

We also comment on the limit $p_T \rightarrow 0$.
As depicted in the right panel of Fig.~\ref{fig:numerics_mpT_small},
the axion assisted Schwinger effect again becomes less significant and eventually dies out in this limit.
Although our formula~\eqref{eq:approx_full_system_app} does not reproduce the numerical results accurately
in this limit as we mentioned above, 
the deviation is at most factor two simply because the suppression due to $p_T$ becomes irrelevant in this limit.

\section{Non-relativistic effective field theory}
\label{app:nreft}

\paragraph{Effective action.}
As demonstrated in the main text, Schwinger pair-production sourced by $\dot \phi$ is enhanced when the mass term cannot be neglected. This implies that it should be also understood in non-relativistic effective field theory of fermions.
Let us take the Dirac basis for the $\gamma$ matrices in this section (and this section only) as they are convenient in the non-relativistic limit:
\begin{align}
	\gamma^0 = \begin{pmatrix}
             1 & 0 \\ 0 &  -1 
            \end{pmatrix} \,, \quad
    \bm{\gamma} = \begin{pmatrix}
             0 & \bm{\sigma} \\ -\bm{\sigma} &  0 
            \end{pmatrix} \,, \quad
	\gamma_5 = \begin{pmatrix}
             0 & 1 \\ 1 &  0 
            \end{pmatrix} \, .
\end{align}
In this basis, the upper and lower components of fermion fields stand for particle and anti-particle:
\begin{align}
	\psi (x) = e^{- i \theta_m (t) \gamma_5} 
	\begin{pmatrix}
             e^{- i m t} \eta (x) \\ e^{i m t} \xi^\dag (x)
            \end{pmatrix} \,.
\end{align}
We have factored out the fast oscillation coming from the mass term and $\eta$ ($\xi$) represents a slowly varying field for the (anti-)particle.
One may derive the non-relativistic effective action of Eq.~\eqref{eq:action_FRW} in terms of $\eta$ and $\xi$ by performing the expansion of $1/m$ systematically, which reads
\begin{align}
	S = \int \dd^4 x\, 
		\left[ \eta^\dag i D_0 \eta 
		- \frac{1}{2m} \eta^\dag \left( i \bm{D} \cdot \bm{\sigma} + \dot \theta_{5 + m} \right)^2 \eta
		+ 
		\xi i D_0 \xi^\dag
		+ \frac{1}{2m} \xi \left( i \bm{D} \cdot \bm{\sigma} + \dot \theta_{5 + m} \right)^2 \xi^\dag \right] + \mathcal{O} (1/m^2).
\end{align}
Here we have assumed that the axion field is homogeneous and slowly varying compared to $1/m$.
Notice that the effective action solely depends on the particular combination $\theta_{5 + m}$ signaling the basis independence under rotations of the fermion fields.

The second (and correspondingly the fourth) term can be rewritten as follows
\begin{align}
	\frac{1}{2m} \eta^\dag \left( i \bm{D} \cdot \bm{\sigma} + \dot \theta_{5 + m} \right)^2 \eta
	= \frac{1}{2m} \eta^\dag \left( \bm{\Pi}^2 - g Q \bm{B} \cdot \bm{\sigma} - 2 \dot \theta_{5 + m} \bm{\Pi} \cdot \bm{\sigma} + \dot \theta_{5 + m}^2 \right) \eta \, ,
\end{align}
where the gauge invariant momentum is defined by $\bm{\Pi} = - i \bm{D} = - i \bm{\nabla} - g Q \bm{A}$.
The first two terms are the ordinary kinetic term and the magnetic dipole interaction, respectively.
The other terms depend on $\dot \theta_{5 + m}$.
In particular, the third term provides the spin-momentum interaction induced by $\dot \theta_{5+m}$.
This operator reduces (increases) the energy of a particle for a given $\Pi = \abs{\bm{\Pi}}$ if its spin is (anti-)parallel to $\bm{\Pi}$.
The same argument holds for the anti-particle $\xi$.
As a result, we have two modes in total that are produced more efficiently than in the case without $\dot \theta_{5+m}$.

\paragraph{Dispersion relation.}
Let us first derive the dispersion relation explicitly without the magnetic field, \textit{i.e.}, $A^\mu = (0, 0, 0, A_z)$.
The case with the magnetic field parallel to the electric field will be mentioned shortly after.
The equation of motion in this case reads
\begin{align}
	0 = \left[ i \partial_0 \mathbb{1}
	- \frac{1}{2m} \begin{pmatrix}
		\left( \Pi_z - \dot \theta_{5+m} \right)^2 + p_T^2 & 
		- 2 \dot \theta_{5 + m} p_T e^{- i \varphi_p } \\
		- 2 \dot \theta_{5 + m} p_T e^{i \varphi_p }  &
		\left( \Pi_z + \dot \theta_{5+m} \right)^2 + p_T^2
	\end{pmatrix}
	\right]
	\begin{pmatrix}
		\eta^+_{\bm{p}} \\ \eta^-_{\bm{p}}
	\end{pmatrix},
\end{align}
with $\varphi_p = \arctan (p_y / p_x)$.
Here we expand the mode as 
\begin{align}
	\eta_{\bm{p}} = \sum_{\sigma = \pm } \eta_{\bm{p}}^\sigma S^z_\sigma,\quad
	\sigma_z S^z_\pm = \pm S^z_\pm.
\end{align}
One may rotate the basis along the $z$-axis so that the phase $\varphi_p$ disappears, \textit{i.e.},
\begin{align}
	0 = \left[ i \partial_0 \mathbb{I}_2
	- \frac{1}{2m} \begin{pmatrix}
		\left( \Pi_z - \dot \theta_{5+m} \right)^2 + p_T^2 & 
		- 2 \dot \theta_{5 + m} p_T \\
		- 2 \dot \theta_{5 + m} p_T  &
		\left( \Pi_z + \dot \theta_{5+m} \right)^2 + p_T^2
	\end{pmatrix}
	\right]
	\begin{pmatrix}
		\tilde \eta^+_{\bm{p}} \\ \tilde \eta^-_{\bm{p}}
	\end{pmatrix},
\end{align}
where $\tilde \eta$ represents a field in the new basis.
For a constant $\Pi_z$, one may solve this equation with two frequencies:
\begin{align}
	\tilde \Omega^\pm = \frac{1}{2 m} \left( \sqrt{\Pi_z^2 + p_T^2} \pm \dot \theta_{5+m} \right)^2.
\end{align}
Now it is clear that the production of the mode with $\tilde \Omega^\pm$ is enhanced for $\dot \theta_{5+m} \lessgtr 0$, since the gap in the corresponding dispersion relation is reduced.
It is straightforward to obtain the same dispersion relation for the anti-particle.

\paragraph{Including the magnetic field.} Once we turn on a magnetic field parallel to electric field, the momentum transverse to the electromagnetic field becomes discretized as the Landau levels.
The equation of motion is obtained by just replacing $m_T$ with $m_B = \sqrt{2n g |Q B|}$ as
\begin{align}
	0 = \left[ i \partial_0 \mathbb{I}_2
	- \frac{1}{2m} \begin{pmatrix}
		\left( \Pi_z - \dot \theta_{5+m} \right)^2 + m_B^2 & 
		- 2 \dot \theta_{5 + m} m_B \\
		- 2 \dot \theta_{5 + m} m_B  &
		\left( \Pi_z + \dot \theta_{5+m} \right)^2 + m_B^2
	\end{pmatrix}
	\right]
	\begin{pmatrix}
		\eta^+_{n} \\ \eta^-_{n-1}
	\end{pmatrix}.
\end{align}
Note that the mode expansion is also modified as
\begin{align}
	\eta_{p_y, p_z} (x) = \sum_{n, \sigma} \eta_n^\sigma h_n S^z_\sigma .
\end{align}
The dispersion relation is now given by
\begin{align}
	\tilde \Omega^\pm_n = 
	\frac{1}{2 m} \left( \sqrt{\Pi_z^2 + m_B^2} \pm \dot \theta_{5+m} \right)^2.
\end{align}
In the same way as the case without the magnetic field, the production of the mode with $\Omega^\pm_n$ is enhanced for $\dot \theta_{5+m} \lessgtr 0$ respectively.

\section{Phase integral method}
\label{app:WKB}

The exponential suppression of Schwinger particle production can be elegantly understood using the phase integral method for quantum mechanical scattering problems (see e.g.~\cite{White}), based on the WKB approximation. Here we briefly review the essence of this method, as it provides useful intuition for understanding the role of the gap in the dispersion relations between particles and antiparticles, as well as for understanding interference effects between several (complex) zero-points of the dispersion relation. See Refs.~\cite{Dumlu:2010ua,Dumlu:2011rr,Dabrowski:2014ica} and references therein for derivations and applications to particle production in scalar and spinor QED.

The computational problem at the heart of particle production in electromagnetic fields can be reduced to solving a harmonic oscillator equation with a time-dependent frequency,
\begin{align}
 \ddot \psi(t) + \Omega^2(t)\,  \psi(t) = 0 \,.
 \label{eq:app_PI_1}
\end{align}
Starting from the Dirac equation, a second order differential equation of the type of Eq.~\eqref{eq:app_PI_1} is obtained by acting with the conjugate differential operator. 
In the WKB approximation, the solution is given by
\begin{equation}
 \psi_k(t) = \frac{\alpha(t)}{\sqrt{2 \Omega}} e^{- i \int \Omega \dd t} + \frac{\beta(t)}{\sqrt{2 \Omega}} e^{i \int \Omega \dd t} \,,
\end{equation}
with the Bogoliubov coefficients $\alpha$ and $\beta$. Determining the asymptotic value $\beta(t \rightarrow \infty)$ for the initial condition $\beta(t \rightarrow - \infty) = 0$ thus requires solving an integral of the type
\begin{align}
 {\cal I}_\pm = \int \dd t \, {\cal C}(t) e^{\mp i \int \dd t'' \Omega(t'')}\,,
 \label{eq:app_PI_2}
\end{align}
with the coefficient ${\cal C}(t)$ determined by the coupled first-order ODEs governing the evolution of the Bogoliubov coefficients. 
This integral can be evaluated by choosing a convenient contour in the complex plane, noting that for $\Omega(t'')$ real (which is e.g.\ typically the case along the real axis), the integrand of the outer integral becomes a highly oscillatory function and hence the contribution to $\cal I_\pm$ is negligible. Moreover, for $i \Omega$ is real and positive, the integrand of ${\cal I}_+$ (${\cal I}_-$) is exponentially suppressed. This qualitatively explains why to good approximation, Eq.~\eqref{eq:app_PI_2} can be evaluated by integrating along the Stokes line\footnote{
The Stokes line is a path in the complex $t$ plane along which $\int \Omega \dd t$ is imaginary. In practice, as long as on pays careful attention to potential branch cuts, on may equally well integrate on a contour parallel to the imaginary axis.
}
connecting the pair of complex zeroes of $\Omega(t)$,
\begin{equation}
 \beta \sim \exp\left( - i \int_{t_0}^{t^*_0} \Omega(t) \dd t \right)  = \exp\left(- \frac{i}{\dot \Pi_z} \int_{\Pi_{z,0}}^{\Pi_{z,0}^*} \Omega(\Pi_z) \dd \Pi_z\right) \quad \text{with} \quad \Omega(\Pi_{z,0}) = \Omega(\Pi_{z,0}^*) = 0 \,,
 \label{eq:app_PI_3}
\end{equation}
where for a constant electric field $\dot \Pi_z = g Q E$.
For gapped dispersion relation, i.e.\ $\Omega(p) > 0$ for all $p \in \mathbb{R}$, the distance $ 2 \text{Im}(\Pi_{z,0})$ between the pair of complex zeroes corresponds to the minimal energy gap which needs to be overcome for the production of a particle from the Dirac sea.

\paragraph{Schwinger production for $\bf{c_5 + c_m = 0}$.} Consider $\Omega = \sqrt{\Pi_z^2 + m_T^2}$, with the zero-point $\Pi_{z,0} = i m_T$. A direct integration yields $ \int_{\Pi_{z,0}}^{\Pi_{z,0}^*} \Omega(\Pi_z) \dd \Pi_z = \pi m_T^2/2$ and hence $|\beta|^2 \sim \exp( - \pi m_T^2/(g |Q| E))$. This reproduces the result for particle production in helical electromagnetic fields obtained in Ref.~\cite{Domcke:2019qmm}.

For the case $c_5 + c_m \ne 0$ considered in the main part of this paper, an additional subtlety is that the eigenvalues of the Dirac equation (in the basis that diagonalizes the Hamiltonian~\eqref{eq:H_app}) do not coincide with the eigenvalues of the equations of motion for the Bogoliubov coefficients derived in App.~\ref{app:bilinear}. However, once we have arrived at the set of coupled equations of motion \eqref{eq:v-eom}, we can again interpret them as coupled oscillators with time-dependent frequencies, similar to Eq.~\eqref{eq:app_PI_1}. Consequently, the particle production is governed by an expression similar to Eq.~\eqref{eq:app_PI_3}, with $\Omega$ replaced by the smallest eigenfrequency of the system.

\paragraph{Interference effects.} In general (and in particular in the situation in the main part of this paper), the function $\Omega(\Pi_z)$ can have multiple complex zero points. Intuitively, this can be understood by recasting  Eq.~\eqref{eq:app_PI_1} as a quantum mechanical scattering problem over a time-dependent potential barrier with a potentially complicated shape~\cite{Dumlu:2010ua}. Partial reflections of different parts of this barrier can lead to interference effects. Eq.~\eqref{eq:app_PI_3} is then modified to~\cite{Dumlu:2011rr},
\begin{align}
 \beta \sim \sum_i  \exp(2 i \theta_i) \exp\left(- \frac{i}{\dot \Pi_z} \int_{\Pi^i_{z,0}}^{(\Pi^i_{z,0})^*} \Omega(\Pi_z) d\Pi_z\right)\,,
\end{align}
with the index $i$ labelling the pairs of complex zeroes and $\theta_i = \dot \Pi_z^{-1} \int_{\text{Re}(\Pi^1_{z,0})}^{\text{Re}(\Pi^i_{z,0})} \Omega(\Pi_z) d\Pi_z$ accounting for the phase accumulated by the integration along the real axis. This leads to interference effects in the final result for $|\beta|^2$.

\small
\bibliographystyle{utphys}
\bibliography{refs}

\providecommand{\href}[2]{#2}\begingroup\raggedright\begin{thebibliography}{10}

\bibitem{Heisenberg:1935qt}
W.~Heisenberg and H.~Euler, ``{Consequences of Dirac's theory of positrons},''
  \href{http://dx.doi.org/10.1007/BF01343663, 10.1007/978-3-642-70078-1_9}{{\em
  Z. Phys.} {\bfseries 98} no.~11-12, (1936) 714--732},
\href{http://arxiv.org/abs/physics/0605038}{{\ttfamily arXiv:physics/0605038
  [physics]}}.

\bibitem{Schwinger:1951nm}
J.~S. Schwinger, ``{On gauge invariance and vacuum polarization},''
  \href{http://dx.doi.org/10.1103/PhysRev.82.664}{{\em Phys. Rev.} {\bfseries
  82} (1951) 664--679}.
[,116(1951)].

\bibitem{Peccei:1977hh}
R.~Peccei and H.~R. Quinn, ``{CP Conservation in the Presence of Instantons},''
  \href{http://dx.doi.org/10.1103/PhysRevLett.38.1440}{{\em Phys. Rev. Lett.}
  {\bfseries 38} (1977) 1440--1443}.

\bibitem{Peccei:1977ur}
R.~Peccei and H.~R. Quinn, ``{Constraints Imposed by CP Conservation in the
  Presence of Instantons},''
  \href{http://dx.doi.org/10.1103/PhysRevD.16.1791}{{\em Phys. Rev. D}
  {\bfseries 16} (1977) 1791--1797}.

\bibitem{Weinberg:1977ma}
S.~Weinberg, ``{A New Light Boson?},''
  \href{http://dx.doi.org/10.1103/PhysRevLett.40.223}{{\em Phys. Rev. Lett.}
  {\bfseries 40} (1978) 223--226}.

\bibitem{Wilczek:1977pj}
F.~Wilczek, ``{Problem of Strong $P$ and $T$ Invariance in the Presence of
  Instantons},'' \href{http://dx.doi.org/10.1103/PhysRevLett.40.279}{{\em Phys.
  Rev. Lett.} {\bfseries 40} (1978) 279--282}.

\bibitem{Pospelov:2008jk}
M.~Pospelov, A.~Ritz, and M.~B. Voloshin, ``{Bosonic super-WIMPs as keV-scale
  dark matter},'' \href{http://dx.doi.org/10.1103/PhysRevD.78.115012}{{\em
  Phys. Rev. D} {\bfseries 78} (2008) 115012},
  \href{http://arxiv.org/abs/0807.3279}{{\ttfamily arXiv:0807.3279 [hep-ph]}}.

\bibitem{Stadnik:2013raa}
Y.~Stadnik and V.~Flambaum, ``{Axion-induced effects in atoms, molecules, and
  nuclei: Parity nonconservation, anapole moments, electric dipole moments, and
  spin-gravity and spin-axion momentum couplings},''
  \href{http://dx.doi.org/10.1103/PhysRevD.89.043522}{{\em Phys. Rev. D}
  {\bfseries 89} no.~4, (2014) 043522},
  \href{http://arxiv.org/abs/1312.6667}{{\ttfamily arXiv:1312.6667 [hep-ph]}}.

\bibitem{Graham:2013gfa}
P.~W. Graham and S.~Rajendran, ``{New Observables for Direct Detection of Axion
  Dark Matter},'' \href{http://dx.doi.org/10.1103/PhysRevD.88.035023}{{\em
  Phys. Rev. D} {\bfseries 88} (2013) 035023},
  \href{http://arxiv.org/abs/1306.6088}{{\ttfamily arXiv:1306.6088 [hep-ph]}}.

\bibitem{Turner:1987bw}
M.~S. Turner and L.~M. Widrow, ``{Inflation Produced, Large Scale Magnetic
  Fields},'' \href{http://dx.doi.org/10.1103/PhysRevD.37.2743}{{\em Phys. Rev.
  D} {\bfseries 37} (1988) 2743}.

\bibitem{Garretson:1992vt}
W.~Garretson, G.~B. Field, and S.~M. Carroll, ``{Primordial magnetic fields
  from pseudoGoldstone bosons},''
  \href{http://dx.doi.org/10.1103/PhysRevD.46.5346}{{\em Phys. Rev. D}
  {\bfseries 46} (1992) 5346--5351},
  \href{http://arxiv.org/abs/hep-ph/9209238}{{\ttfamily arXiv:hep-ph/9209238}}.

\bibitem{Anber:2006xt}
M.~M. Anber and L.~Sorbo, ``{N-flationary magnetic fields},''
  \href{http://dx.doi.org/10.1088/1475-7516/2006/10/018}{{\em JCAP} {\bfseries
  10} (2006) 018}, \href{http://arxiv.org/abs/astro-ph/0606534}{{\ttfamily
  arXiv:astro-ph/0606534}}.

\bibitem{Nikishov:1969tt}
A.~I. Nikishov, ``{Pair production by a constant external field},''
{\em Zh. Eksp. Teor. Fiz.} {\bfseries 57} (1969) 1210--1216.

\bibitem{Bunkin:1969if}
F.~V. Bunkin and I.~I. Tugov, ``{The possibility of electron-positron pair
  production in vacuum when laser radiation is focussed},''
{\em Dokl. Akad. Nauk Ser. Fiz.} {\bfseries 187} (1969) 541--544.

\bibitem{Adler:1969gk}
S.~L. Adler, ``{Axial vector vertex in spinor electrodynamics},''
  \href{http://dx.doi.org/10.1103/PhysRev.177.2426}{{\em Phys. Rev.} {\bfseries
  177} (1969) 2426--2438}.
[,241(1969)].

\bibitem{Bell:1969ts}
J.~S. Bell and R.~Jackiw, ``{A PCAC puzzle: $\pi^0 \to \gamma \gamma$ in the
  $\sigma$ model},''
\href{http://dx.doi.org/10.1007/BF02823296}{{\em Nuovo Cim.} {\bfseries A60}
  (1969) 47--61}.

\bibitem{Nielsen:1983rb}
H.~B. Nielsen and M.~Ninomiya, ``{ADLER-BELL-JACKIW ANOMALY AND WEYL FERMIONS
  IN CRYSTAL},'' \href{http://dx.doi.org/10.1016/0370-2693(83)91529-0}{{\em
  Phys. Lett. B} {\bfseries 130} (1983) 389--396}.

\bibitem{Warringa:2012bq}
H.~J. Warringa, ``{Dynamics of the Chiral Magnetic Effect in a weak magnetic
  field},'' \href{http://dx.doi.org/10.1103/PhysRevD.86.085029}{{\em Phys. Rev.
  D} {\bfseries 86} (2012) 085029},
  \href{http://arxiv.org/abs/1205.5679}{{\ttfamily arXiv:1205.5679 [hep-th]}}.

\bibitem{Domcke:2018eki}
V.~Domcke and K.~Mukaida, ``{Gauge Field and Fermion Production during Axion
  Inflation},'' \href{http://dx.doi.org/10.1088/1475-7516/2018/11/020}{{\em
  JCAP} {\bfseries 11} (2018) 020},
  \href{http://arxiv.org/abs/1806.08769}{{\ttfamily arXiv:1806.08769
  [hep-ph]}}.

\bibitem{Copinger:2018ftr}
P.~Copinger, K.~Fukushima, and S.~Pu, ``{Axial Ward identity and the Schwinger
  mechanism -- Applications to the real-time chiral magnetic effect and
  condensates},'' \href{http://dx.doi.org/10.1103/PhysRevLett.121.261602}{{\em
  Phys. Rev. Lett.} {\bfseries 121} no.~26, (2018) 261602},
  \href{http://arxiv.org/abs/1807.04416}{{\ttfamily arXiv:1807.04416
  [hep-th]}}.

\bibitem{Domcke:2019qmm}
V.~Domcke, Y.~Ema, and K.~Mukaida, ``{Chiral Anomaly, Schwinger Effect,
  Euler-Heisenberg Lagrangian, and application to axion inflation},''
  \href{http://dx.doi.org/10.1007/JHEP02(2020)055}{{\em JHEP} {\bfseries 02}
  (2020) 055}, \href{http://arxiv.org/abs/1910.01205}{{\ttfamily
  arXiv:1910.01205 [hep-ph]}}.

\bibitem{Schutzhold:2008pz}
R.~Schutzhold, H.~Gies, and G.~Dunne, ``{Dynamically assisted Schwinger
  mechanism},'' \href{http://dx.doi.org/10.1103/PhysRevLett.101.130404}{{\em
  Phys. Rev. Lett.} {\bfseries 101} (2008) 130404},
  \href{http://arxiv.org/abs/0807.0754}{{\ttfamily arXiv:0807.0754 [hep-th]}}.

\bibitem{Dumlu:2010ua}
C.~K. Dumlu and G.~V. Dunne, ``{The Stokes Phenomenon and Schwinger Vacuum Pair
  Production in Time-Dependent Laser Pulses},''
  \href{http://dx.doi.org/10.1103/PhysRevLett.104.250402}{{\em Phys. Rev.
  Lett.} {\bfseries 104} (2010) 250402},
  \href{http://arxiv.org/abs/1004.2509}{{\ttfamily arXiv:1004.2509 [hep-th]}}.

\bibitem{Freese:1990rb}
K.~Freese, J.~A. Frieman, and A.~V. Olinto, ``{Natural inflation with pseudo -
  Nambu-Goldstone bosons},''
  \href{http://dx.doi.org/10.1103/PhysRevLett.65.3233}{{\em Phys. Rev. Lett.}
  {\bfseries 65} (1990) 3233--3236}.

\bibitem{Nomura:2017ehb}
Y.~Nomura, T.~Watari, and M.~Yamazaki, ``{Pure Natural Inflation},''
  \href{http://dx.doi.org/10.1016/j.physletb.2017.11.052}{{\em Phys. Lett. B}
  {\bfseries 776} (2018) 227--230},
  \href{http://arxiv.org/abs/1706.08522}{{\ttfamily arXiv:1706.08522
  [hep-ph]}}.

\bibitem{Adshead:2012kp}
P.~Adshead and M.~Wyman, ``{Chromo-Natural Inflation: Natural inflation on a
  steep potential with classical non-Abelian gauge fields},''
  \href{http://dx.doi.org/10.1103/PhysRevLett.108.261302}{{\em Phys. Rev.
  Lett.} {\bfseries 108} (2012) 261302},
  \href{http://arxiv.org/abs/1202.2366}{{\ttfamily arXiv:1202.2366 [hep-th]}}.

\bibitem{Maleknejad:2012fw}
A.~Maleknejad, M.~Sheikh-Jabbari, and J.~Soda, ``{Gauge Fields and
  Inflation},'' \href{http://dx.doi.org/10.1016/j.physrep.2013.03.003}{{\em
  Phys. Rept.} {\bfseries 528} (2013) 161--261},
  \href{http://arxiv.org/abs/1212.2921}{{\ttfamily arXiv:1212.2921 [hep-th]}}.

\bibitem{Domcke:2018rvv}
V.~Domcke, B.~Mares, F.~Muia, and M.~Pieroni, ``{Emerging chromo-natural
  inflation},'' \href{http://dx.doi.org/10.1088/1475-7516/2019/04/034}{{\em
  JCAP} {\bfseries 04} (2019) 034},
  \href{http://arxiv.org/abs/1807.03358}{{\ttfamily arXiv:1807.03358
  [hep-ph]}}.

\bibitem{Domcke:2019lxq}
V.~Domcke and S.~Sandner, ``{The Different Regimes of Axion Gauge Field
  Inflation},'' \href{http://dx.doi.org/10.1088/1475-7516/2019/09/038}{{\em
  JCAP} {\bfseries 09} (2019) 038},
  \href{http://arxiv.org/abs/1905.11372}{{\ttfamily arXiv:1905.11372
  [astro-ph.CO]}}.

\bibitem{Domcke:2018gfr}
V.~Domcke, Y.~Ema, K.~Mukaida, and R.~Sato, ``{Chiral Anomaly and Schwinger
  Effect in Non-Abelian Gauge Theories},''
  \href{http://dx.doi.org/10.1007/JHEP03(2019)111}{{\em JHEP} {\bfseries 03}
  (2019) 111},
\href{http://arxiv.org/abs/1812.08021}{{\ttfamily arXiv:1812.08021 [hep-ph]}}.

\bibitem{Adshead:2018oaa}
P.~Adshead, L.~Pearce, M.~Peloso, M.~A. Roberts, and L.~Sorbo, ``{Phenomenology
  of fermion production during axion inflation},''
  \href{http://dx.doi.org/10.1088/1475-7516/2018/06/020}{{\em JCAP} {\bfseries
  1806} no.~06, (2018) 020},
\href{http://arxiv.org/abs/1803.04501}{{\ttfamily arXiv:1803.04501
  [astro-ph.CO]}}.

\bibitem{Cheng:2015oqa}
S.-L. Cheng, W.~Lee, and K.-W. Ng, ``{Numerical study of pseudoscalar inflation
  with an axion-gauge field coupling},''
  \href{http://dx.doi.org/10.1103/PhysRevD.93.063510}{{\em Phys. Rev. D}
  {\bfseries 93} no.~6, (2016) 063510},
  \href{http://arxiv.org/abs/1508.00251}{{\ttfamily arXiv:1508.00251
  [astro-ph.CO]}}.

\bibitem{Notari:2016npn}
A.~Notari and K.~Tywoniuk, ``{Dissipative Axial Inflation},''
  \href{http://dx.doi.org/10.1088/1475-7516/2016/12/038}{{\em JCAP} {\bfseries
  12} (2016) 038}, \href{http://arxiv.org/abs/1608.06223}{{\ttfamily
  arXiv:1608.06223 [hep-th]}}.

\bibitem{DallAgata:2019yrr}
G.~Dall'Agata, S.~Gonz\'alez-Mart\'\i{}n, A.~Papageorgiou, and M.~Peloso,
  ``{Warm dark energy},''
  \href{http://dx.doi.org/10.1088/1475-7516/2020/08/032}{{\em JCAP} {\bfseries
  08} (2020) 032}, \href{http://arxiv.org/abs/1912.09950}{{\ttfamily
  arXiv:1912.09950 [hep-th]}}.

\bibitem{Domcke:2020zez}
V.~Domcke, V.~Guidetti, Y.~Welling, and A.~Westphal, ``{Resonant backreaction
  in axion inflation},''
  \href{http://dx.doi.org/10.1088/1475-7516/2020/09/009}{{\em JCAP} {\bfseries
  09} (2020) 009}, \href{http://arxiv.org/abs/2002.02952}{{\ttfamily
  arXiv:2002.02952 [astro-ph.CO]}}.

\bibitem{Barnaby:2011qe}
N.~Barnaby, E.~Pajer, and M.~Peloso, ``{Gauge Field Production in Axion
  Inflation: Consequences for Monodromy, non-Gaussianity in the CMB, and
  Gravitational Waves at Interferometers},''
  \href{http://dx.doi.org/10.1103/PhysRevD.85.023525}{{\em Phys. Rev.}
  {\bfseries D85} (2012) 023525},
\href{http://arxiv.org/abs/1110.3327}{{\ttfamily arXiv:1110.3327
  [astro-ph.CO]}}.

\bibitem{Cook:2011hg}
J.~L. Cook and L.~Sorbo, ``{Particle production during inflation and
  gravitational waves detectable by ground-based interferometers},''
  \href{http://dx.doi.org/10.1103/PhysRevD.86.069901,
  10.1103/PhysRevD.85.023534}{{\em Phys. Rev.} {\bfseries D85} (2012) 023534},
  \href{http://arxiv.org/abs/1109.0022}{{\ttfamily arXiv:1109.0022
  [astro-ph.CO]}}.
[Erratum: Phys. Rev.D86,069901(2012)].

\bibitem{Anber:2012du}
M.~M. Anber and L.~Sorbo, ``{Non-Gaussianities and chiral gravitational waves
  in natural steep inflation},''
  \href{http://dx.doi.org/10.1103/PhysRevD.85.123537}{{\em Phys. Rev.}
  {\bfseries D85} (2012) 123537},
\href{http://arxiv.org/abs/1203.5849}{{\ttfamily arXiv:1203.5849
  [astro-ph.CO]}}.

\bibitem{Linde:2012bt}
A.~Linde, S.~Mooij, and E.~Pajer, ``{Gauge field production in supergravity
  inflation: Local non-Gaussianity and primordial black holes},''
  \href{http://dx.doi.org/10.1103/PhysRevD.87.103506}{{\em Phys. Rev. D}
  {\bfseries 87} no.~10, (2013) 103506},
  \href{http://arxiv.org/abs/1212.1693}{{\ttfamily arXiv:1212.1693 [hep-th]}}.

\bibitem{Garcia-Bellido:2016dkw}
J.~Garcia-Bellido, M.~Peloso, and C.~Unal, ``{Gravitational waves at
  interferometer scales and primordial black holes in axion inflation},''
  \href{http://dx.doi.org/10.1088/1475-7516/2016/12/031}{{\em JCAP} {\bfseries
  12} (2016) 031}, \href{http://arxiv.org/abs/1610.03763}{{\ttfamily
  arXiv:1610.03763 [astro-ph.CO]}}.

\bibitem{Domcke:2017fix}
V.~Domcke, F.~Muia, M.~Pieroni, and L.~T. Witkowski, ``{PBH dark matter from
  axion inflation},''
  \href{http://dx.doi.org/10.1088/1475-7516/2017/07/048}{{\em JCAP} {\bfseries
  07} (2017) 048}, \href{http://arxiv.org/abs/1704.03464}{{\ttfamily
  arXiv:1704.03464 [astro-ph.CO]}}.

\bibitem{Stadnik:2015upa}
Y.~Stadnik and V.~Flambaum, {\em {Manifestations of dark matter and variations
  of fundamental constants in atoms and astrophysical phenomena}}.
\newblock 9, 2015.
\newblock \href{http://arxiv.org/abs/1509.00966}{{\ttfamily arXiv:1509.00966
  [physics.atom-ph]}}.

\bibitem{Ringwald:2001cp}
A.~Ringwald, ``{Fundamental physics at an x-ray free electron laser},'' in {\em
  {Workshop on Electromagnetic Probes of Fundamental Physics}}, pp.~63--74.
\newblock 12, 2001.
\newblock \href{http://arxiv.org/abs/hep-ph/0112254}{{\ttfamily
  arXiv:hep-ph/0112254}}.

\bibitem{Dunne:2008kc}
G.~V. Dunne, ``{New Strong-Field QED Effects at ELI: Nonperturbative Vacuum
  Pair Production},'' \href{http://dx.doi.org/10.1140/epjd/e2009-00022-0}{{\em
  Eur. Phys. J. D} {\bfseries 55} (2009) 327--340},
  \href{http://arxiv.org/abs/0812.3163}{{\ttfamily arXiv:0812.3163 [hep-th]}}.

\bibitem{Tajima}
T.~Tajima, ``Prospects for extreme field science,''
  \href{http://dx.doi.org/10.1140/epjd/e2009-00107-8}{{\em Eur. Phys. J. D}
  {\bfseries 55} (2009) 519 -- 529}.

\bibitem{Berry:1984jv}
M.~V. Berry, ``{Quantal phase factors accompanying adiabatic changes},''
  \href{http://dx.doi.org/10.1098/rspa.1984.0023}{{\em Proc. Roy. Soc. Lond. A}
  {\bfseries 392} (1984) 45--57}.

\bibitem{White}
R.~White, ``{Phase Integral Methods},''.
  \url{https://w3.pppl.gov/~rwhite/wkb.pdf}.

\bibitem{Dumlu:2011rr}
C.~K. Dumlu and G.~V. Dunne, ``{Interference Effects in Schwinger Vacuum Pair
  Production for Time-Dependent Laser Pulses},''
  \href{http://dx.doi.org/10.1103/PhysRevD.83.065028}{{\em Phys. Rev. D}
  {\bfseries 83} (2011) 065028},
  \href{http://arxiv.org/abs/1102.2899}{{\ttfamily arXiv:1102.2899 [hep-th]}}.

\bibitem{Dabrowski:2014ica}
R.~Dabrowski and G.~V. Dunne, ``{Superadiabatic particle number in Schwinger
  and de Sitter particle production},''
  \href{http://dx.doi.org/10.1103/PhysRevD.90.025021}{{\em Phys. Rev. D}
  {\bfseries 90} no.~2, (2014) 025021},
  \href{http://arxiv.org/abs/1405.0302}{{\ttfamily arXiv:1405.0302 [hep-th]}}.

\end{thebibliography}\endgroup
  
\end{document}